\documentclass[12pt]{article}
\usepackage{epsf,epsfig,cite,amsmath,arydshln}

\graphicspath{{figs/}}

\input paperdef

\begin{document}
\begin{titlepage}
\pagestyle{empty}
\baselineskip=21pt
\begin{flushright}
CERN--PH--TH/2007-138\hfill
DCPT/07/102, IPPP/07/51\\
MPP--2007--117\hfill
UMN--TH--2615/07, FTPI--MINN--07/25 \\
arXiv:0709.0098 [hep-ph]
\end{flushright}
\vskip 0.2in
\begin{center}
{\Large\sc {\bf WMAP-Compliant Benchmark Surfaces for\\[.3em] 
                MSSM Higgs Bosons}}
\end{center}
\begin{center}
\vskip 0.05in
{{\bf J.~Ellis}$^1$,
{\bf T.~Hahn}$^2$,
{\bf S.~Heinemeyer}$^3$,
{\bf K.A.~Olive}$^{4}$
and {\bf G.~Weiglein}$^{5}$}\\
\vskip 0.05in
{\it
$^1${TH Division, Physics Department, CERN, Geneva, Switzerland}\\
$^2${Max-Planck-Institut f\"ur Physik, % (Werner-Heisenberg-Institut),\\ 
F\"ohringer Ring 6, D--80805 Munich, Germany} \\
$^3$Instituto de Fisica de Cantabria (CSIC-UC), 
Santander,  Spain \\
$^4${William I.\ Fine Theoretical Physics Institute,\\
University of Minnesota, Minneapolis, MN~55455, USA}\\
$^5${IPPP, University of Durham, Durham DH1~3LE, UK}\\
}
\vskip 0.1in
{\bf Abstract}
\end{center}
\baselineskip=18pt \noindent
%%%%%%%%%%%%%%%%%%%%%%%%%%%%%%%%%%%%%%%%%%%%%%%

{%\small
We explore `benchmark surfaces' suitable for studying the phenomenology
of Higgs bosons in the minimal supersymmetric extension of the Standard
Model (MSSM), which are chosen so that the supersymmetric relic density
is generally compatible with the range of cold dark matter density
preferred by WMAP and other observations. These benchmark surfaces are
specified assuming that gaugino masses $m_{1/2}$, soft trilinear
supersymmetry-breaking parameters $A_0$ and the soft
supersymmetry-breaking contributions $m_0$ to the squark and slepton
masses are universal, but not those associated with the Higgs multiplets
(the NUHM framework).  The benchmark surfaces may be presented as $(\MA,
\tb)$ planes with fixed or systematically varying values of the other
NUHM parameters, such as $m_0$, $m_{1/2}$, $A_0$ and the Higgs mixing
parameter $\mu$. We discuss the prospects for probing experimentally
these benchmark surfaces at the Tevatron collider, the LHC, the ILC, 
in $B$ physics and in direct dark-matter detection experiments. 
An Appendix documents developments in the {\tt FeynHiggs}
code that enable the user to explore for her/himself the WMAP-compliant
benchmark surfaces.} 

%%%%%%%%%%%%%%%%%%%%%%%%%%%%%%%%%%%%%%%%%%%%%%%%%%%

\vskip 0.15in
\leftline{CERN--PH--TH/2007-138}
\leftline{\today}
\end{titlepage}
\baselineskip=18pt

%%%%%%%%%%%%%%%%%%%%%%%%%%%%%%%%%%%%%%%%%%%%%%%%%%%

%%%%%%%%%%%%%%%%%%%%%%%%%%%%%%%%%%%%%%%%%%%%%%%%%%%%%%%%%%%%%%%%%%%%%%%%%%%%%%%
%%%%%%%%%%%%%%%%%%%%%%%%%%%%%%%%%%%%%%%%%%%%%%%%%%%%%%%%%%%%%%%%%%%%%%%%%%%%%%%

\section{Introduction}

Some of the best prospects for probing the minimal supersymmetric
extension of the Standard Model (MSSM)~\cite{susy,susy2} might be
offered by searches for 
the bosons appearing in its extended Higgs sector. It may be challenging
to distinguish between the lightest MSSM Higgs boson and a Standard
Model (SM) Higgs boson with the same mass, and searches for MSSM Higgs
bosons are, in many ways, complementary to searches for supersymmetric
particles as avenues to establish the existence of physics beyond the
SM.  

Searches at the Tevatron collider are closing in on the possible
existence of an SM-like Higgs boson over a limited range of low
masses~\cite{TevHiggsSM,CDFHiggsSM,D0HiggsSM}, 
and are also starting to encroach significantly on the options for
heavier  MSSM Higgs bosons, particularly at large
$\tb$~\cite{CDFHiggsMSSM,D0HiggsMSSM,CDFHiggsCharged,CDFHiggsMSSMnew,D0HiggsMSSMnew}.  
Studies have 
shown that experiments at the LHC will be able to establish the
existence or otherwise of an SM-like Higgs boson over all its possible
mass range, and also explore many options for the heavier MSSM Higgs
bosons~\cite{atlastdr,atlasrev,cms,CMS-TDR}. On the other hand, the LHC
might well be unable to distinguish 
between the lightest MSSM Higgs boson and an SM Higgs boson of the same
mass. The ILC would have better chances of making such a
distinction~\cite{teslatdr,orangebook,acfarep,Snowmass05Higgs,djouadi2,deschi,ehow2,asbs2}, 
and might also be able to produce the other MSSM Higgs bosons if they
are not too
heavy~\cite{teslatdr,orangebook,acfarep,Snowmass05Higgs,djouadi2}. CLIC would
also be able to study a light SM-like 
Higgs boson,  as well as extend the search for MSSM Higgs bosons to much
higher masses~\cite{clic}. Searches for new phenomena in $B$ physics, 
including rare decays such as $b \to s \gamma$, $B_s \to \mu^+ \mu^-$ and
$B_u \to \tau \nu$, also have good potential to explore the MSSM Higgs
sector and, at least in some specific MSSM scenarios, electroweak
precision observables (EWPO) may also provide interesting
constraints~\cite{PomssmRep,ehoww}. 
In parallel to these accelerator searches for MSSM Higgs bosons and
their effects, non-accelerator searches for supersymmetric dark
matter~\cite{CDMS,Xenon} 
will also be able to explore significant regions of the MSSM Higgs
parameter space~\cite{NUHMad,cdmDD,ehow5}, since the exchanges of massive MSSM
Higgs bosons have significant impacts on dark matter scattering cross
sections.  

In order to correlate the implications of searches at hadron colliders
and linear colliders, in $B$ physics, in dark matter searches and
elsewhere, it is desirable to define MSSM Higgs benchmark scenarios that
are suitable for comparing and assessing the relative scopes of
different search strategies, see, e.g.,
\citeres{benchmark,benchmark2,benchmark3,cpx,preSnowmass,sps,postLEP,bench3}.  

Since the MSSM Higgs sector is governed by the two parameters $\MA$ (or
$\MHp$) and $\tb$ at lowest order, aspects of MSSM Higgs-boson phenomenology
such as current exclusion bounds and the sensitivities of future searches
are usually displayed in
terms of these two parameters. The other MSSM parameters enter via
higher-order corrections, and are conventionally fixed according to certain
benchmark definitions~\cite{benchmark,benchmark2,benchmark3,cpx}.
The benchmark scenarios commonly used in the literature encompass a range of
different possibilities for the amount of mixing between the scalar top
quarks, which have
significant implications for MSSM Higgs phenomenology, and also include
the possibility of radiatively-induced $\cp$ violation. The best-known
example is the so-called ``$\mhmax$
scenario''~\cite{benchmark,benchmark2,benchmark3}, which allows the search
for the light $\cp$-even Higgs boson
to be translated into conservative bounds on $\tb$ for fixed values of the
top-quark mass and the scale of the supersymmetric
particles~\cite{tbexcl}. The existing benchmark scenarios designed for the
MSSM Higgs sector are formulated entirely in terms of low-scale parameters,
i.e., they are not related to any particular SUSY-breaking scheme and
make no provision for a possible unification of the
SUSY-breaking parameters at some high mass scale, as occurs
in generic supergravity and string scenarios.

In applications of the existing benchmark
scenarios for the MSSM Higgs
sector~\cite{benchmark,benchmark2,benchmark3,cpx}, one is normally
concerned only with the phenomenology of the Higgs sector itself.
Besides the direct searches for supersymmetric particles, other
constraints arising from EWPO, $B$-physics observables (BPO) and the
possible supersymmetric origin of the astrophysical
cold dark matter (CDM) are not usually taken into account.
This may be motivated by the fact that the additional constraints from EWPO,
BPO and CDM can depend sensitively on soft-supersymmetry breaking parameters
that otherwise have minor impacts on Higgs phenomenology. For example,
the presence of small flavour-mixing terms in the MSSM Lagrangian would
severely affect the predictions for the BPO while leaving Higgs
phenomenology essentially unchanged (see also \citere{sps} for a
discussion of this issue).

In this paper we follow a different approach and adopt specific
universality assumptions about the soft
SUSY-breaking parameters, restricting our analysis of the MSSM 
to a well-motivated subspace of manageable dimensionality. It is
frequently assumed that the
scalar masses $m_0$ are universal at some high unification scale, as are
the gaugino masses $m_{1/2}$ and the trilinear parameters $A_0$, a
framework known as the constrained MSSM (CMSSM). In such a scenario,
the heavier MSSM Higgs boson masses are fixed in terms of the input
parameters and $\tb$, so that $\MA$ is not an independent parameter, and
consequently this scenario is too restrictive for our purposes.
However, there is no good phenomenological or theoretical reason why the
soft supersymmetry-breaking contributions to the Higgs masses should not
be non-universal, a scenario termed the
NUHM~\cite{NUHM1,NUHM2,NUHMother}. Within the NUHM, $\MA$
and $\mu$ can be treated as free parameters for any specified values of
$m_0, m_{1/2}$, $A_0$ and $\tb$, so that this scenario provides a suitable
framework for studying the phenomenology of the MSSM Higgs sector. 
Since the low-scale parameters in this scenario are derived from a small
set of input quantities in a meaningful way, it is of interest to take
into account other experimental constraints. 

The main purpose of this paper is to explore new benchmark
surfaces for MSSM Higgs phenomenology that are compatible with the
cosmological density of cold dark matter inferred from a combination of
WMAP and other observations~\cite{WMAP}. While in the CMSSM only narrow
strips in $(m_{1/2}, m_0)$ planes are
compatible with WMAP et al.~\cite{WMAPstrips,wmapothers} for given
values of $A_0$ and $\tb$, the NUHM offers the attractive possibility to
specify $(\MA, \tb)$ planes such that essentially the whole plane is
allowed by the constraints from WMAP and other observations~\cite{ehoww}.
This is done assuming that
$R$~parity is conserved, that the lightest supersymmetric particle (LSP)
is the lightest neutralino $\neu{1}$, and that it furnishes most of the
cold dark matter required~\cite{EHNOS}. 
As we discuss in more detail below, compatibility with WMAP et al.\
cannot be maintained while keeping all the other NUHM parameters
fixed. Accordingly, we discuss two examples of WMAP-compliant benchmark
surfaces that are 
specified for fixed $m_0$, $\mu$ and $A_0 = 0$ but varying $m_{1/2}$, and two
surfaces that are specified for fixed $m_{1/2}, m_0$ and $A_0 = 0$ but varying
$\mu$. 
For the first two benchmark surfaces, a simple linear relation between
$m_{1/2}$ and $\MA$ is imposed as the $(\MA, \tb)$ plane is scanned,
whereas for the other two surfaces $\mu$ is varied through a relatively
narrow range.

Following the specifications of these NUHM benchmark surfaces, we then
explore the possibilities for studies of the MSSM Higgs bosons and other
supersymmetric signatures across these $(\MA, \tb)$ planes. We consider
the electroweak precision observables, principally 
$a_\mu \equiv \frac{1}{2}(g-2)_\mu$ and $\Mh$, prospects for the search for 
$H/A \to \tau^+ \tau^-$ at the Tevatron,
prospects at the LHC -- including searches for 
$h \to \ga \ga$ and $\tau^+\tau^-$, 
$H/A \to \tau^+ \tau^-$
and $H^\pm \to \tau^\pm \nu$, and measurements of the ratio of 
$h \to \tau^+\tau^-$ and $WW^*$ 
branching ratios, prospects at the ILC -- including ways of
distinguishing between the light MSSM $h$ boson and an SM Higgs boson of the
same mass by measuring (ratios of) branching ratios,
prospects in $B$ physics -- including $B_s \to \mu^+ \mu^-$, $b \to s \ga$
and $B_u \to \tau \nu$, and the direct detection of
supersymmetric cold dark matter. In an Appendix we introduce developments
in the {\tt FeynHiggs} code that enable the user to explore for
her/himself the WMAP-compliant benchmark surfaces. These include the
concept of a {\tt FeynHiggs} record, a new data type that captures the
entire content of a parameter file in the native format of {\tt FeynHiggs}.

%%%%%%%%%%%%%%%%%%%%%%%%%%%%%%%%%%%%%%%%%%%%%%%%%%%%%%%%%%%%%%%%%%%%%%%%%%%%%%%
%%%%%%%%%%%%%%%%%%%%%%%%%%%%%%%%%%%%%%%%%%%%%%%%%%%%%%%%%%%%%%%%%%%%%%%%%%%%%%%

\section{Specification of the Benchmark Surfaces}
\label{sec:benchmarks}

\begin{figure}[htb!]
%\vspace{10mm}
\begin{center}
\includegraphics[width=.48\textwidth]{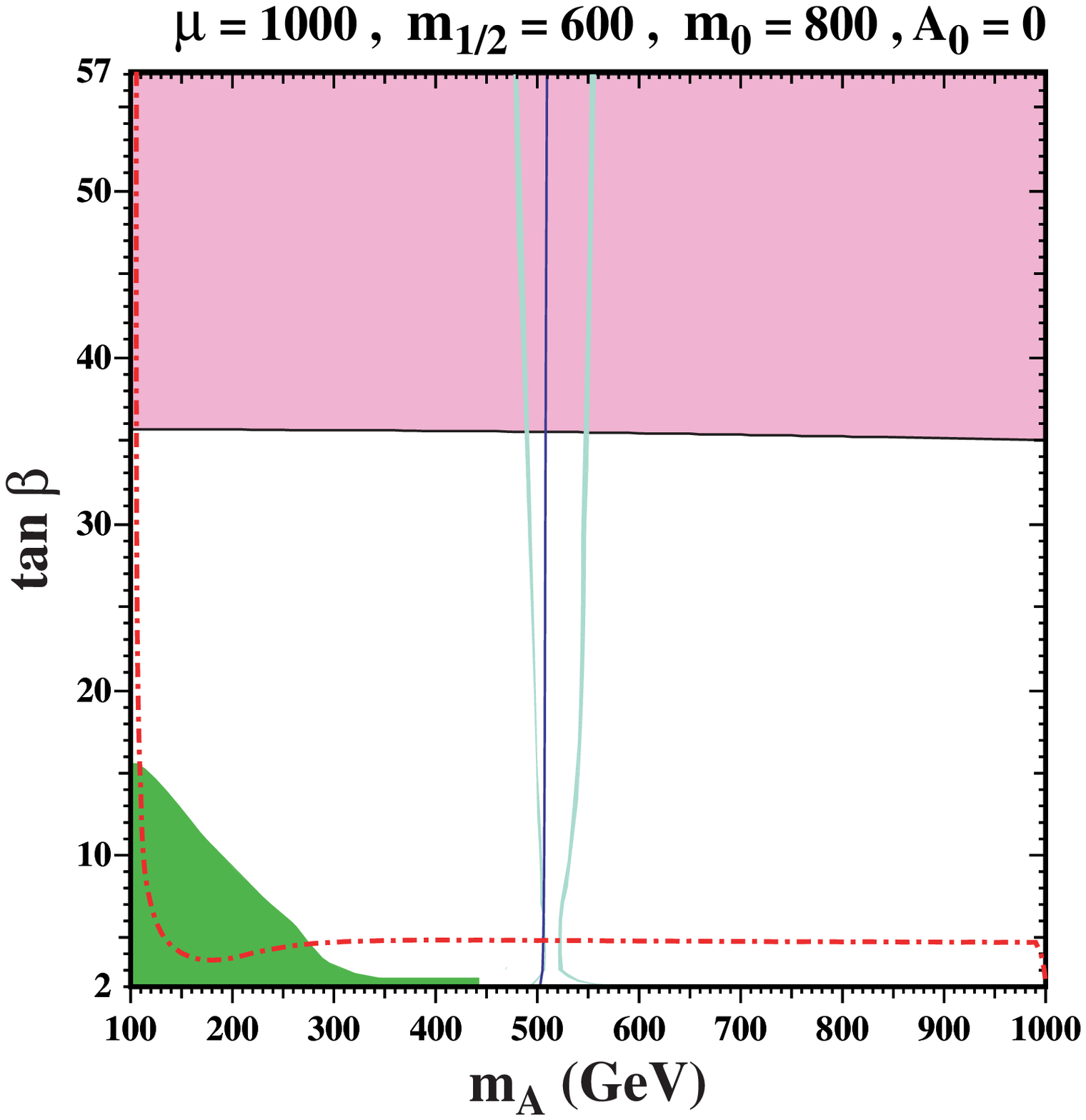}
\includegraphics[width=.49\textwidth]{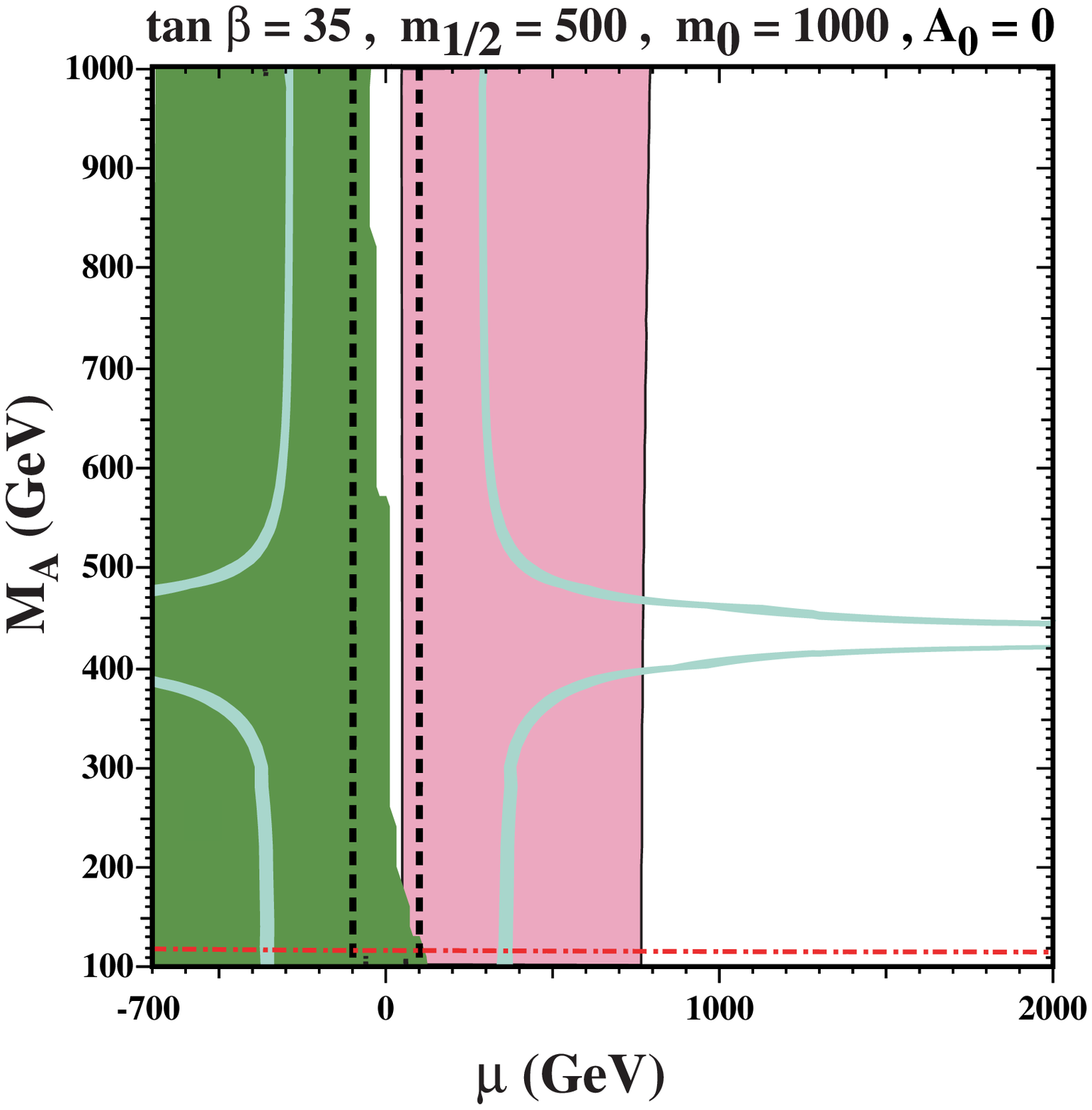}\\ 
\caption{
Sample NUHM parameter planes with two parameters varied and the other four
fixed, adapted from \protect{\citeres{ourBmumu2,eoss}}. The left plot
displays a \plane{\MA}{\tb} with 
$m_{1/2} = 600 \gev$, 
$m_0 = 800 \gev$,
$\mu = 1000 \gev$
and $A_0 = 0$. The range of cold dark matter density preferred by WMAP
and other observations is attained in two narrow (pale blue) strips, one
on either 
side of the vertical solid (blue) line where $\mneu{1} = \MA/2$. 
The dark (green) shaded region at low $\MA$ and low $\tb$ is excluded by
$b \to s \ga$, and
the medium (pink) shaded region at $\tb > 36$ is favoured by $a_\mu$.
The region below the (red) dot-dashed line is
excluded by the LEP bounds on $\Mh$.
The right plot displays a \plane{\mu}{\MA} with $m_{1/2} = 500 \gev$, 
$m_0 = 1000 \gev$,
$\tb = 35$
and $A_0 = 0$. 
Here the WMAP range of cold dark matter density is attained in two
narrow strips at roughly constant positive and negative values of
$\mu$, which are swept apart by rapid  annihilation when $\MA \sim 2
\mneu{1}$. The dark (green) shaded region at $\mu < 0$  is excluded by
$b \to s \ga$, and the $0 < \mu < 760\gev$ strip (pink shading) is
favoured by $a_\mu$. The region below the (red) dot-dashed line again is
excluded by the LEP bounds on $\Mh$, and the region between the vertical
(black) dashed lines has $\mcha{1} < 104\gev$.
}
\label{fig:fixed}
\end{center}
%\vspace{1em}
\end{figure}

As an introduction to the specification of the benchmark surfaces in the NUHM, 
we first consider a generic $(\MA, \tb)$ plane for fixed $m_{1/2}, m_0, A_0$
and $\mu$, adapted from \citere{ourBmumu2}. 
As we see in \reffi{fig:fixed}(a), in the $(\MA, \tb)$ plane for
$m_{1/2} = 600 \gev$, $m_0 = 800 \gev$, $\mu = 1000 \gev$ and $A_0 = 0$,
the relic LSP density satisfies the WMAP constraint only in narrow,
near-vertical (pale blue) shaded strips 
crossing the plane. These lie to either side of the vertical (purple)
line where 
$\mneu{1} = \MA/2$. Within the narrow unshaded strip straddling this
line, the relic density is suppressed by rapid direct-channel annihilations to
a value below the lower limit of the range
for the cold dark matter density indicated by
WMAP et al. This strip would be acceptable for cosmology if there were some
additional component of cold dark matter. Outside the
shaded WMAP-compatible strips, at both larger and smaller values of $\MA$,
the relic LSP density is too high, and these regions are
unacceptable%
\footnote{We note in passing that the LEP lower limit on $\Mh$
  excludes a strip of this plane at low $\MA$ and/or $\tb$ indicated
  by the dash-dotted (red) line, that
  $a_\mu$ (pink shading) prefers relatively large $\tb > 36$, 
  that $b \to s \ga$
  excludes a (green shaded) region at low $\MA$ and $\tb$, and that the other 
  BPO disfavour a region at low $\MA$ and high $\tb$ (not shown).}%
.

It is clear from this example that one may arrange for the relic LSP density
to remain within the preferred WMAP range over (essentially) the entire 
$(\MA, \tb)$ plane if one adjusts $m_{1/2}$ continuously as a function of
$\MA$ so as to remain within one of the narrow WMAP strips as $\MA$
increases. Accordingly, we study a benchmark \plane{\MA}{\tb}
\Athree\ with the same values of $m_0 = 800 \gev$, $\mu = 1000 \gev$ and
$A_0 = 0$, 
but with varying $m_{1/2} \sim \frac{9}{8} \MA$. Since we evaluate observables
using a 
discrete sampling of the NUHM parameter space, we consider values of
$m_{1/2}$ lying within the small range of this central value:
\begin{equation}
\frac{9}{8} \MA  - 12.5  \gev \le m_{1/2} \;  \le \frac{9}{8} \MA + 37.5\gev.
\label{A3}
\end{equation}
The observables that we study do not vary significantly as $m_{1/2}$ is
varied across this range. Specifically, we use the $m_{1/2}$ that gives
the value of the cold dark matter density that is closest to the central value
within the allowed range,
$0.0882 < \Omega_{\rm CDM} h^2 < 0.1204$~\cite{WMAP} (see below).

%%%%%%%%%%%%%%%%%%%%%%%%% F I G U R E %%%%%%%%%%%%%%%%%%%%%%%%%%%%%%%%%%%%%%%%%

%%%%%%%%%%%%%%%%%%%%%%%%% F I G U R E %%%%%%%%%%%%%%%%%%%%%%%%%%%%%%%%%%%%%%%%%

Previous analyses of the CMSSM indicated that values of 
$m_{1/2}$ and $m_0$ below 1~TeV are
preferred, in particular by the EWPO~\cite{ehow3,ehow4,ehoww} (see also
\citere{other}). Accordingly, we 
study also a benchmark \plane{\MA}{\tb}\ \Afive\ with the fixed values 
$m_0 = 300 \gev$, $\mu = 800 \gev$ and $A_0 = 0$, 
with $m_{1/2} \sim 1.2 \MA$ again varying
continuously across the plane so as to maintain the WMAP relationship
with $\MA$. As before, because of our discrete sampling of the NUHM parameter
space, we consider values of $m_{1/2}$ lying within a small range
of this central value: 
\begin{equation}
1.2 \MA  - 40 \gev \le m_{1/2} \;  \le 1.2 \MA + 40 \gev.
\label{A5}
\end{equation}
Again, the observables that we study do not vary significantly as $m_{1/2}$ is
varied across this range. 

More examples could be chosen with different fixed values of 
$m_0$, $\mu$ and $A_0$ but, 
as long as $m_{1/2}$ is the parameter being varied to
keep the LSP density within the WMAP range, a similar relationship between
$m_{1/2}$ and $\MA$ will always apply. The only flexibility in the choice of
$m_{1/2}$ is whether one wishes to stay 
within the left or right near-vertical shaded strip. However, the
corresponding values of $m_{1/2}$ do not differ greatly, and neither do the
corresponding phenomenological signatures, though the lightest Higgs
boson mass can be somewhat sensitive to this choice. The values of $m_0$
and (to a 
lesser extent) $\mu$ have far more impact on the phenomenology, and the
benchmark choices we have made: $m_0 = 800 \gev$ for \Athree\ and 
$m_0 = 300 \gev$ for \Afive, provide significant and 
interesting differences worthy of examination.

We also study two other $(\MA, \tb)$ planes, whose motivation can be gained
from examination of the $(\mu, \MA)$ plane shown in Fig.~\ref{fig:fixed}(b),
which is adapted from \citere{eoss}. 
We see that, for a fixed choice of values of
$m_{1/2} = 500 \gev$, $m_0 = 1000 \gev$ and $A_0 = 0$, 
there is a narrow strip of values of $\mu \sim 300 \to 350 \gev$ where 
the relic density lies within the WMAP range for almost all values of $\MA$.
The exception is a narrow strip centred on $\MA \sim 430 \gev$,
namely the rapid-annihilation funnel where
$\mneone \sim \MA/2$, which would be acceptable if there is some other
source of cold dark matter. 
This funnel is narrower
(wider) for smaller (larger) values of $\tb$, but its location in $\mu$
does not vary much as a function of  $\tan \beta$.%
\footnote{We note in passing that the LEP lower limit on $\Mh$
  excludes a strip of this plane at low $\MA$ indicated by the (red)
  dash-dotted line, and the LEP lower limit on the chargino mass
  excludes values  
  of $\mu$ between the two vertical (black) dashed lines.}%

Motivated by this example, we explore two benchmark surfaces with
different fixed values of $m_{1/2}$ and $m_0$, and $\mu$ varying within a
restricted range chosen to maintain the LSP density within or below the WMAP
range. The first example of such a benchmark plane, \Atwo, has fixed 
$m_{1/2} = 500 \gev$, $m_0 = 1000 \gev$ and $A_0 = 0$, 
with $\mu$ in the range
\begin{equation}
\mu \; =  250 - 400 \gev.
\label{A2}
\end{equation}
In the following, we evaluate observables for a discrete sampling of
$\mu$ values within this range. Since the corresponding variation of
the particle mass spectrum is quite small, the impact of the variation
of $\mu$ on the observables discussed below is negligible.

The other example of such a benchmark plane, \Afour, has fixed 
$m_{1/2} = 300 \gev$, $m_0 = 300 \gev$ and $A_0 = 0$, with $\mu$ in the range
\begin{equation}
\mu \; = \; 200 - 350 \gev.
\label{A4}
\end{equation}
As in the previous case, the LSP density lies within the WMAP range except for
a small range of $\MA \sim 2 \mneone$ where the density is below the preferred
range. However, again this is acceptable if there is some other
component of cold 
dark matter. The parameter choices for this and the other NUHM benchmark
surfaces are summarized in \refta{tab:nuhmscen}%
\footnote{
A minor change in the best-fit point and the $\chi^2_{\rm min}$ ocurred for
the \Afive\ scenario in comparison with \citere{ehoww} due to a slightly
different choice of the $m_{1/2}$ values.
}%
.

%%%%%%%%%%%%%%%%%%%%%%%%%%%%%%% T A B L E %%%%%%%%%%%%%%%%%%%%%%%%%%%%%%%%%%%%%
\begin{table}[tbh!]
\renewcommand{\arraystretch}{1.5}
\BC
\begin{tabular}{|c||c|c|c|c||c|}
\cline{2-6} \multicolumn{1}{c||}{}
 & $m_{1/2}$ & $m_0$ & $A_0$ & $\mu$ & $\chi^2_{\rm min}$ \\ \hline\hline
\Athree\  & $\sim\frac{9}{8} \MA$ & 800  & 0 & 1000 & 7.1  \\ \hline
\Afive\   & $\sim 1.2 \MA$ & 300  & 0 & 800 & 3.1  \\ \hline
\Atwo\    & 500    & 1000 & 0 & 250 ... 400 & 7.4 \\ \hline
\Afour\   & 300    & 300  & 0 & 200 ... 350 & 5.6 \\ 

\hline\hline
\end{tabular}
\EC
\renewcommand{\arraystretch}{1}
\caption{
The four NUHM benchmark surfaces are specified by the above fixed and varying
parameters, allowing $\MA$ and $\tb$ to vary freely. All mass parameters are
in GeV. The rightmost column shows the minimum $\chi^2$ value found in
each plane at the points labelled as the best fits in the plots.
}
\label{tab:nuhmscen}
\end{table}
%%%%%%%%%%%%%%%%%%%%%%%%%%%%%%% T A B L E %%%%%%%%%%%%%%%%%%%%%%%%%%%%%%%%%%%%%

A likelihood analysis of these four NUHM benchmark surfaces, including the
EWPO $\MW$, $\sweff$, $\Ga_Z$, $(g-2)_\mu$
and $\Mh$ and the BPO 
$\br(b \to s \ga)$, $\br(B_s \to \mu^+\mu^-)$, 
$\br(B_u \to \tau \nu_\tau)$ and $\De M_{B_s}$ was performed recently
in \citere{ehoww}. The lowest $\chi^2$ value in each plane, denoted as 
$\chi^2_{\rm min}$, is shown in the rightmost column of
\refta{tab:nuhmscen}, corresponding to the points labeled as the best fits
in the plots below. We display in each of the following figures the
locations of these best-fit points by a (red) cross and 
the $\Delta \chi^2 = 2.30$ and 4.61 contours around the best-fit points 
in the \plane{\MA}{\tb}s for each
of these benchmark surfaces. These contours would correspond to the
68~\% and 95~\% C.L. contours in the \plane{\MA}{\tb}s {\it if} the
overall likelihood distribution, $\cL \propto e^{-\chi^2/2}$,
were Gaussian. This is clearly only approximately true, 
but these contours nevertheless give interesting indications
on the regions in the \plane{\MA}{\tb}s that are currently
preferred. 
The varied parameter in each scenario (i.e.\ $m_{1/2}$ in \Athree, \Afive\ and
$\mu$ in \Atwo, \Afour) is chosen such that the cold dark matter density
is closest to the central value
within the allowed range,
$0.0882 < \Omega_{\rm CDM} h^2 < 0.1204$~\cite{WMAP}. 

\bigskip
On surfaces \Athree\ and \Afive, where $m_{1/2}$ scales with
$\MA$ so as to remain in the funnel region, much of the mass spectrum 
scales with $\MA$. Specifically, the lightest neutralino and chargino
masses simply scale in direct proportion to $\MA$ for these surfaces.
The light squark masses and stau masses also scale with $m_{1/2}$ (and
hence $\MA$), though the latter are also slightly dependent on $\tb$ as
well. In the range $\MA \le 1$~TeV displayed in these planes, the light
squark masses range up to $\sim 2.3$~TeV for surface \Athree, within
reach of the 
LHC. However, because of the relatively large values of $m_0$, the light
squarks are beyond the current reach of the Tevatron collider even at
low $\MA$ (and hence $m_{1/2}$). For \Afive, the light squark masses range up
to $\sim 1.7\tev$.

Turning to surfaces \Atwo\ and \Afour, because they have fixed values of
$m_{1/2}$ and $m_0$, there are very small variations in the sparticle
mass spectra across these planes.  For example, the lightest  neutralino
and chargino masses  
are determined primarily by $m_{1/2}$, and so they both take almost
constant
values on the benchmark surfaces.  Similarly, the light squark masses
are determined by a combination of $m_{1/2}$ and $m_0$
and show little dependence on either $\MA$ or $\tb$.  On the other hand, 
the lightest stau mass has a slight dependence on $\tb$, due
to the variable splitting of the third-generation sparticle masses.  
These mass splittings increase at large $\tb$, leading to smaller stau
masses.   

We display in each plane the
region excluded (black shaded) at the 95~\% C.L.\ by the LEP Higgs
searches in the channel
$e^+e^- \to Z^* \to Z h, H$~\cite{LEPHiggsSM,LEPHiggsMSSM}.
For a SM-like Higgs boson we use a bound of $\Mh > 113 \gev$. The
difference from the nominal LEP mass limit
allows for the estimated theoretical uncertainty in
the calculation of $\Mh$ for specific values of the input MSSM
parameters~\cite{mhiggsAEC}. In the region of small $\MA$ and large
$\tb$, where the coupling of the light $\cp$-even Higgs boson to gauge
bosons is suppressed, the bound on $\Mh$ is reduced to 
$\Mh > 91 \gev$~\cite{LEPHiggsMSSM}.

%%%%%%%%%%%%%%%%%%%%%%%%%%%%%%%%%%%%%%%%%%%%%%%%%%%%%%%%%%%%%%%%%%%%%%%%%%%%%%%
%%%%%%%%%%%%%%%%%%%%%%%%%%%%%%%%%%%%%%%%%%%%%%%%%%%%%%%%%%%%%%%%%%%%%%%%%%%%%%%

\section{Electroweak precision observables}
\label{sec:ewpo}

In this Section we summarize key predictions for electroweak precision
observables (EWPO) over the four benchmark surfaces. In \citere{ehoww} it was
shown that $\MW$, $\sweff$ and $\Ga_Z$ agree within $\sim 1\,\si$ with
the current experimental value over all the benchmark
surfaces. Since their variations are relatively small, we do not display these
observables in this paper, though they are included in the overall $\chi^2$
function. Here we focus on two other EWPO, namely the mass of the
lightest Higgs  boson, $\Mh$, and the anomalous magnetic moment of the muon, 
$a_\mu \equiv \frac{1}{2} (g-2)_\mu$. 

\begin{figure}[htb!]
%\vspace{10mm}
\begin{center}
\includegraphics[width=.47\textwidth]{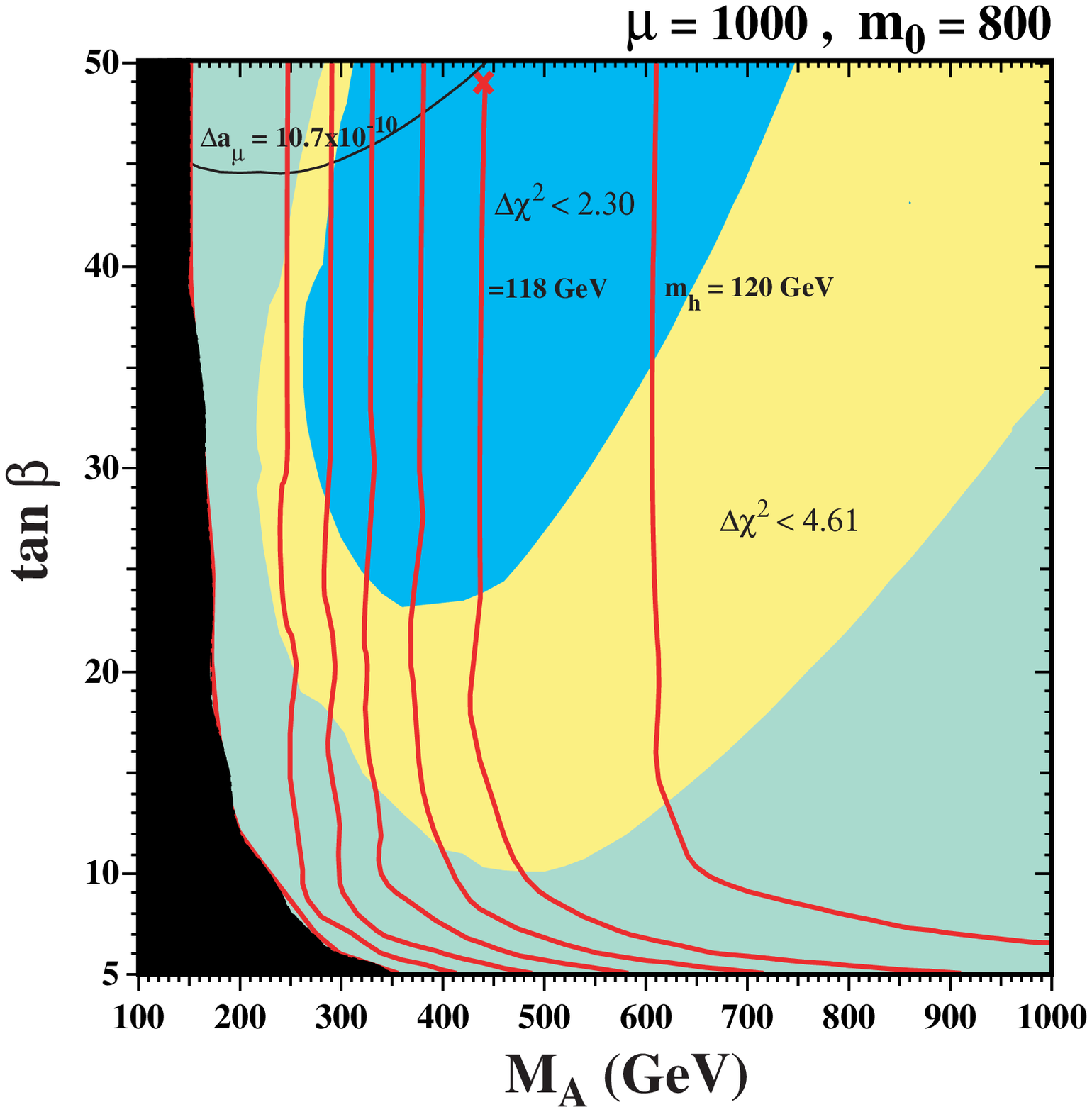}
%\hspace{-25mm}
\includegraphics[width=.47\textwidth]{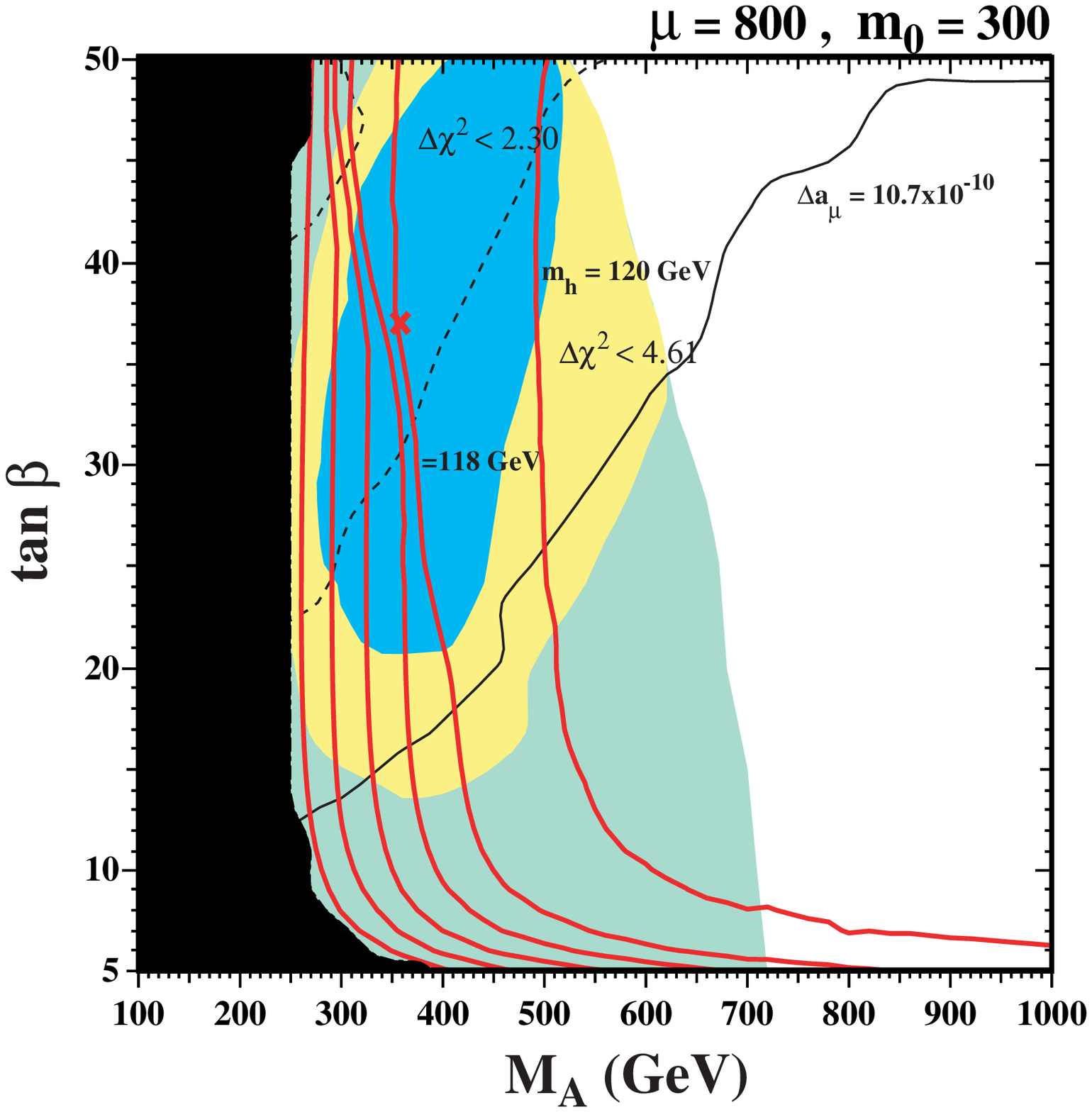}\\ %[2em]
\includegraphics[width=.47\textwidth]{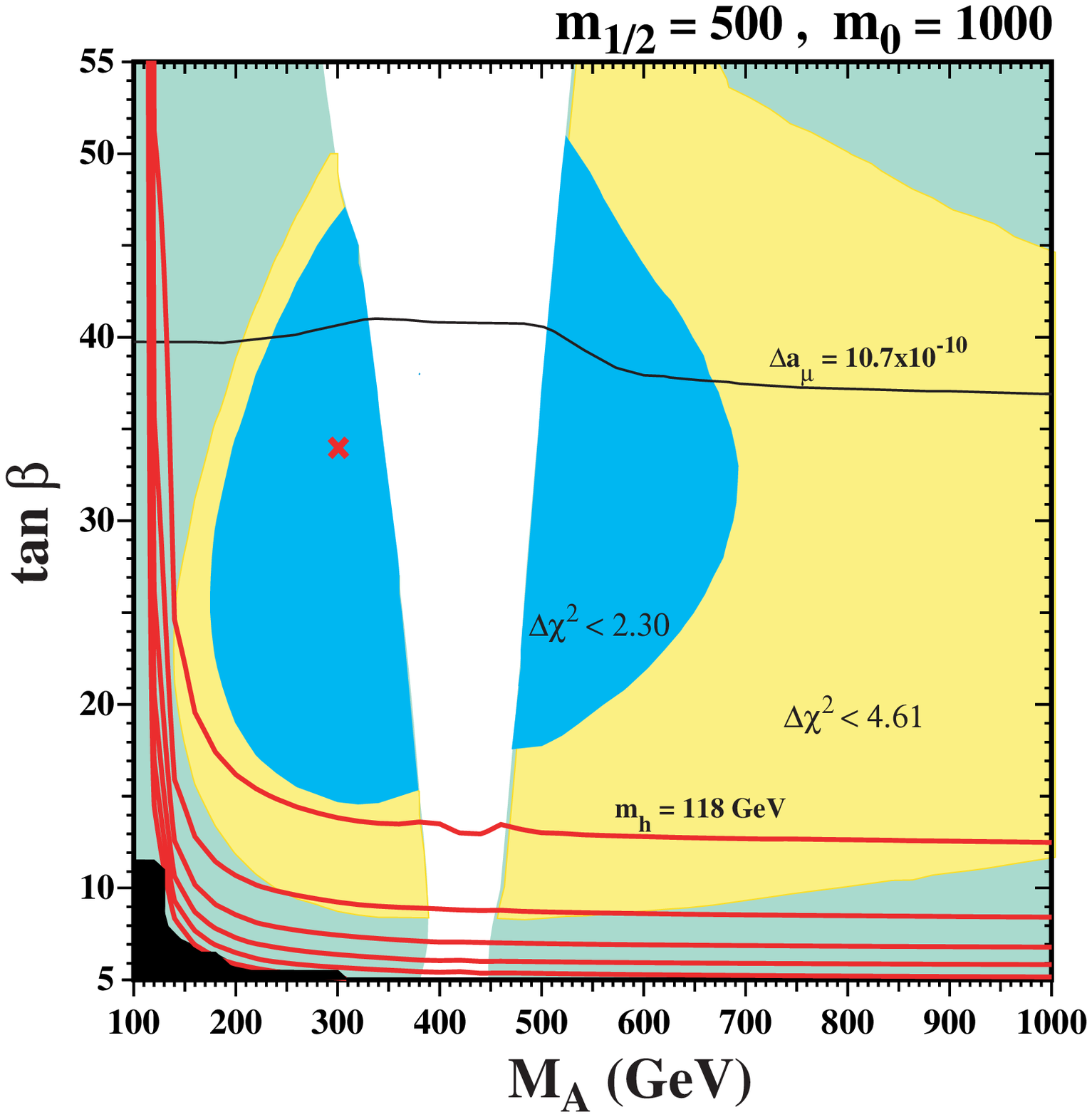}
%\hspace{-25mm}
\includegraphics[width=.47\textwidth]{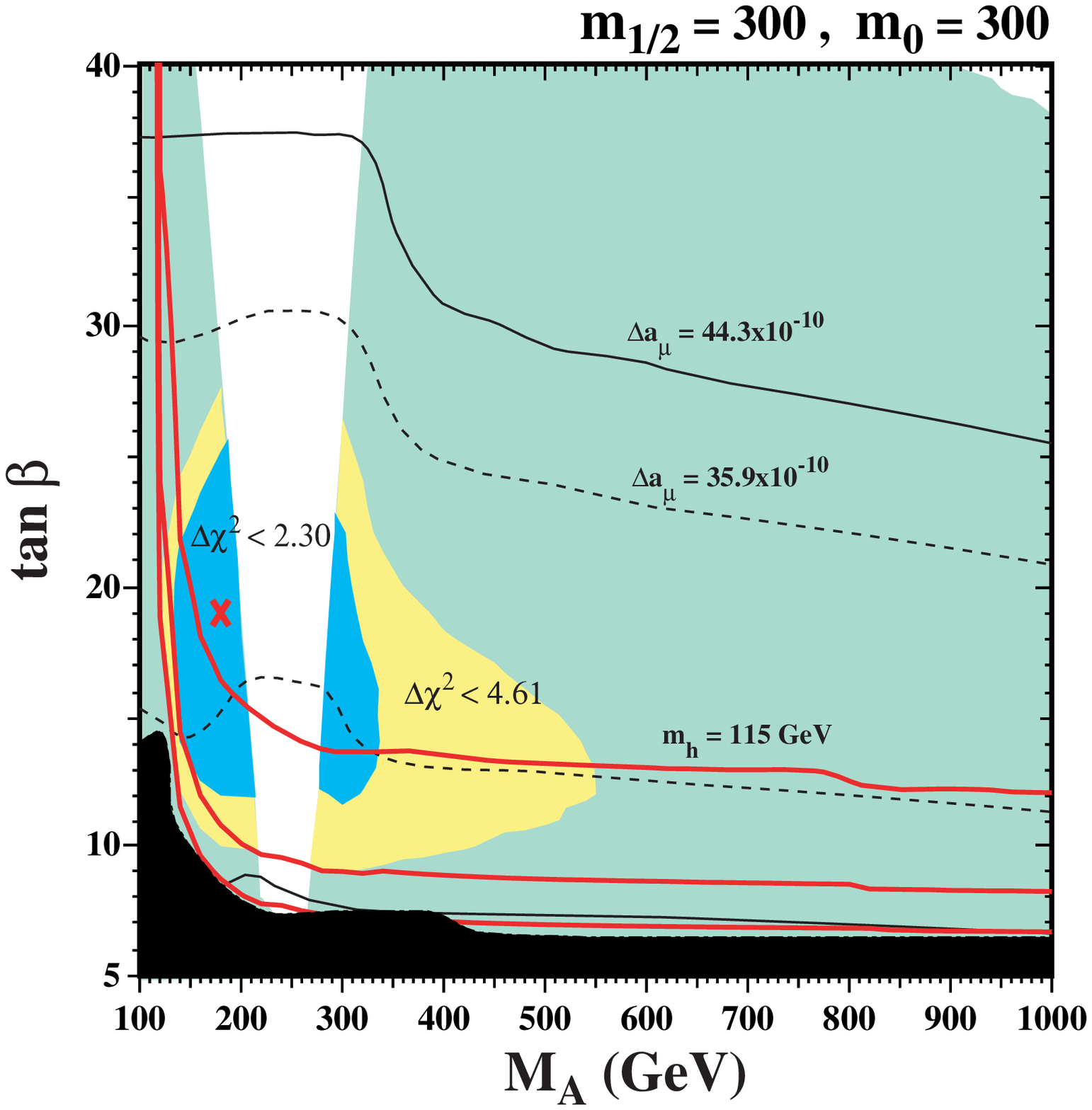}
%\hspace{-5mm}
%\vspace{-25mm}
\caption{%
The $(\MA, \tb)$ planes for the NUHM benchmark surfaces  (a) \Athree, (b)
\Afive, (c) \Atwo\ and (d) \Afour, displaying the contours of $\De \chi^2$
found in a recent global fit to EWPO and BPO~\protect\cite{ehoww}.
All surfaces have $A_0 = 0$.
We also display individually the contours of $\Mh$ found using 
{\tt FeynHiggs}~\protect\cite{feynhiggs,mhiggslong,mhiggsAEC,mhcMSSMlong}
and the contours of  $a_\mu$ found using 
\protect\citeres{g-2MSSMf1l,g-2MSSMlog2l,g-2FSf,g-2CNH}.
The 1(2)-$\si$ range for $a_\mu$ is demarcated by dashed (solid) lines.
The dark shaded (black)
region corresponds to the parameter region that
is excluded by the LEP Higgs
searches in the channel
$e^+e^- \to Z^* \to Z h, H$~\cite{LEPHiggsSM,LEPHiggsMSSM}.
}
\label{fig:Mh}
\end{center}
\vspace{1em}
\end{figure}

The evaluation of $\Mh$ is performed using
{\tt FeynHiggs}~\cite{feynhiggs,mhiggslong,mhiggsAEC,mhcMSSMlong}. 
In \reffi{fig:Mh} we show the contours for 
$\Mh = 113, 114, 115, 116, 117, 118$ and 120~GeV.  
As discussed in the previous Section, the boundary of the region
excluded by the LEP searches for the lightest MSSM Higgs boson does
not coincide with the nominal limit $\Mh = 114.4$~GeV on the mass of a
Standard Model Higgs boson. 
Nevertheless, it can be seen in \reffi{fig:Mh} that the
$\Delta \chi^2 = 2.30$ and 4.61 contours are highly correlated with the
$\Mh$ contours at low values of $\MA$ and $\tb$. This is a consequence
of the fact that
the full likelihood information from the LEP Higgs exclusion
limit (as well as the theoretical uncertainty) is incorporated into the
overall $\chi^2$ function (see \citere{ehoww}).
Note that for the plane \Afour\  (and to a lesser extent \Atwo)
the maximum value for the Higgs mass is limited by the relatively low
value of $m_{1/2}$.

%%%%%%%%%%%%%%%%%%%%%%%%% F I G U R E %%%%%%%%%%%%%%%%%%%%%%%%%%%%%%%%%%%%%%%%%
%%%%%%%%%%%%%%%%%%%%%%%%% F I G U R E %%%%%%%%%%%%%%%%%%%%%%%%%%%%%%%%%%%%%%%%%

Concerning $a_\mu$, we recall that, according to a recent evaluation of
the Standard Model contribution based on low-energy $e^+ e^-$
data, there is a discrepancy with the experimental measurement by the
E821 Collaboration~\cite{g-2exp,g-2exp2}. It would be premature to
regard this deviation as solid evidence for new physics. However,
within the SUSY framework we explore here, this discrepancy does
impose a significant constraint on the parameter space, and makes an
important contribution to the global $\chi^2$ function whose contours 
are shown in Fig.~\ref{fig:Mh}.
Our evaluation of $a_\mu$ is based on
\citeres{g-2MSSMf1l,g-2MSSMlog2l,g-2FSf,g-2CNH}, which
yields~\cite{DDDD,g-2SEtalk}: 
\BE
\amuexp-\amutheo = (27.5 \pm 8.4) \times 10^{-10},
\label{delamu}
\EE
equivalent to a 3.3-$\si$ effect%
\footnote{Three other recent evaluations yield slightly different
  numbers~\cite{g-2HMNT2,g-2reviewMRR,g-2reviewFJ,TW}, 
  but similar discrepancies with the SM prediction.}%
. In \reffi{fig:Mh} we show the contours 
$\Delta a_\mu = 10.7, 19.1, 35.9, 44.3 \times 10^{-10}$ 
for the net supersymmetric contribution to $a_\mu$.

In the case of surface \Athree, we see that the best-fit point
corresponds to $\Mh \sim 118 \gev$ and 
$\De a_\mu \sim 10.7 \times 10^{-10}$. In most of the 
displayed region of the surface that is favoured at the global 
$\De \chi^2 < 4.61$ level, $\De a_\mu$ is considerably {\it lower} than
the range favoured in \refeq{delamu}. In the case of surface \Afive,
the best-fit point has $\Mh \sim 118\gev$, and 
$\De a_\mu$ is within the 1-$\si$ range given by \refeq{delamu}. In the
case of surface \Atwo, the best-fit point has $\Mh > 118 \gev$ and
again a low value of $\De a_\mu$. Finally, the best-fit point in
surface \Afour\  has $\Mh \sim 115 \gev$ and an excellent value of
$\De a_\mu$, according to \refeq{delamu}. The fact that the
best-fit points do not always have favoured values of $\De a_\mu$
reflects the importance of other precision observables, notably the
BPO discussed later. 

%%%%%%%%%%%%%%%%%%%%%%%%%%%%%%%%%%%%%%%%%%%%%%%%%%%%%%%%%%%%%%%%%%%%%%%%%%%%%%%
%%%%%%%%%%%%%%%%%%%%%%%%%%%%%%%%%%%%%%%%%%%%%%%%%%%%%%%%%%%%%%%%%%%%%%%%%%%%%%%

\section{Tevatron Phenomenology}
\label{sec:tev}

\begin{figure}[htb!]
\vspace{10mm}
\begin{center}
\includegraphics[width=.49\textwidth]{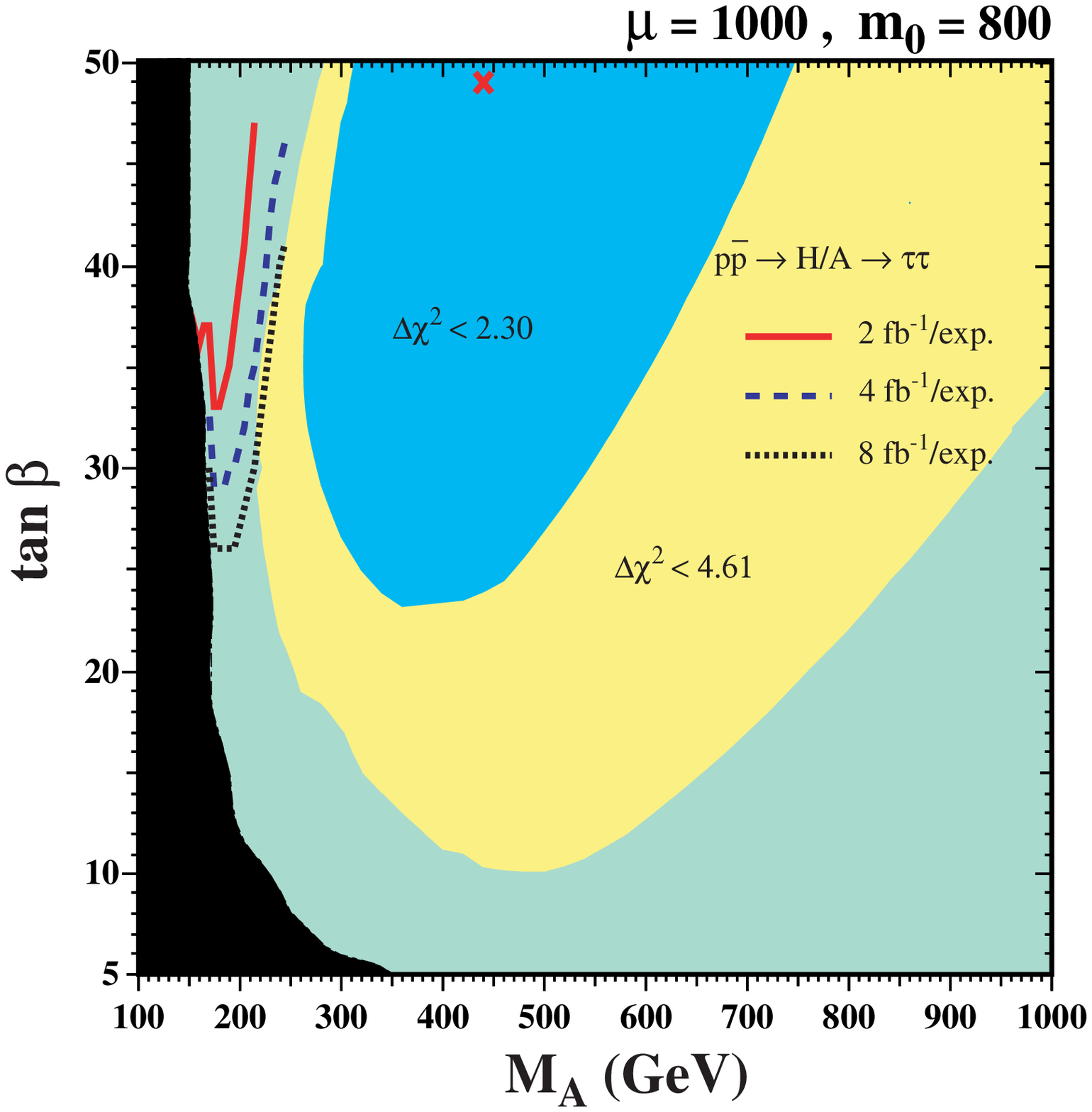}
%\hspace{-25mm}
\includegraphics[width=.49\textwidth]{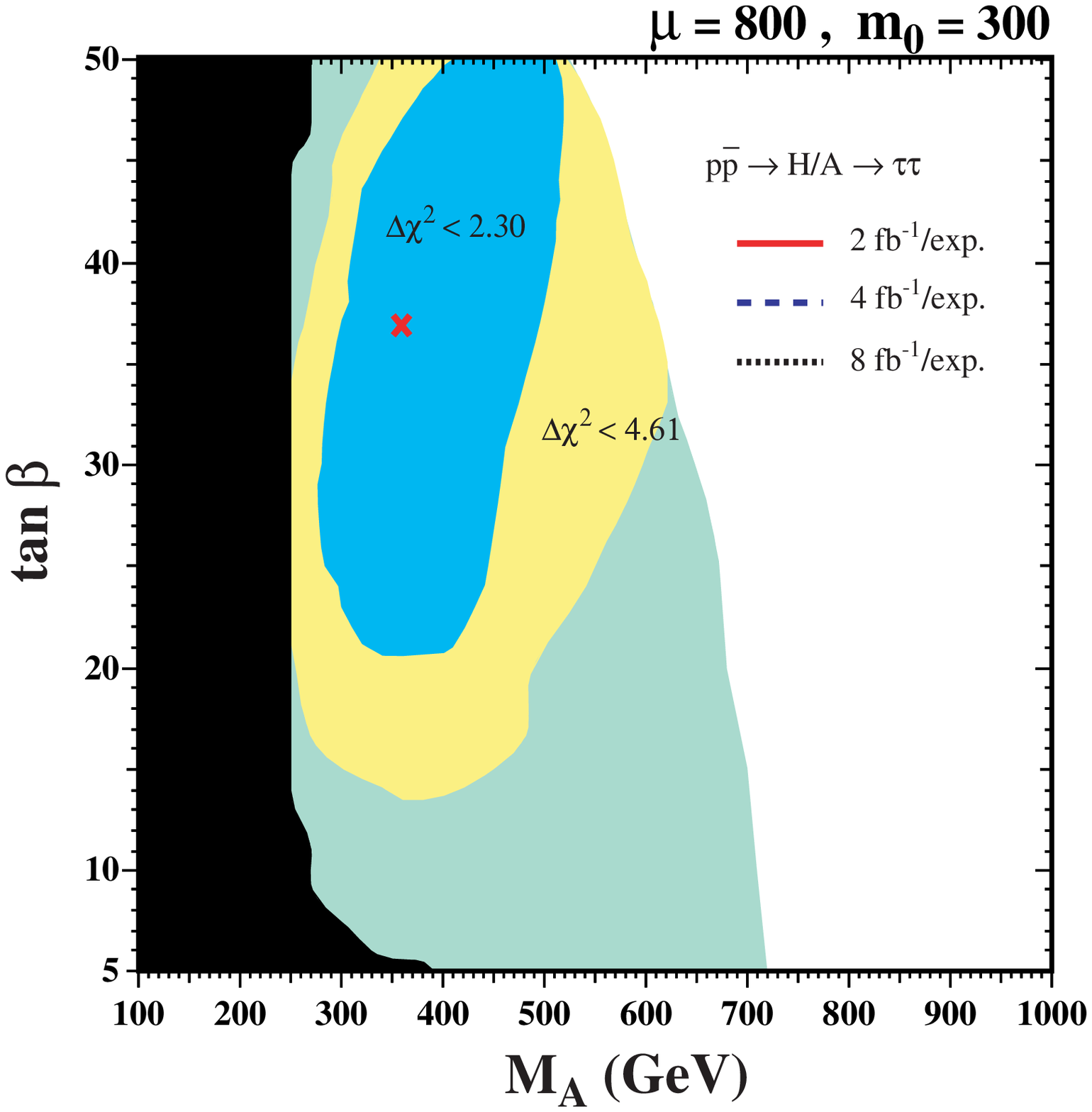}\\ %[2em]
\includegraphics[width=.49\textwidth]{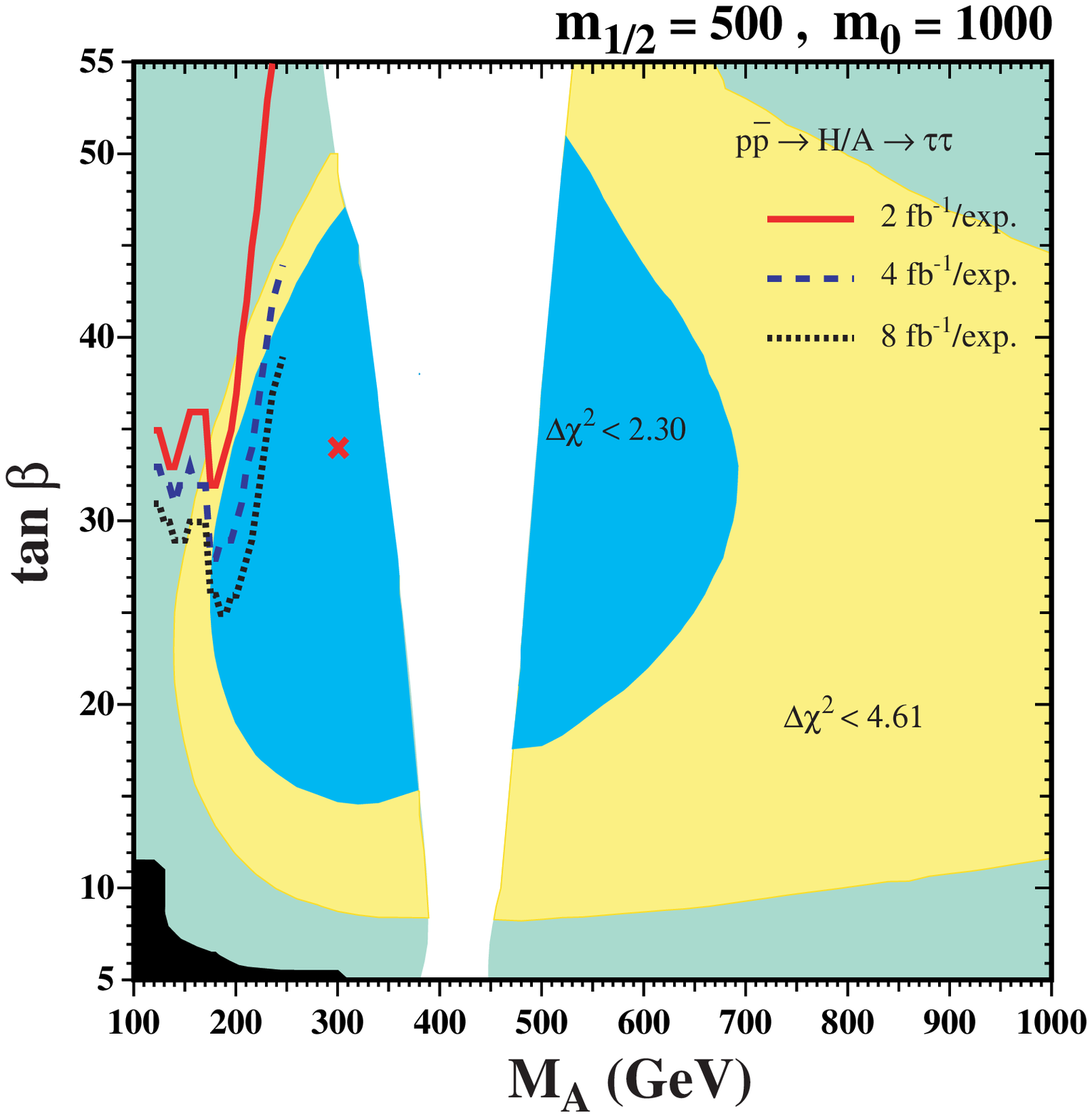}
%\hspace{-25mm}
\includegraphics[width=.49\textwidth]{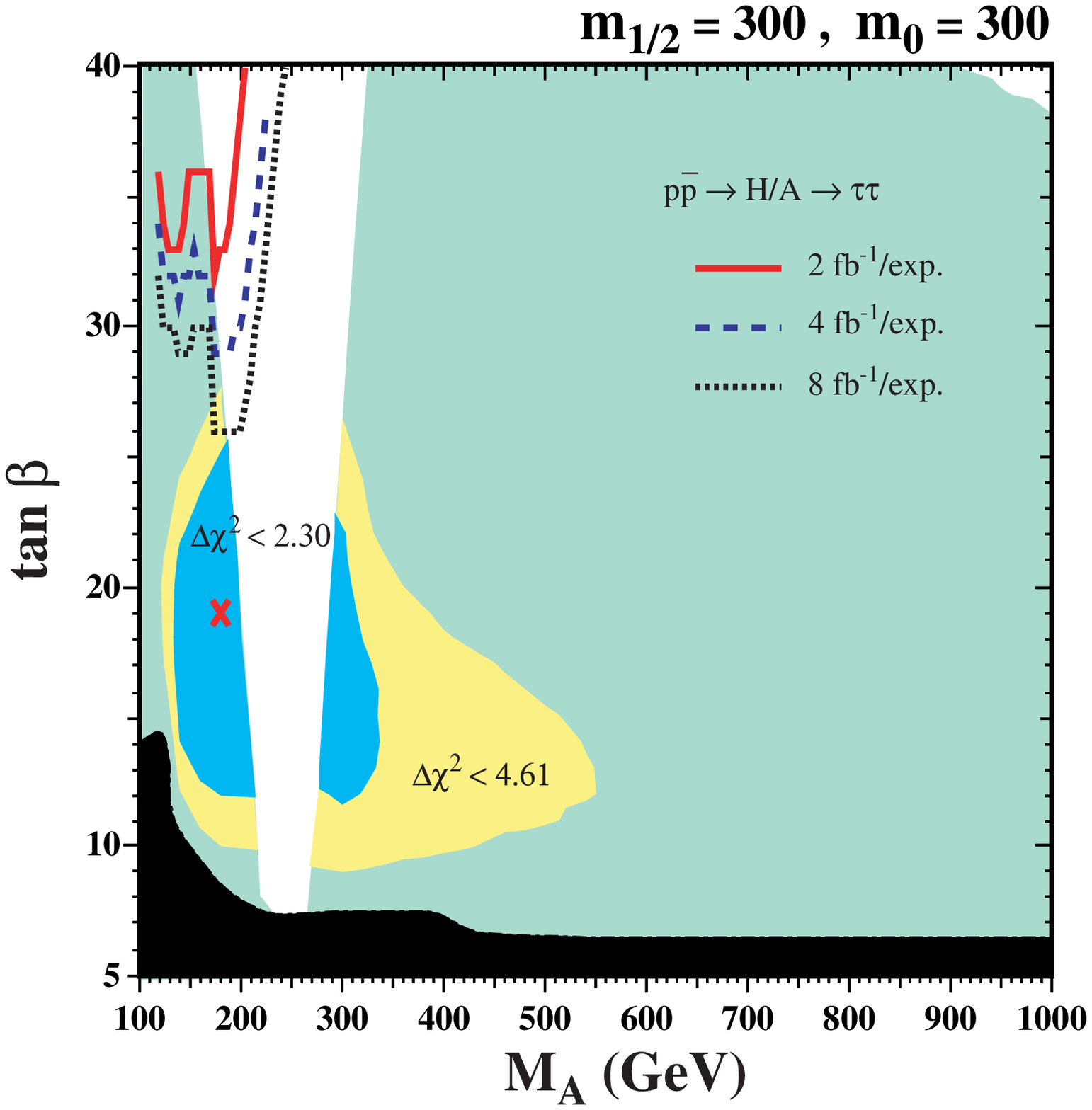}
%\hspace{-5mm}
%\vspace{-25mm}
\caption{%
The same $(\MA, \tb)$ planes for the NUHM benchmark surfaces (a)
\Athree, (b) \Afive, (c) \Atwo\ and (d) \Afour\ as in
Fig.~\protect\ref{fig:Mh}, displaying also the expected 95\% C.L.\ exclusion
sensitivities 
of searches for $H/A \to \tau^+ \tau^-$ at the Tevatron collider with
2, 4, 8~fb$^{-1}$ in each of the CDF and D0 experiments (see text).
}
\label{fig:Tev02}
\end{center}
\vspace{1em}
\end{figure}

We first consider how experiments at the Tevatron collider in the next
years could probe the
benchmark surfaces \Athree, \Afive, \Atwo\ and \Afour. We consider one
possible Tevatron signature for the MSSM Higgs sector, namely
$H/A \to \tau^+ \tau^-$, for which expectations are evaluated
using the results from~\citere{CDFprojections}. They are based on the
expectation of a 30\% 
improvement in the sensitivity with respect to~\citere{CDFHiggsMSSM}. 
We see in \reffi{fig:Tev02} that, at the Tevatron with 2 (4, 8)~fb$^{-1}$ of
integrated and analyzed luminosity per experiment%
\footnote{
We note that both CDF and D0 have already recorded more than 
2.5~fb$^{-1}$ of integrated luminosity.},
the channel $H/A \to \tau^+ \tau^-$ 
would provide a 95\% C.L.\ exclusion sensitivity to $\tb \sim 35 (30, 25)$
when $\MA \sim 200 \gev$, and the sensitivity decreases slowly (rapidly)
at smaller (larger) $\MA$. In the case of the benchmark surface
\Athree, 8~fb$^{-1}$ would start accessing the region with 
$\De \chi^2 < 4.61$. 
For \Afive, however, the area accessible to the Tevatron is not visible
in the figure since it is completely covered by the excluded region
from the LEP Higgs searches.
The region $\De \chi^2 < 4.61$
could be accessed already with 2~fb$^{-1}$ in case \Atwo, and 
8~fb$^{-1}$ would give access to the region with $\De \chi^2 < 2.30$. 
However, even the $\De \chi^2 < 4.61$ region of the \Afour\ surface
would be inaccessible with 8~fb$^{-1}$.

%%%%%%%%%%%%%%%%%%%%%%%%% F I G U R E %%%%%%%%%%%%%%%%%%%%%%%%%%%%%%%%%%%%%%%%%
%%%%%%%%%%%%%%%%%%%%%%%%% F I G U R E %%%%%%%%%%%%%%%%%%%%%%%%%%%%%%%%%%%%%%%%%

We note that the CDF Collaboration has recently reported a 
$\sim 2$-$\si$ excess
of candidate $H/A \to \tau^+ \tau^-$ events~\cite{CDFHiggsMSSMnew}, which
would correspond to $\MA \sim 160 \gev$ and $\tb > 45$. As discussed
in \citere{ehow5}, taking into account all the available experimental
constraints, this possible excess could be accommodated within the NUHM 
only for rather different values of the parameters from those considered in
the benchmark scenarios, namely $m_{1/2} \sim 650 \gev$, $m_0 \sim 1000 \gev$,
$A_0 \sim -1900 \gev$, $\mu \sim 385 \gev$. A likelihood analysis
yields values of $\chi^2 \sim 9$--10, somewhat higher than the values
for the benchmark surfaces. 
Within the four benchmark scenarios here, the precision observables are
not in good agreement with low $\MA$ and large $\tb$, 
reflecting the fact that the points with $\MA \sim 160 \gev$ and $\tb > 45$
lie well outside the regions with $\De \chi^2 < 4.61$ on all of these
benchmark surfaces.

%%%%%%%%%%%%%%%%%%%%%%%%%%%%%%%%%%%%%%%%%%%%%%%%%%%%%%%%%%%%%%%%%%%%%%%%%%%%%%%
%%%%%%%%%%%%%%%%%%%%%%%%%%%%%%%%%%%%%%%%%%%%%%%%%%%%%%%%%%%%%%%%%%%%%%%%%%%%%%%

\section{LHC Phenomenology}
\label{sec:lhc}

In this Section we present and compare the sensitivities of various LHC
searches for MSSM Higgs bosons as functions of $\MA$ and $\tb$ in the
benchmark surfaces \Athree, \Afive, \Atwo\ and \Afour. 
The Higgs bosons can either be produced 'directly' or via cascades, starting
with gluino or squark production~\cite{HCascades}. We focus here on the
first possibility, but it should be kept in mind that the production via
cascades could offer additional channels for the Higgs detection.
A full evaluation of these channels across the benchmark surfaces must
await a more complete evaluation of the experimental sensitivities to
such decay modes.

%%%%%%%%%%%%%%%%%%%%%%%%% F I G U R E %%%%%%%%%%%%%%%%%%%%%%%%%%%%%%%%%%%%%%%%%
\begin{figure}[htb!]
%\vspace{10mm}
\begin{center}
\includegraphics[width=.49\textwidth]{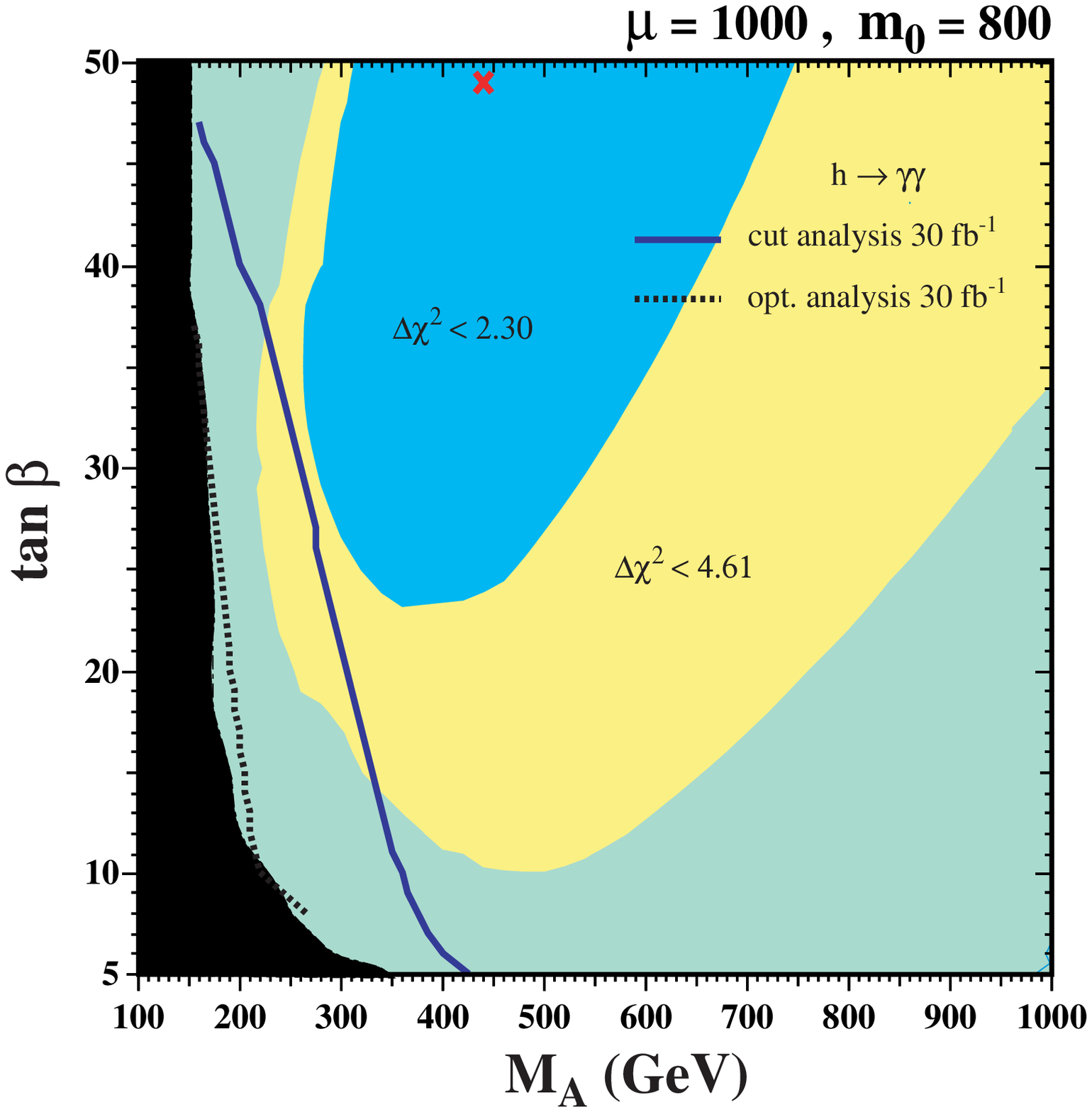}
%\hspace{-25mm}
\includegraphics[width=.49\textwidth]{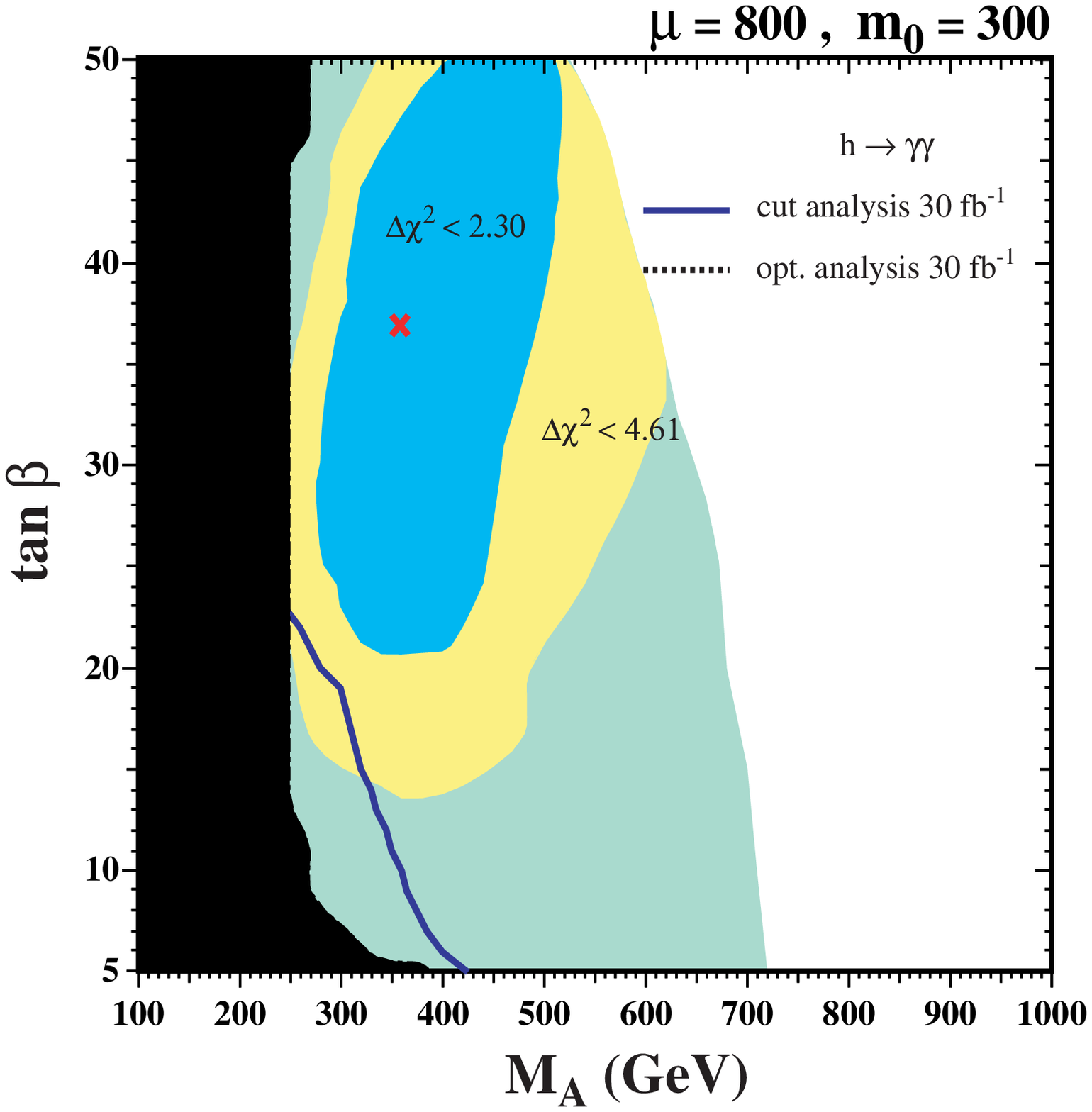}\\ %[2em]
\includegraphics[width=.49\textwidth]{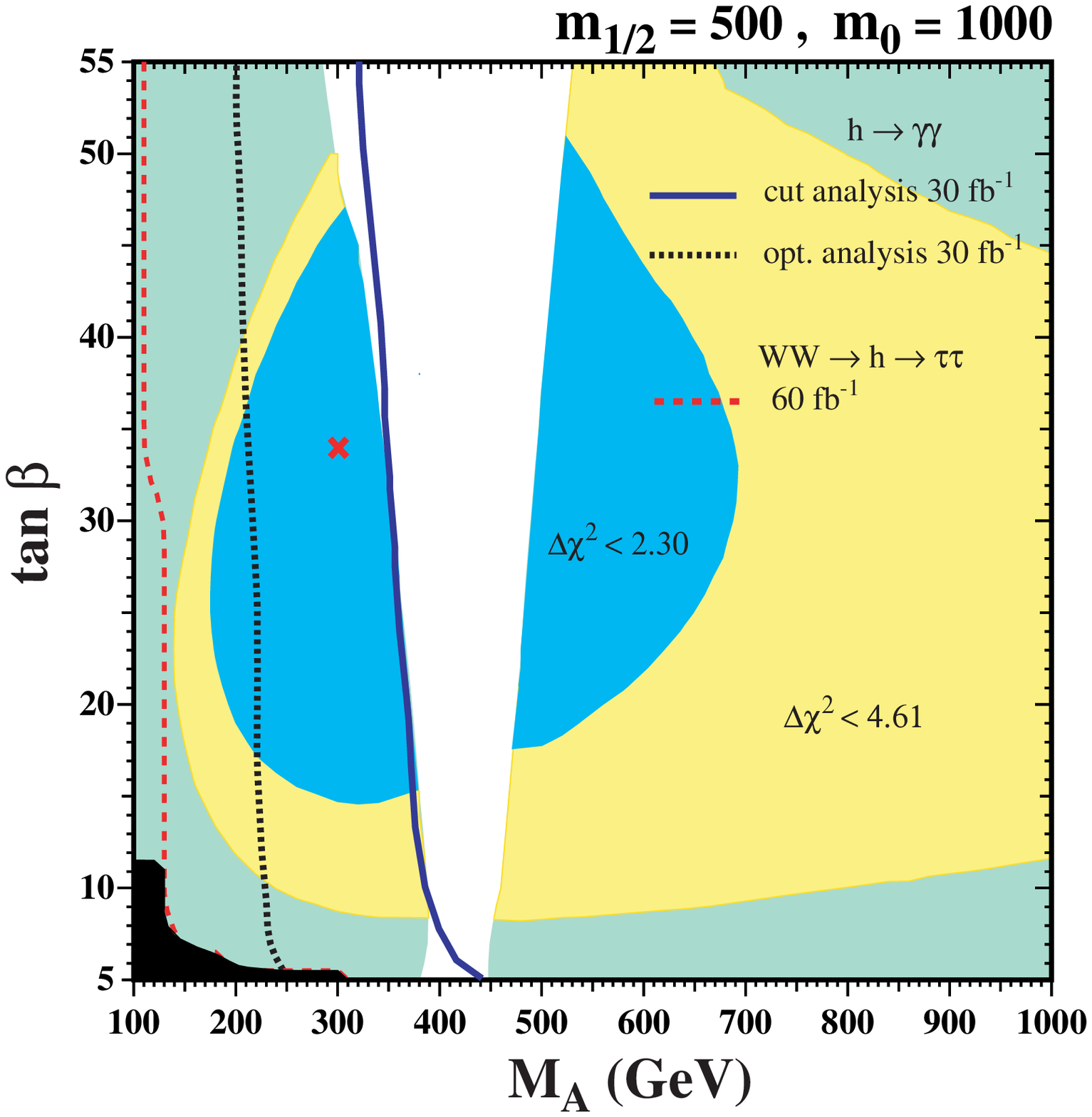}
%\hspace{-25mm}
\includegraphics[width=.49\textwidth]{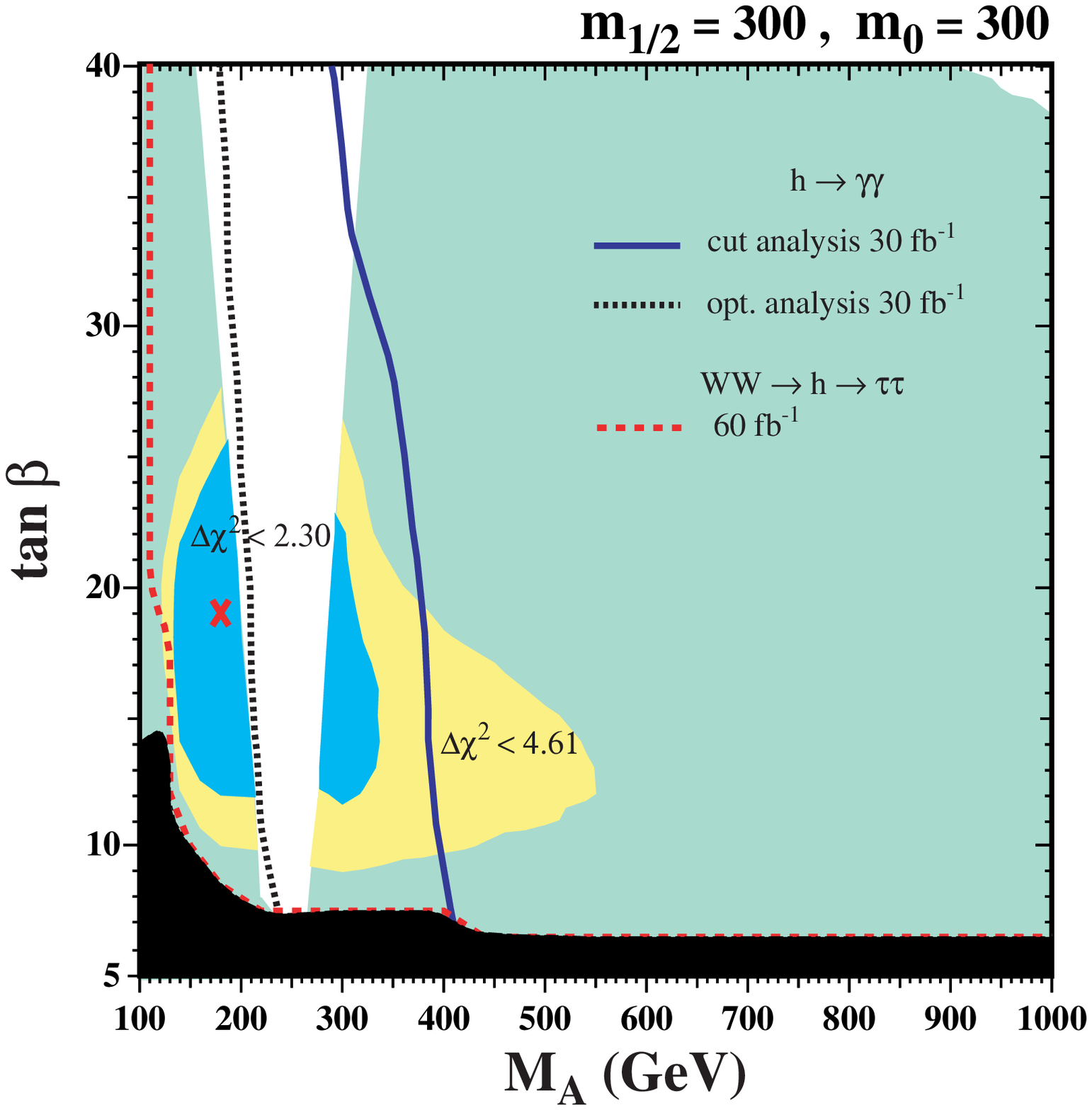}
%\hspace{-5mm}
%\vspace{-25mm}
\caption{%
The same $(\MA, \tb)$ planes for the NUHM benchmark surfaces (a)
\Athree, (b) \Afive, (c) \Atwo\ and (d) \Afour\ as in
Fig.~\protect\ref{fig:Mh}, displaying the expected sensitivities of
searches for $pp \to h \to \ga \ga$ at the LHC with 30~fb$^{-1}$ in the CMS
detector using a cut analysis or an ``optimized'' analysis (see text)
as well as the searches for $W^+ W^- \to h \to \tau^+\tau^-$ with 
60~fb$^{-1}$ in the CMS detector. The parameter regions to the right of
the contours are covered at the 5-$\si$ level. For \Athree\ and \Afive\ 
the $W^+ W^- \to h \to \tau^+\tau^-$ channel covers the whole region of
the $(\MA, \tb)$ plane that is unexcluded by LEP.
}
\label{fig:LHC12}
\end{center}
\vspace{-1em}
\end{figure}
%%%%%%%%%%%%%%%%%%%%%%%%% F I G U R E %%%%%%%%%%%%%%%%%%%%%%%%%%%%%%%%%%%%%%%%%

We start the analysis with the light MSSM Higgs boson that 
behaves like the SM Higgs boson for $\MA \gg \MZ$. 
As a consequence, the region $\MA \gg \MZ$ can be covered in all 
benchmark scenarios if a SM Higgs with $\MHSM = \Mh$ is accessible at
the LHC~\cite{atlastdr,atlasrev,CMS-TDR}.
In \reffi{fig:LHC12} we display on the WMAP-compatible $(\MA, \tb)$ planes the
5-$\si$ discovery contours for $pp \to h \to \ga \ga$ at the LHC with
30~fb$^{-1}$ in the CMS detector~\cite{CMS-TDR}, where the areas to the
right of the lines (i.e.\ for larger $\MA$) are covered by the 
$pp \to h \to \ga\ga$ search. This channel is
particlarly important for a precise mass measurement of the lightest MSSM
Higgs boson. We show separately the
sensitivities for a cut-based analysis (blue solid line) and for an
``optimized'' analysis (black dotted line), see \citere{CMS-TDR} for details. 
The cut-based analysis should be regarded as a
conservative result, while the ``optimized'' analysis should perhaps be
regarded as an optimistic expectation~\cite{hgagaOPT}.
In the cases of surfaces \Athree\ and \Afive, the LHC cut analysis for the
$pp \to h \to \ga\ga$ search covers all of the $\De\chi^2 < 2.30$ region and
the optimized analysis nearly the whole parameter plane. For \Atwo\ only
parts of the preferred region can be covered, while for \Afour\ even
with the optimized analysis the best-fit point as well as large parts of
the $\De\chi^2 < 2.30$ area remain uncovered. In this region, more
luminosity would need to be accumulated
in order to see a
5-$\si$ signal in the $pp \to h \to \ga\ga$ channel.

We turn next to the reaction $W^+ W^- \to h \to \tau^+\tau^-$. 
On the WMAP-compatible $(\MA, \tb)$ planes in~\reffi{fig:LHC12} we display  the
5-$\si$ discovery contours for $W^+ W^- \to h \to \tau^+\tau^-$ at the LHC with
60~fb$^{-1}$ in the CMS detector~\cite{CMS-TDR}, where the areas to the
right of the lines (i.e.\ for larger $\MA$) are covered by this 
search. In the cases of surfaces \Athree\ and \Afive, the 5-$\si$
discovery contours lie within the region already excluded by LEP, so
this search covers all the unexcluded parts of the surfaces. In the
cases of surfaces \Atwo\ and \Afour, however, the 
$W^+ W^- \to h \to \tau^+\tau^-$ discovery contours leave uncovered narrow 
strips at low $\MA$ for $\tan \beta > 11, 14$, respectively. In this
part of the parameter space the search for $H \to \tau^+\tau^-$ should
be investigated.
In all cases, the 5-$\si$ discovery contours cover the entire 
$\Delta \chi^2 < 4.61$
regions. However, we note that this channel does not permit a very accurate
measurement of $\Mh$, unlike the $pp \to h \to \ga \ga$ channel.

%%%%%%%%%%%%%%%%%%%%%%%%% F I G U R E %%%%%%%%%%%%%%%%%%%%%%%%%%%%%%%%%%%%%%%%%
\begin{figure}[htb!]
\vspace{10mm}
\begin{center} %xxxxx
\includegraphics[width=.49\textwidth]{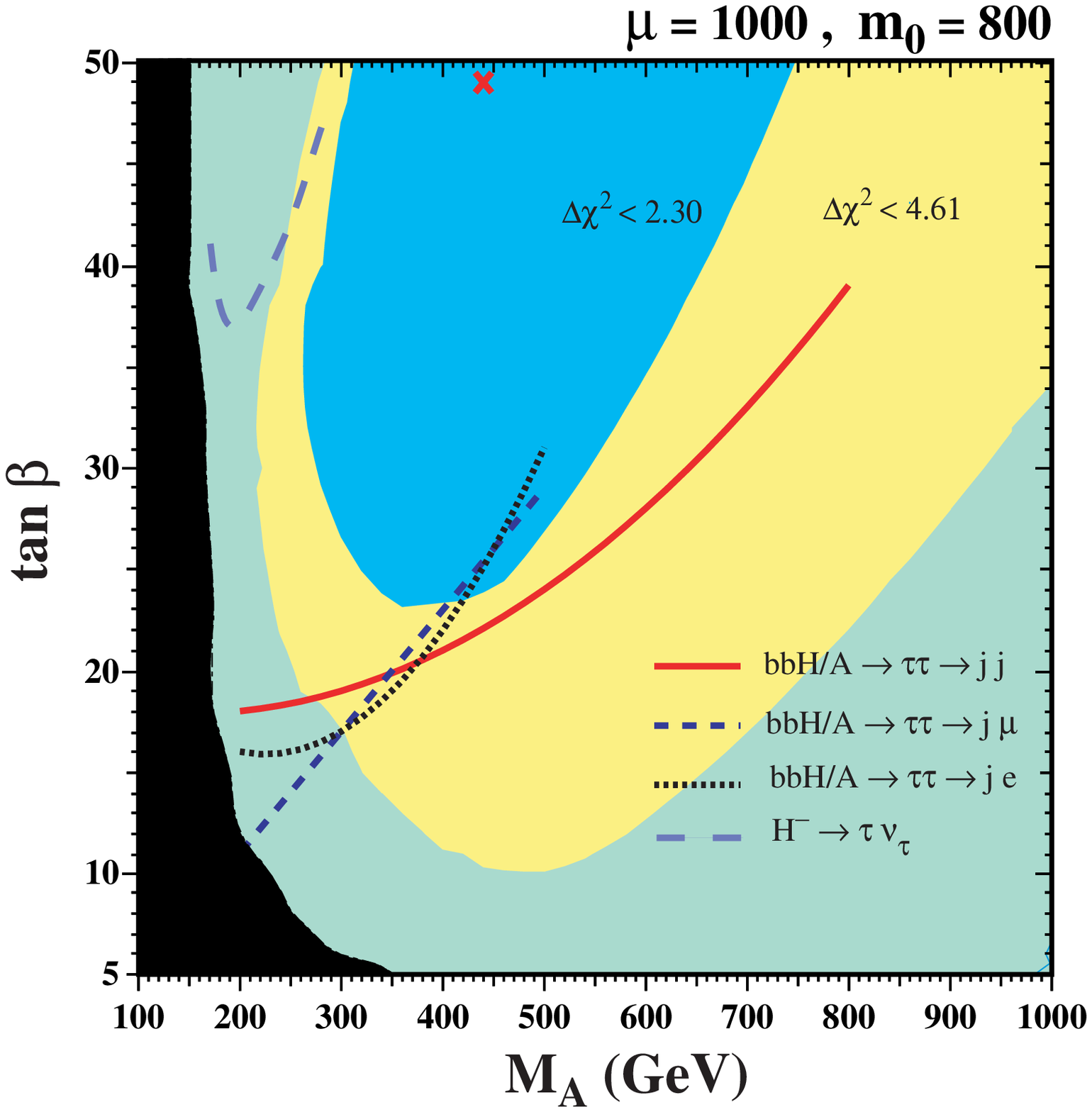}
%\hspace{-25mm}
\includegraphics[width=.49\textwidth]{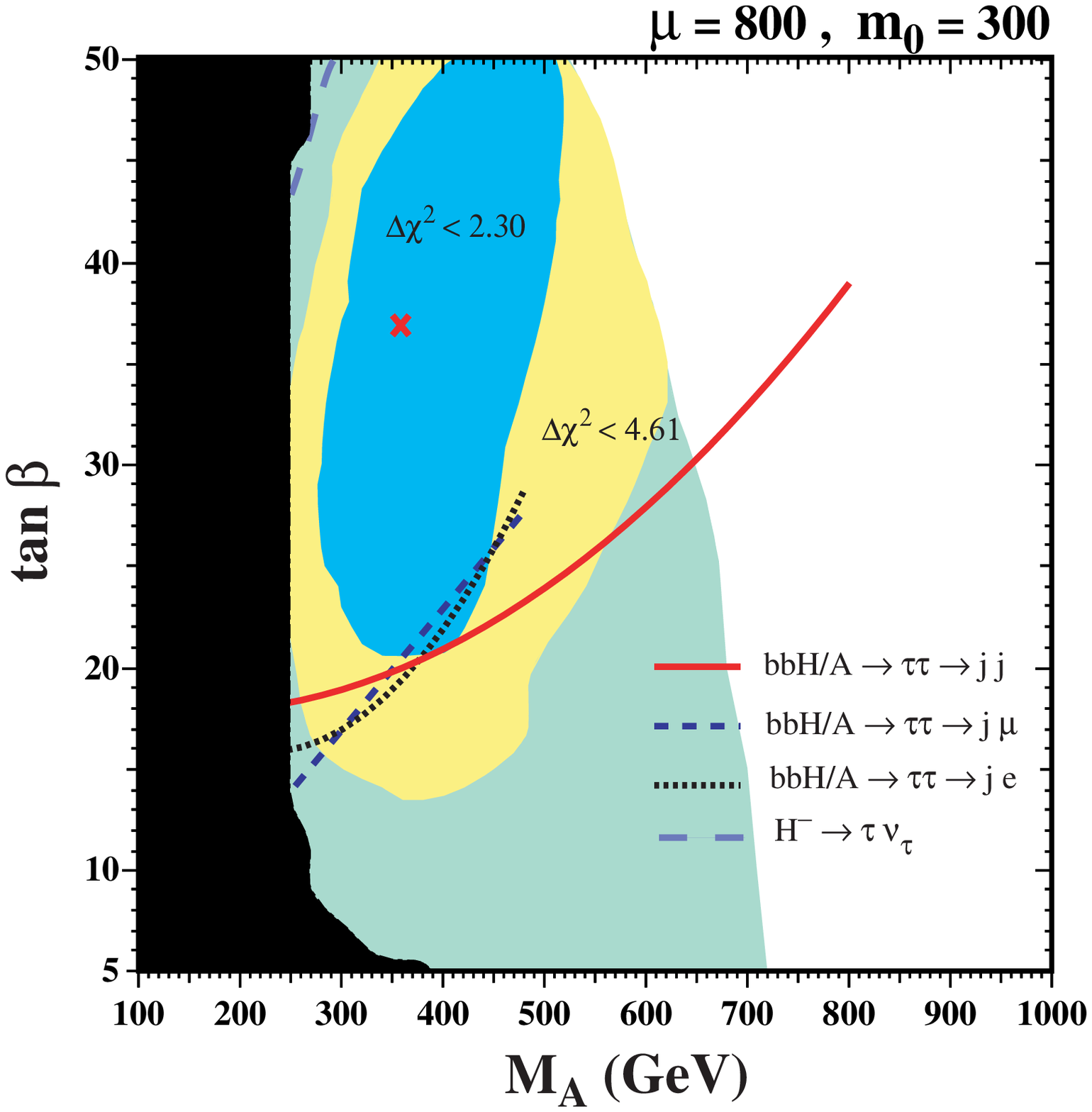}
\includegraphics[width=.49\textwidth]{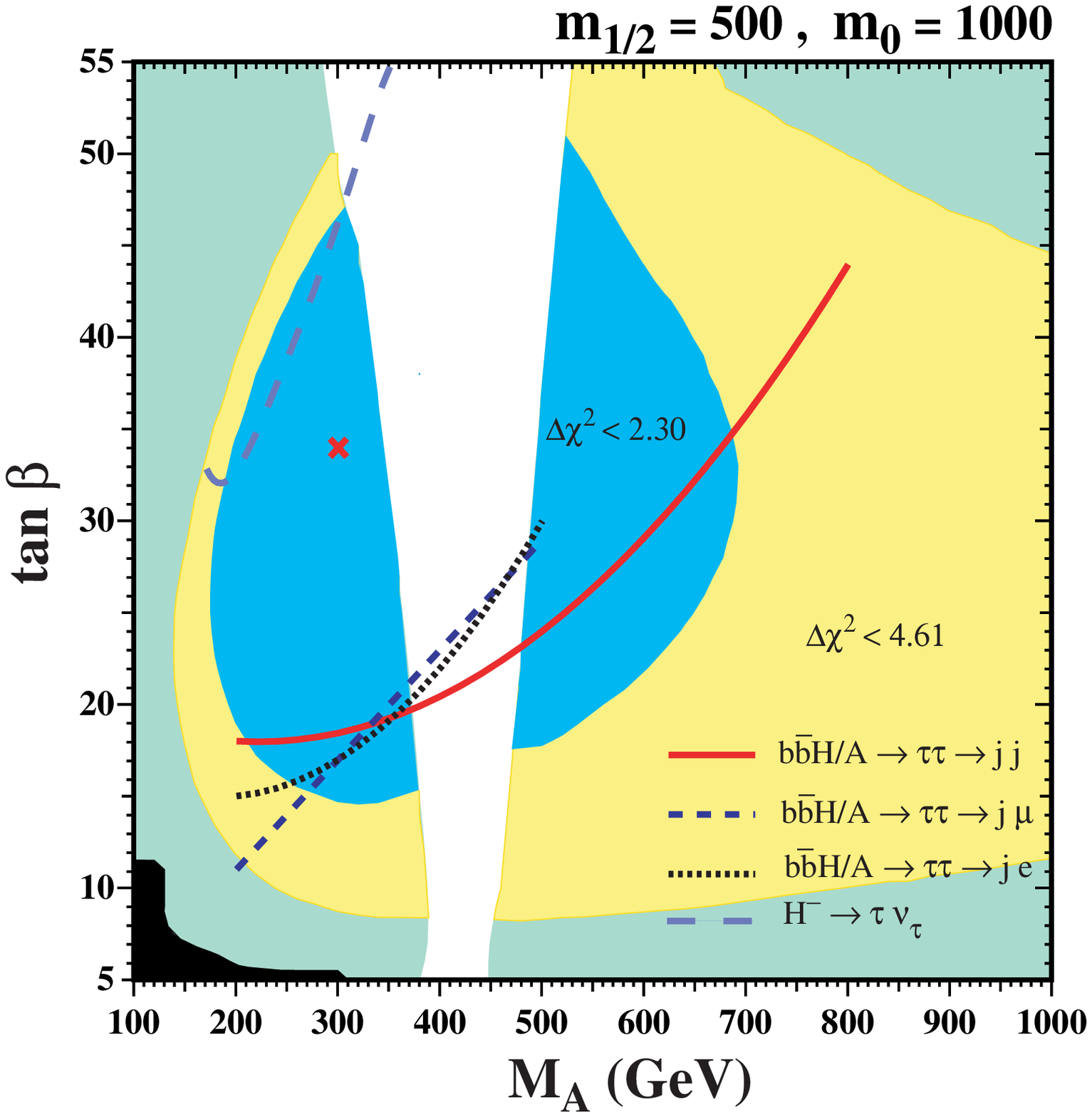}
%\hspace{-25mm}
\includegraphics[width=.49\textwidth]{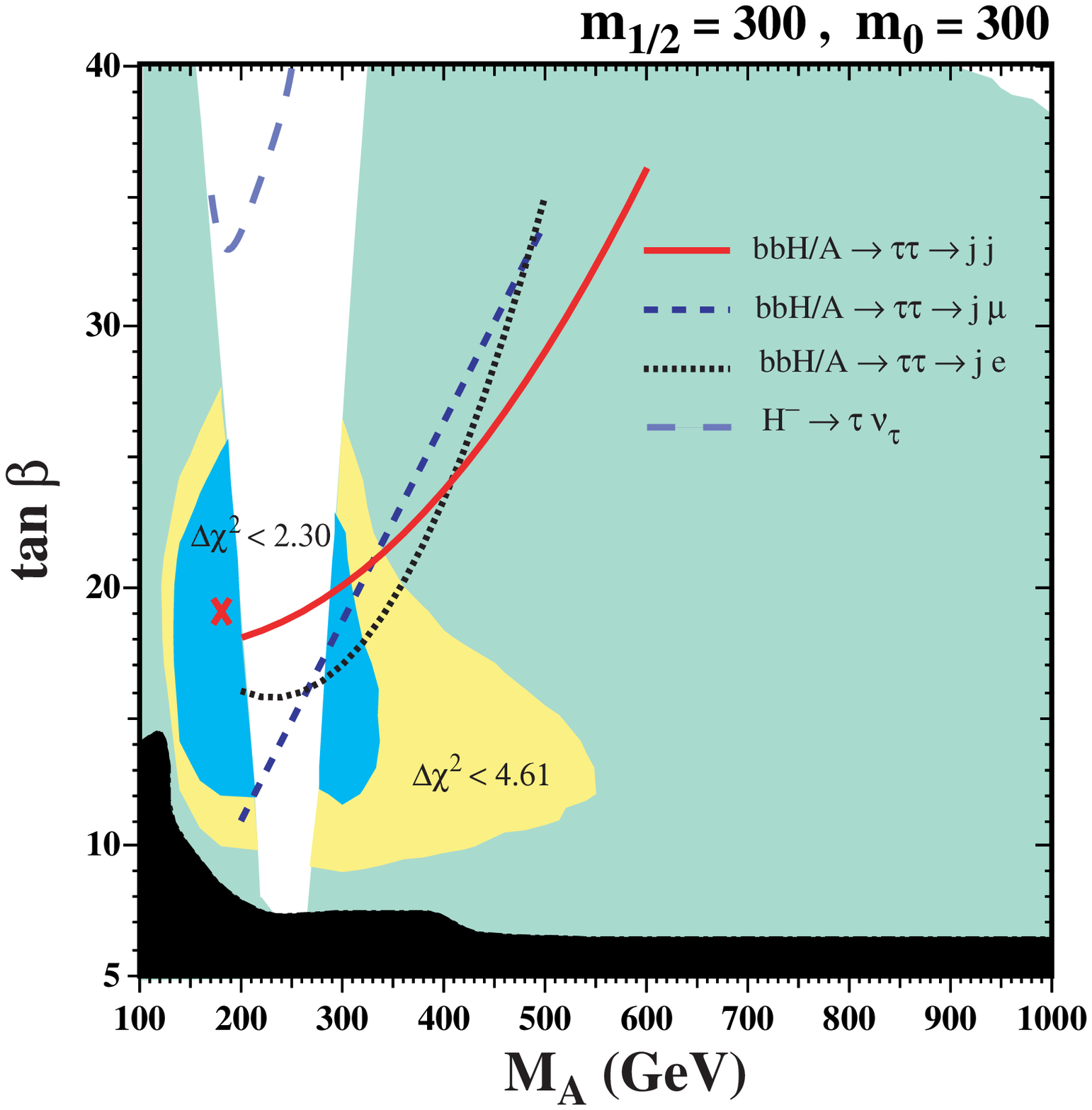}
\caption{%
The same $(\MA, \tb)$ planes for the NUHM benchmark surfaces (a)
\Athree, (b) \Afive, (c) \Atwo\ and (d) \Afour\ as in
Fig.~\protect\ref{fig:Mh}, displaying the 5-$\,\si$ discovery contours 
for $H/A \to \tau^+ \tau^-$ at the LHC with 60~or 30~fb$^{-1}$
(depending on the $\tau$ decay channels) and for $H^\pm \to \tau^\pm \nu$
detection in the
CMS detector when $M_{H^\pm} > m_t$ (see text).}
\label{fig:LHC02}
\end{center}
\vspace{1em}
\end{figure}
%%%%%%%%%%%%%%%%%%%%%%%%% F I G U R E %%%%%%%%%%%%%%%%%%%%%%%%%%%%%%%%%%%%%%%%%

We now turn to the heavy MSSM Higgs bosons. 
In \reffi{fig:LHC02} we display in the $(\MA, \tb)$ planes the
5-$\si$ discovery contours for 
$b\bar b \to H/A \to \tau^+ \tau^-$ at the LHC, where
the $\tau$'s decay to jets and electrons or muons (in the BR evaluation
for the heavy Higgs bosons possible decays to SUSY
particles~\cite{HtoSUSYearly,HtoSUSYexp,HtoSUSYrecent} have also
been taken into account). The analysis is based
on 60~\ifb\ for the final state 
$\tau^+\tau^- \to \,\mbox{jets}$~\cite{CMSPTDRjj} 
and on 30~\ifb\ for the $\tau^+\tau^- \to e + \,\mbox{jet}$~\cite{CMSPTDRej}
and $\tau^+\tau^- \to \mu + \mbox{jet}$~\cite{CMSPTDRmj} channels, collected
with the CMS detector. As shown in  \citere{HNW}, the impact of the
supersymmetric parameters other than $\MA$ and $\tb$ on the discovery
contours is relatively small in this channel, and the decays of $H/A$ to
SUSY particles~\cite{HtoSUSYearly,HtoSUSYexp,HtoSUSYrecent} are
in general 
suppressed by large sparticle masses. Only in \Afour\ the decay to the
lightest neutralinos and charginos is possible over nearly the whole
plane (see also \refse{sec:benchmarks}). Including such decays in the
evaluation of the discovery reach could increase the coverage for heavy
Higgs bosons somewhat.  
As a consequence of the relatively small impact of the other SUSY
parameters, the discovery contours  
in the four benchmark surfaces are similar to each other and
to those in the ``conventional'' benchmark
scenarios~\cite{HNW}.
The 5-$\si$ discovery contours for the various 
$\tau$ decay modes are shown separately: they may
each be scaled individually for different values of the jet (j), $\mu$ and
electron ($e$) detection efficiencies, see \citere{HNW}. The
sensitivities of the three 
different search strategies could in principle be combined, but information
required for making such a  combination is not yet available from the CMS
Collaboration. Nor is the information available that would be needed to
extend the discovery contours to small $\MA < 200$~GeV or to large 
$\MA > 500$ to 800~GeV. Nevertheless, we see that the whole
$\De \chi^2 < 2.30$ regions of the
surfaces \Athree\ and \Afive\ would be covered by the LHC $H/A \to \tau^+
\tau^-$ searches, and most of the corresponding regions of the surfaces
\Atwo\ and \Afour. Comparing the LHC sensitivities shown in
\reffi{fig:LHC02} with  the Tevatron sensitivities shown in
\reffi{fig:Tev02}, we see that the LHC provides access to
considerably heavier $H/A$, up to about 800~GeV, and that the
covered region extends to lower values of $\tb$, 
reaching $\tb \sim 10$ at low
$\MA$. Comparing with~\reffi{fig:LHC12}, we see that the $H/A \to \tau^+
\tau^-$ searches presumably also cover the regions at $\MA < 150$~GeV and
$\tan \beta > 11, 14$ that were left uncovered in the \Atwo\ and \Afour\
surfaces, 
respectively, by the $W^+ W^- \to h \to \tau^+\tau^-$ searches. It
would be interesting to verify this by means of an
extension of the available CMS analysis.

We also show in Fig.~\ref{fig:LHC02} the 5-$\si$ contours
for discovery of the $H^\pm$ via its $\tau^\pm \nu$ decay mode
at the LHC, in the case
$\MHp > \mt$. We see that the coverage is
limited in each of the scenarios \Athree, \Afive, \Atwo\ and \Afour\
to $\MA < 300 \gev$ and $\tb > 30$, reaching a small part of the $\De
\chi^2 < 2.30$ region of surface \Atwo, only a small part of the $\De
\chi^2 < 4.61$ region of surface \Athree, and not even reaching this
region in scenarios \Afive\ and \Afour. One may also search for
$H^\pm \to \tau^\pm \nu$ for lighter $\MHp < \mt$, but in the
cases of surfaces \Athree\ and \Afive\ this would be useful only in the
regions already excluded by LEP, and the accessible regions in surfaces
\Atwo\ and \Afour\ would also be quite limited.

Another class of possible measurements at the LHC comprises the precise
determinations of $h$ decay branching ratios~\cite{HcoupLHCSM}, 
and using their ratios to search for deviations from the SM predictions
for a Higgs boson of the same mass. 
Such deviations may arise in the MSSM due to
differences in the tree-level couplings and due to additional (loop)
corrections. 
The most sensitive observable is likely to be the ratio of 
$\br(h \to \tau^+\tau^-)/\br(h \to WW^*)$.
We display in \reffi{fig:brLHC} the 1-, 2-, 3-
and 5-$\,\si$ contours (2-$\,\si$ in bold) for SUSY induced deviations
of this ratio of branching ratios from the SM prediction 
(with $\MHSM = \Mh$). The contours
correspond to an integrated luminosity at the LHC of 30~or
300~fb$^{-1}$~\cite{LHCbrR} (assuming SM decay rates). 
An experimental resolution for 
$\br(h \to \tau^+\tau^-)/\br(h \to WW^*)$ between 30\% (28\%) and 45\%
(33\%) can be achieved for 30 (300)~\ifb. For $\Mh = 120 \gev$ the
corresponding precision is 38\% (29\%).
The most promising surfaces are \Atwo\ and \Afour, and we 
see that over essentially all the left lobe of the $\De \chi^2 < 4.61$
region for \Afour\ a 
5-$\si$ discrepancy with the SM should be detectable%
\footnote{ 
It should be kept in mind that the actual experimental precision on the
ratio $\br(h \to \tau^+\tau^-)/\br(h \to WW^*)$ will be different in
this parameter region from the numbers quoted above which assume SM
rates.}%
. On the other
hand, only partial coverage of the left lobe of surface \Atwo\ would
be possible, and the sensitivities in the right lobes of \Afour\ and
\Atwo\ and in the \Athree\ and \Afive\
surfaces are considerably less promising. Nevertheless, measuring 
$\br(h \to \tau^+\tau^-)/\br(h \to WW^*)$ does offer the prospect
of distinguishing  between the NUHM and the SM 
in the low $\MA$ regions of surfaces \Atwo\ and \Afour.

 %%%%%%%%%%%%%%%%%%%%%%%%% F I G U R E %%%%%%%%%%%%%%%%%%%%%%%%%%%%%%%%%%%%%%%%
\begin{figure}[htb!]
\vspace{10mm}
\begin{center}
\includegraphics[width=.47\textwidth]{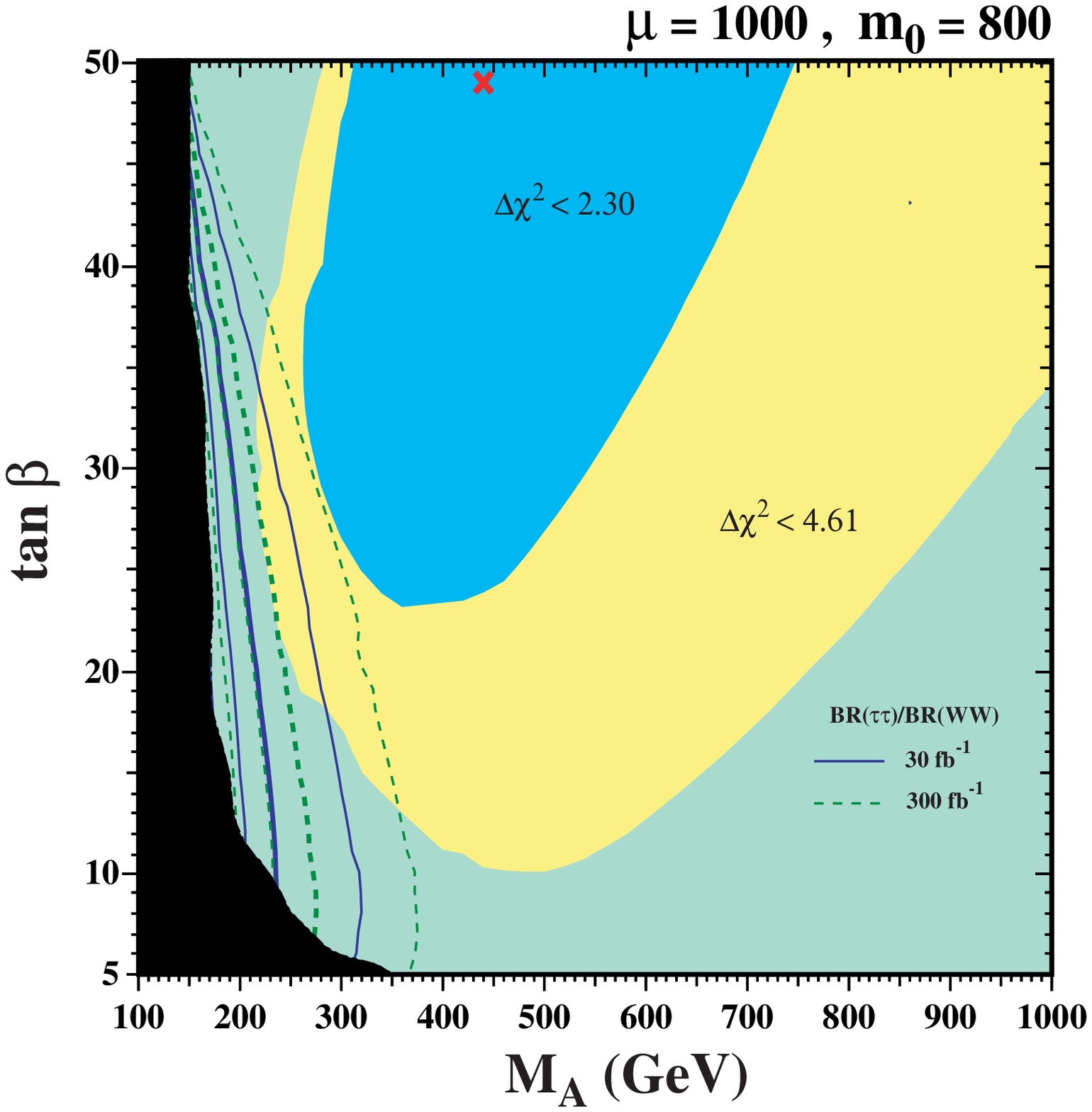}
%\hspace{-25mm}
\includegraphics[width=.47\textwidth]{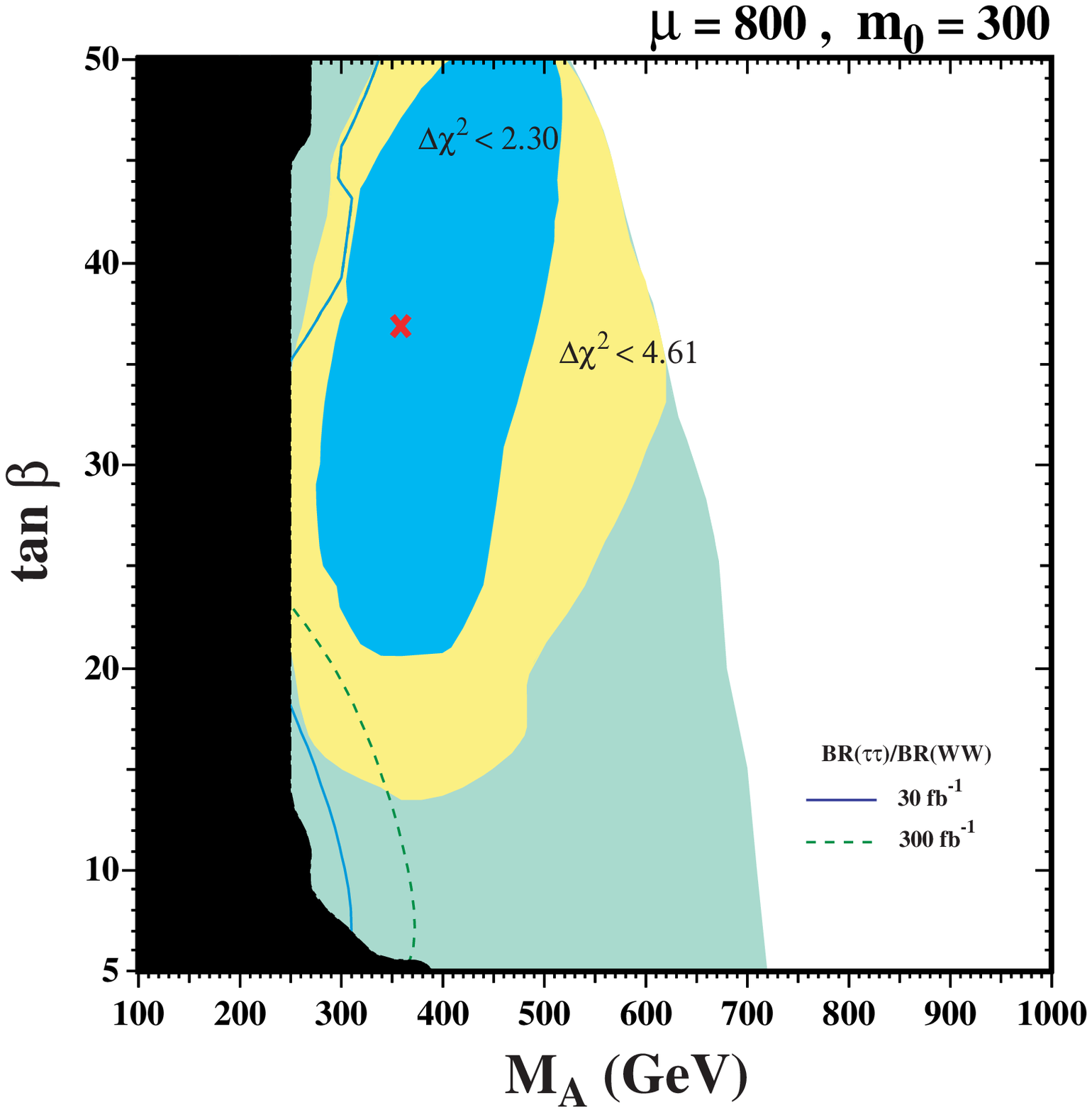}
\includegraphics[width=.47\textwidth]{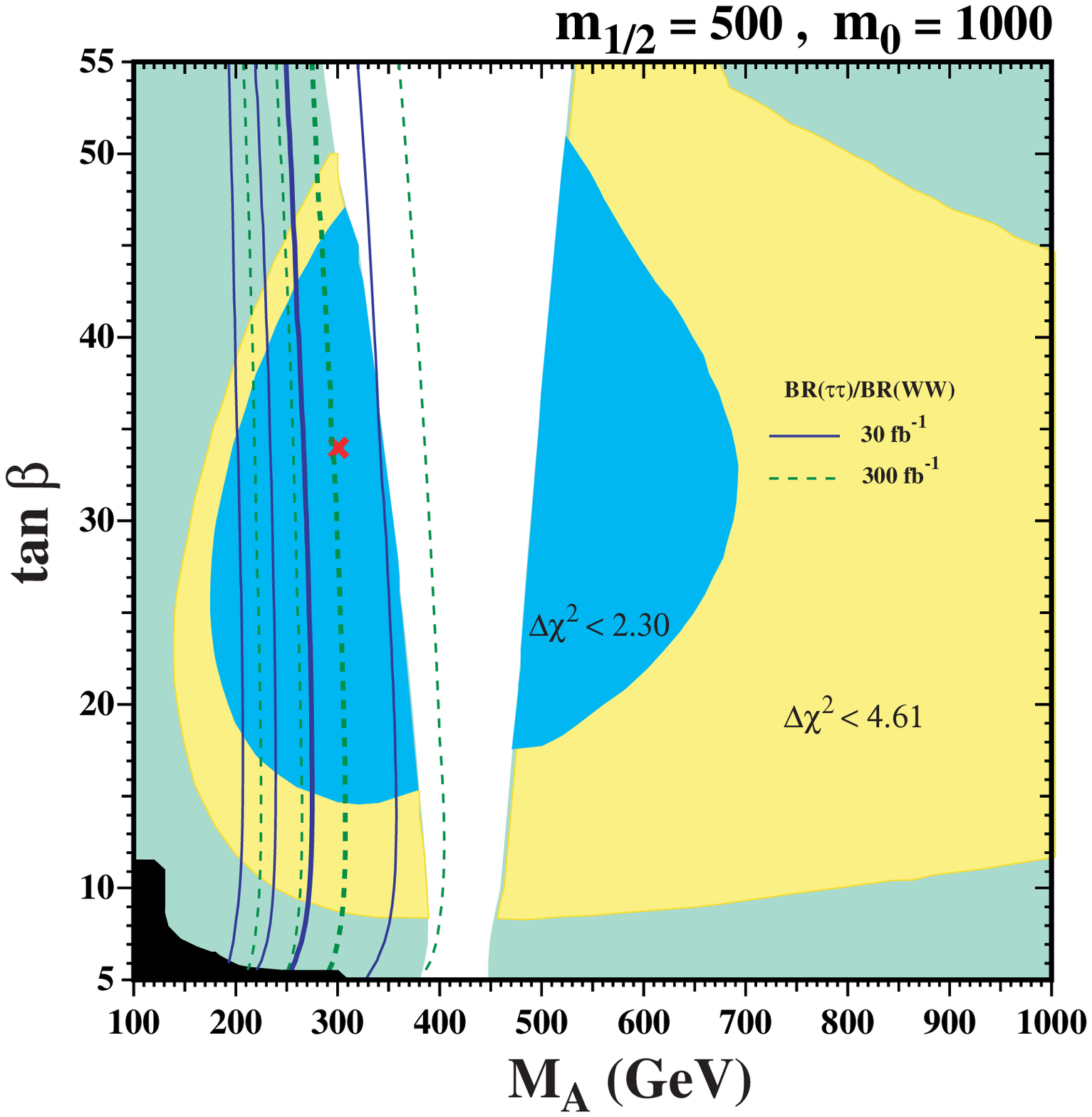}
%\hspace{-25mm}
\includegraphics[width=.47\textwidth]{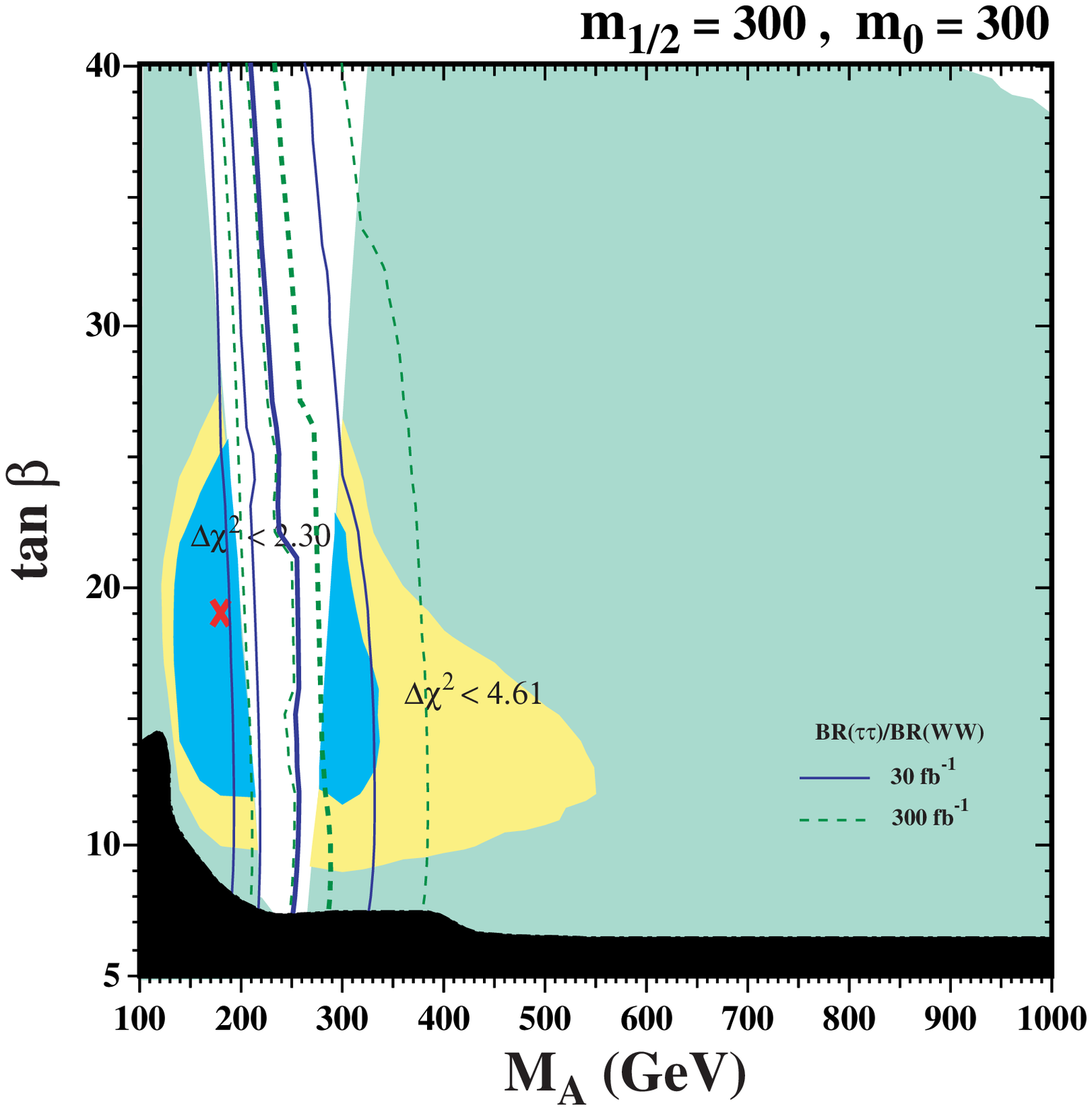}
%\hspace{-5mm}
%\vspace{-25mm}
\caption{%
The same $(\MA, \tb)$ planes for the NUHM benchmark surfaces (a)
\Athree, (b) \Afive, (c) \Atwo\ and (d) \Afour\ as in
Fig.~\protect\ref{fig:Mh}, displaying the 1-, 2-, 3- and 5-$\,\si$
contours (2-$\si$ in bold) 
for SUSY-induced deviations on the ratio
$\br(h \to \tau^+\tau^-)/\br(h \to WW^*)$ at the LHC with 30~or 300~fb$^{-1}$
(see text). In the case of surface \Afive, only 1-$\,\si$ curves are
seen in the lower part of the figure. 
The upper curves correspond to 0-$\,\si$.
}
\label{fig:brLHC}
\end{center}
\vspace{1em}
\end{figure}
%%%%%%%%%%%%%%%%%%%%%%%%% F I G U R E %%%%%%%%%%%%%%%%%%%%%%%%%%%%%%%%%%%%%%%%%

%%%%%%%%%%%%%%%%%%%%%%%%%%%%%%%%%%%%%%%%%%%%%%%%%%%%%%%%%%%%%%%%%%%%%%%%%%%%%%%
%%%%%%%%%%%%%%%%%%%%%%%%%%%%%%%%%%%%%%%%%%%%%%%%%%%%%%%%%%%%%%%%%%%%%%%%%%%%%%%

\section{ILC Phenomenology}
\label{sec:ilc}

In this section we analyze the deviations in the branching ratios of
the lightest MSSM Higgs boson to SM fermions and gauge bosons in
comparison with a SM 
Higgs boson of the same mass that could be measured at the ILC (see also
\citere{ehow2}). 
The experimental precisions for the branching ratios we
analyze are summarized in \refta{tab:brexp}.

%%%%%%%%%%%%%%%%%%%%%%%%%%%%%%% T A B L E %%%%%%%%%%%%%%%%%%%%%%%%%%%%%%%%%%%%%
\begin{table}[tbh!]
\renewcommand{\arraystretch}{1.5}
\BC
\begin{tabular}{|c|c|c|}
\hline\hline
collider & channel & exp.\ precision [\%] \\ 
\hline\hline
ILC(500)  & $\br(h \to b \bar b)$ & 1.5 \\ \hline
ILC(500)  & $\br(h \to \tau^+\tau^-)$ & 4.5 \\ \hline
ILC(500)  & $\br(h \to WW^*)$ & 3.0 \\ \hline
ILC(1000) & $\br(h \to b \bar b)/\br(h \to WW^*)$ & 1.5 \\
\hline\hline
\end{tabular}
\EC
\renewcommand{\arraystretch}{1}
\caption{
Experimental precisions at the ILC for various branching ratios of
the lightest MSSM Higgs
boson (assuming SM decay rates)~\cite{talkbrient,Snowmass05Higgs,barklow}.
The experimental precision in the last column corresponds to $1\,\si$ in
the plots below. 
ILC(500,1000) refers to a center-of-mass energy of $500, 1000 \gev$,
respectively. 
}
\label{tab:brexp}
\end{table}
%%%%%%%%%%%%%%%%%%%%%%%%%%%%%%% T A B L E %%%%%%%%%%%%%%%%%%%%%%%%%%%%%%%%%%%%%

We show in \reffi{fig:hbb} the prospective sensitivity of an
ILC measurement of the $\br(h \to b \bar b)$ in the four
$(\MA, \tb)$ planes. The experimental precision is anticipated to be
1.5\%, see \refta{tab:brexp}. 
We display as solid (blue) lines the contours of the 
$+5, +3, +2, +1, 0\,\si$ deviations (with $+2\,\si$ in bold)
of the MSSM result from the corresponding SM result
(for low $\MA$ and large $\tb$ in \Afive\ we also find
contours for $-2, -1\, \sigma$, with $-2\,\si$ in bold). 
The separations between the contours indicate how sensitively the SUSY
results depend on variations of $\MA$ and $\tb$.
Also shown in \reffi{fig:hbb} via dashed (green) lines is the
sensitivity to SUSY effects of the ILC measurement of the ratio 
of branching ratios $\br(h \to b \bar b)/\br(h \to WW^*)$
(for low $\MA$ and large $\tb$ in \Afive\ we also find
contours for $-5, -3, -2, -1\, \sigma$).
The precision measurement of the ratio $\br(h \to b \bar b)/\br(h \to WW^*)$
clearly provides a much higher sensitivity to SUSY effects than the 
measurement of $\br(h \to b \bar b)$ alone (see also \citere{deschi}). 

For the ILC measurement of the $\br(h \to b \bar b)$, in the cases of 
\Athree\ and \Afive\ we see that the prospective sensitivities are less
than 3~$\si$ throughout almost all the regions with
$\De \chi^2 < 4.61$. The situations are different, however, for the
planes \Atwo\ and \Afour. In each case, the cosmologically-favoured
region is divided into separate lobes at low and high $\MA$. In the
\Atwo\ case, the measurement of $\br(h \to b \bar b)$ would be
sufficient to establish a SUSY effect with more than five
$\si$ throughout most of the low-$\MA$ lobe, and all of it in the
\Afour\ case. A precision measurement of 
$\br(h \to b \bar b)/\br(h \to WW^*)$ yields a significant improvement
for all benchmark surfaces. 
We see that, in case \Athree, the sensitivity
already exceeds 5~$\si$ in much of the region with $\De \chi^2
< 2.30$, and the fraction of this region covered at the 5-$\si$ level
is even larger
in the case \Afive. Even more encouragingly, in the case \Atwo\ the
sensitivity exceeds 5~$\si$ throughout the $\De \chi^2 < 2.30$
region, and in the case \Afour\ it exceeds 5~$\si$ by a
substantial amount throughout the $\De \chi^2 < 4.61$ region.

%%%%%%%%%%%%%%%%%%%%%%%%% F I G U R E %%%%%%%%%%%%%%%%%%%%%%%%%%%%%%%%%%%%%%%%%
\begin{figure}[htb!]
\vspace{10mm}
\begin{center}
\includegraphics[width=.47\textwidth]{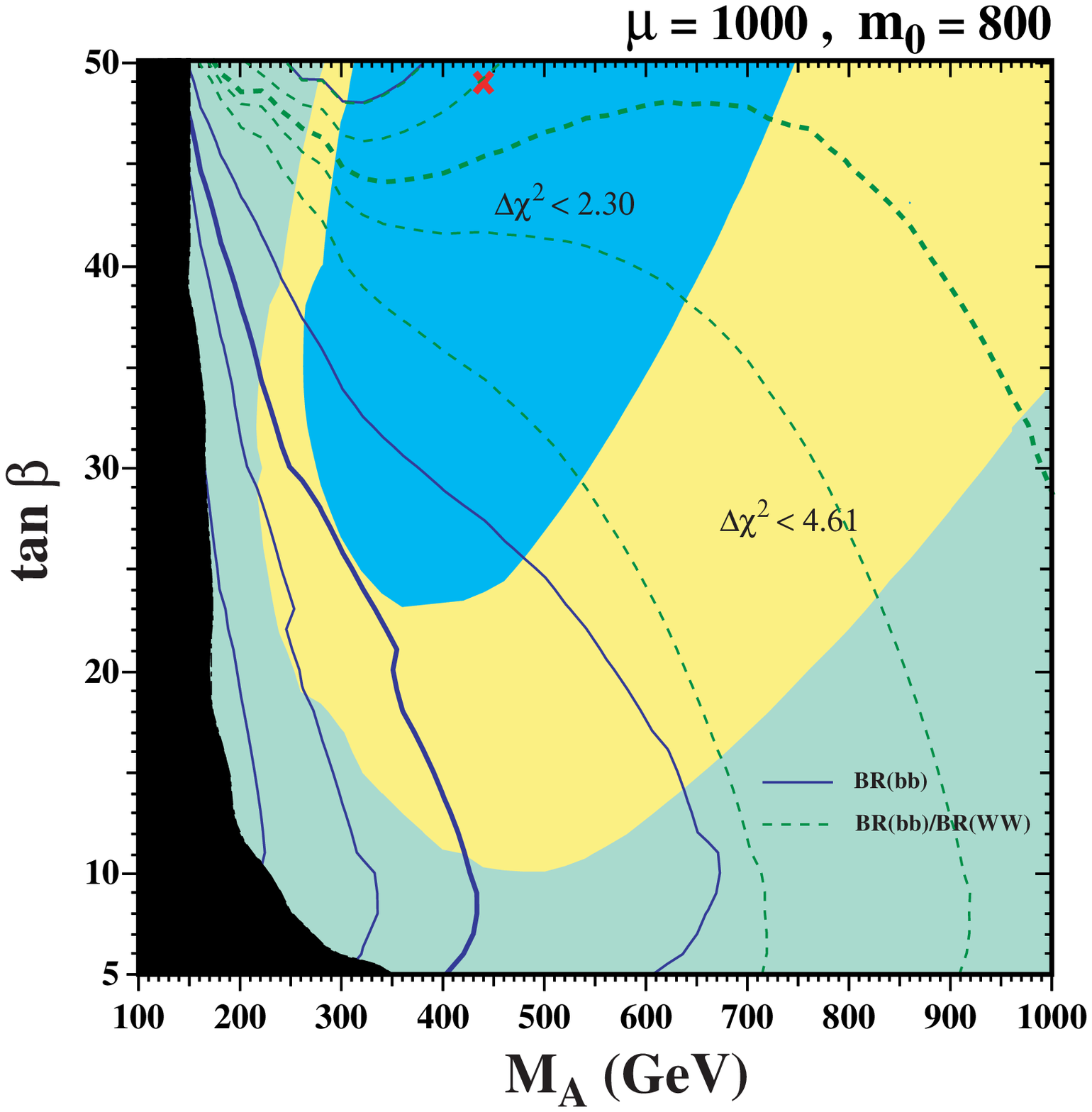}
%\hspace{-25mm}
\includegraphics[width=.47\textwidth]{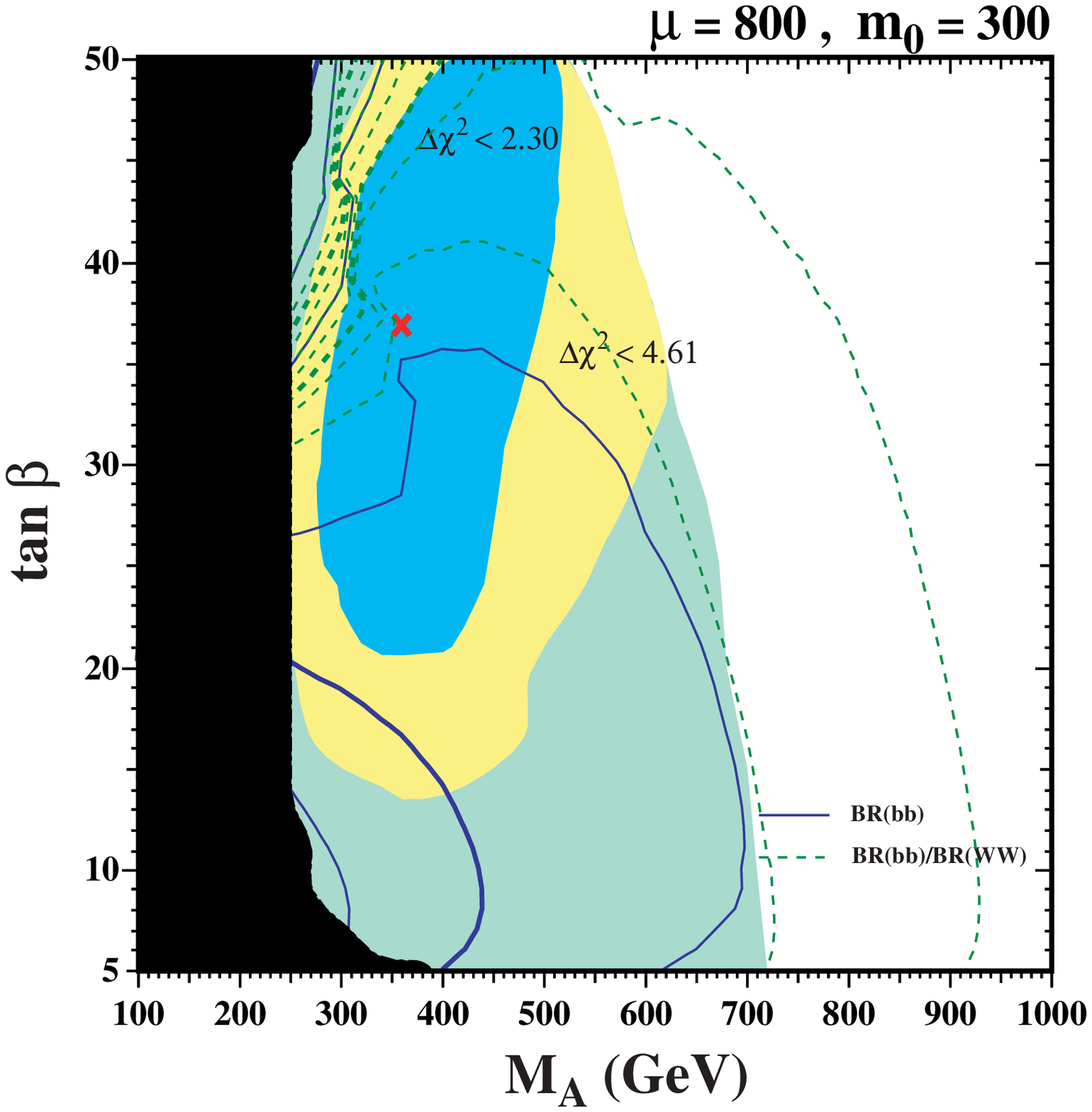}
\includegraphics[width=.47\textwidth]{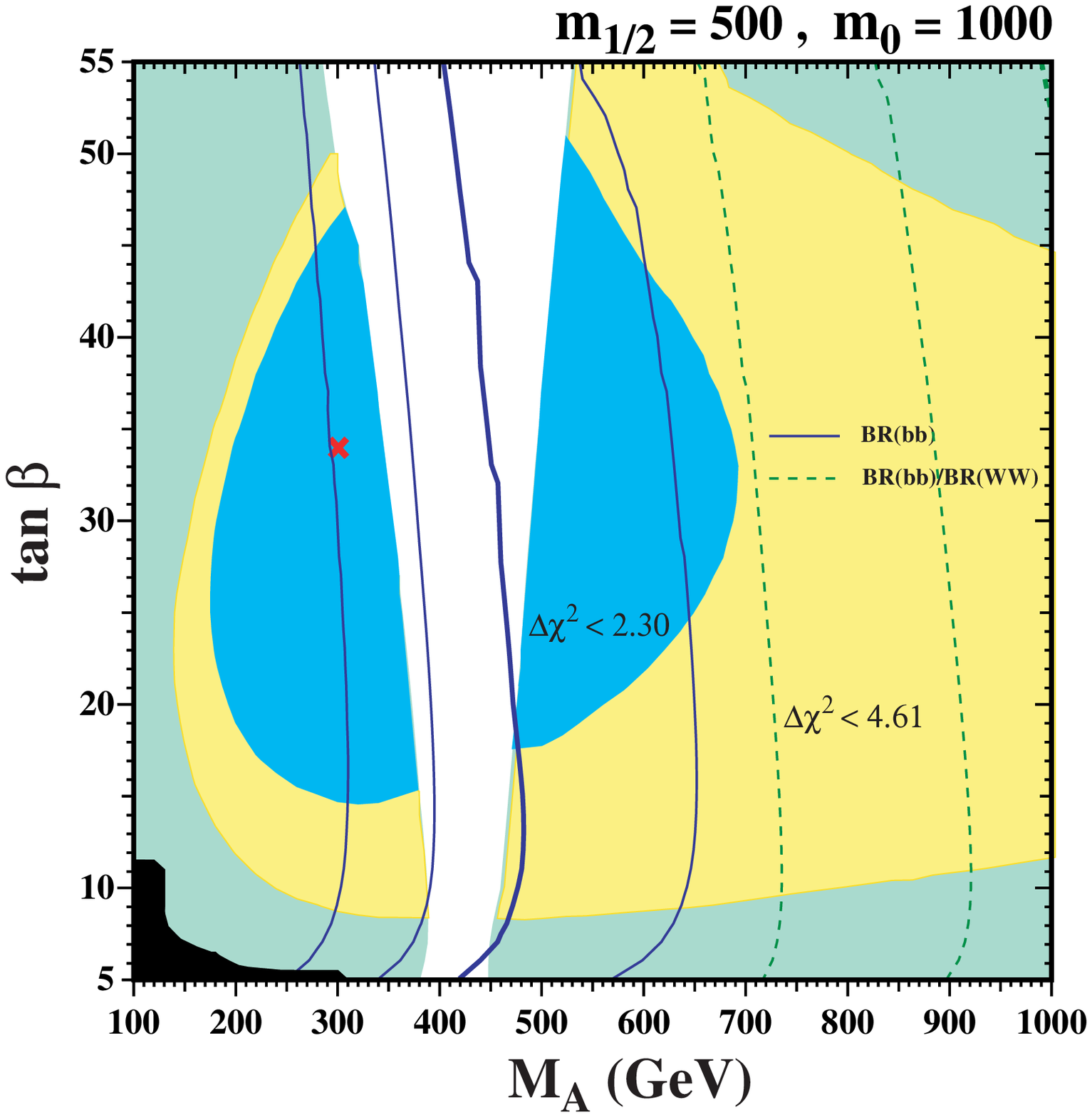}
%\hspace{-25mm}
\includegraphics[width=.47\textwidth]{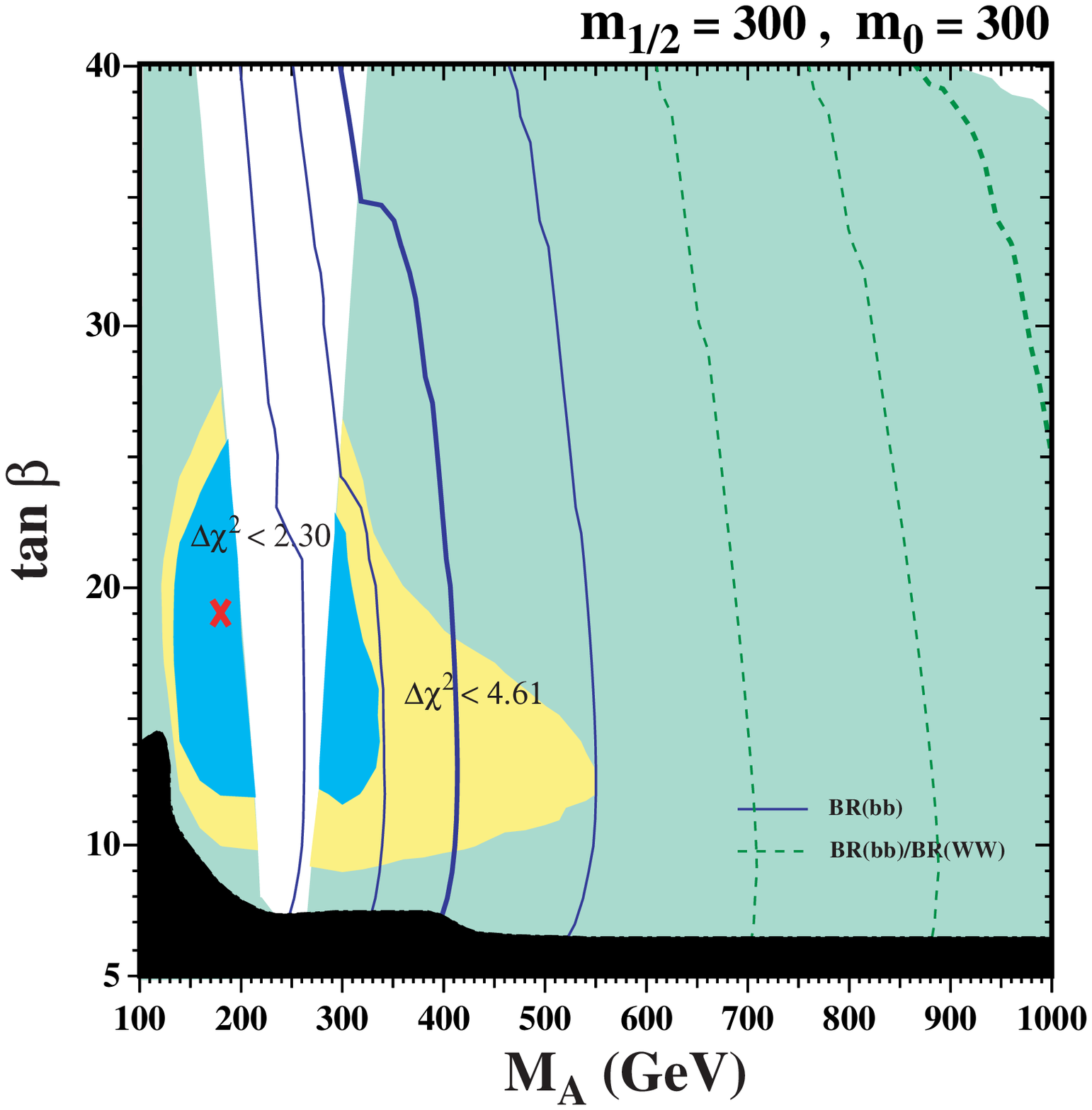}
%\hspace{-5mm}
%\vspace{-25mm}
\caption{%
The same $(\MA, \tb)$ planes for the NUHM benchmark surfaces (a)
\Athree, (b) \Afive, (c) \Atwo\ and (d) \Afour\ as in
Fig.~\protect\ref{fig:Mh}, displaying $5,3,2,1,0$-$\si$ sensitivity contours 
(2-$\si$ in bold) for SUSY effects on $\br(h \to b \bar b)$ (solid blue
lines) and $\br(h \to b \bar b)/\br(h \to WW^*)$ (dashed green lines)
at the ILC (see text). Note that for surface \Afive\ for low $\MA$ and
large $\tb$ also $-2, -1$-$\si$
are shown for $\br(h \to \bar b b)$, and 
$-5, -3, -2, -1$-$\si$
%and $0$-$\si$ 
are shown for $\br(h \to \bar b b)/\br(h \to WW^*)$.}
\label{fig:hbb}
\end{center}
%\vspace{1em}
\end{figure}
%%%%%%%%%%%%%%%%%%%%%%%%% F I G U R E %%%%%%%%%%%%%%%%%%%%%%%%%%%%%%%%%%%%%%%%%

Next, we show in \reffi{fig:htt} the prospective sensitivity of an ILC
measurement of the $\br(h \to \tau^+\tau^-)$ in the four
\plane{\MA}{\tb}s, using solid (red) contours. In the cases of
\Athree\ and \Afive, we again see that the prospective sensitivities are
less than 3~$\si$ throughout almost all the regions with 
$\De \chi^2 < 4.61$.
In the cases of planes \Atwo\ and \Afour, the
sensitivities are greater, but less than the corresponding sensitivities
to the $\br(h \to b \bar b)$ shown previously in
\reffi{fig:hbb}. Of all the single ILC measurements, the one with the greatest
sensitivity to SUSY effects is that of the $\br(h \to WW^*)$,
which is also shown in \reffi{fig:htt} using dashed
(black) lines. In the cases \Athree\ and \Afive, we see that the
sensitivity may rise above 5~$\si$ already within the $\De
\chi^2 < 4.61$ region. In the case of \Atwo, the sensitivity is well
above 5~$\si$ throughout the low-$\MA$ region. In the case of \Afour,
a 5-$\si$ significance is exceeded already in much of the
high-$\MA$ lobe, where the sensitivity never falls as low as 
3~$\si$ in the $\chi^2$ favored region.

%%%%%%%%%%%%%%%%%%%%%%%%% F I G U R E %%%%%%%%%%%%%%%%%%%%%%%%%%%%%%%%%%%%%%%%%
\begin{figure}[htb!]
\vspace{10mm}
\begin{center}
\includegraphics[width=.46\textwidth]{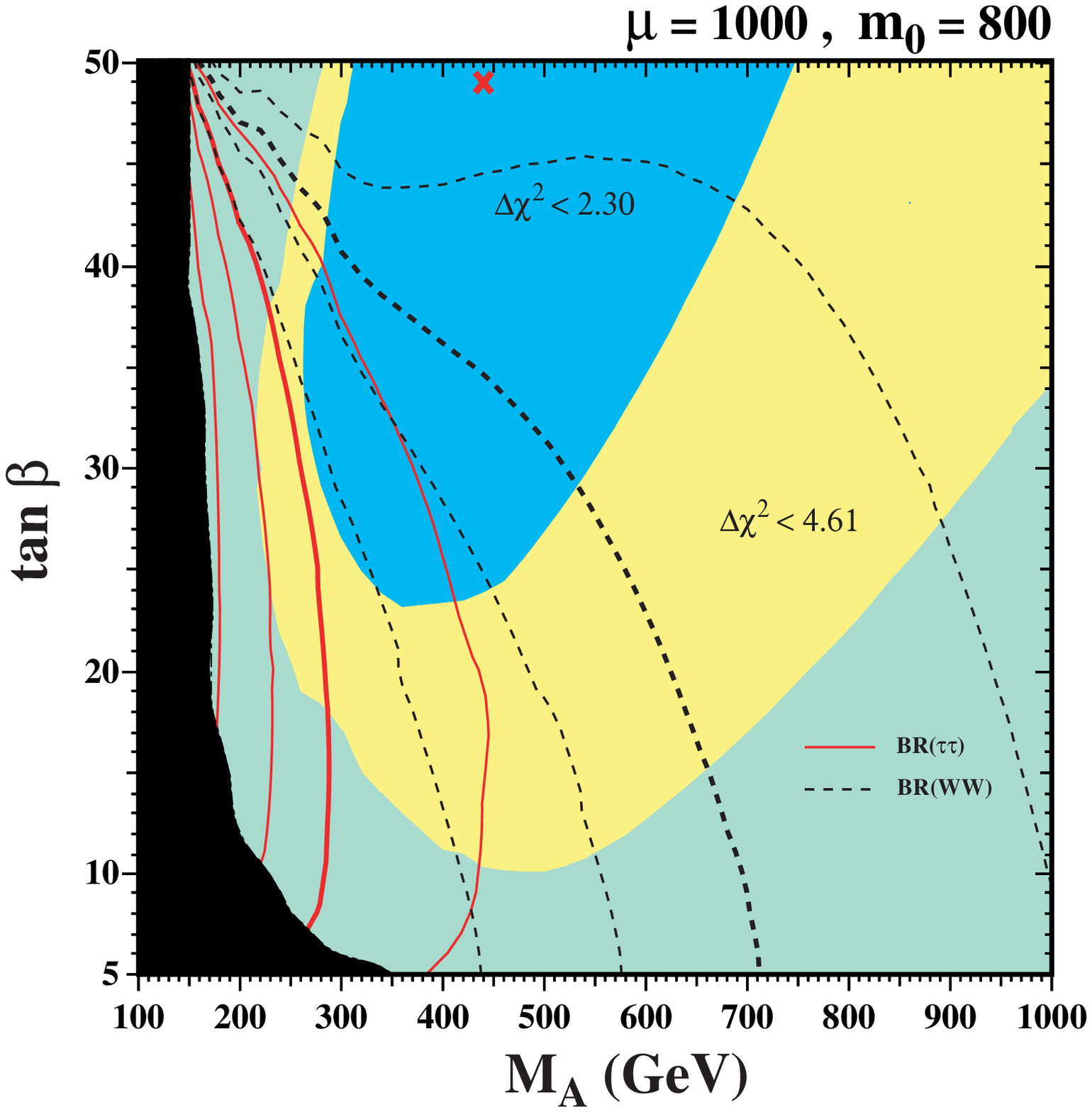}
%\hspace{-25mm}
\includegraphics[width=.46\textwidth]{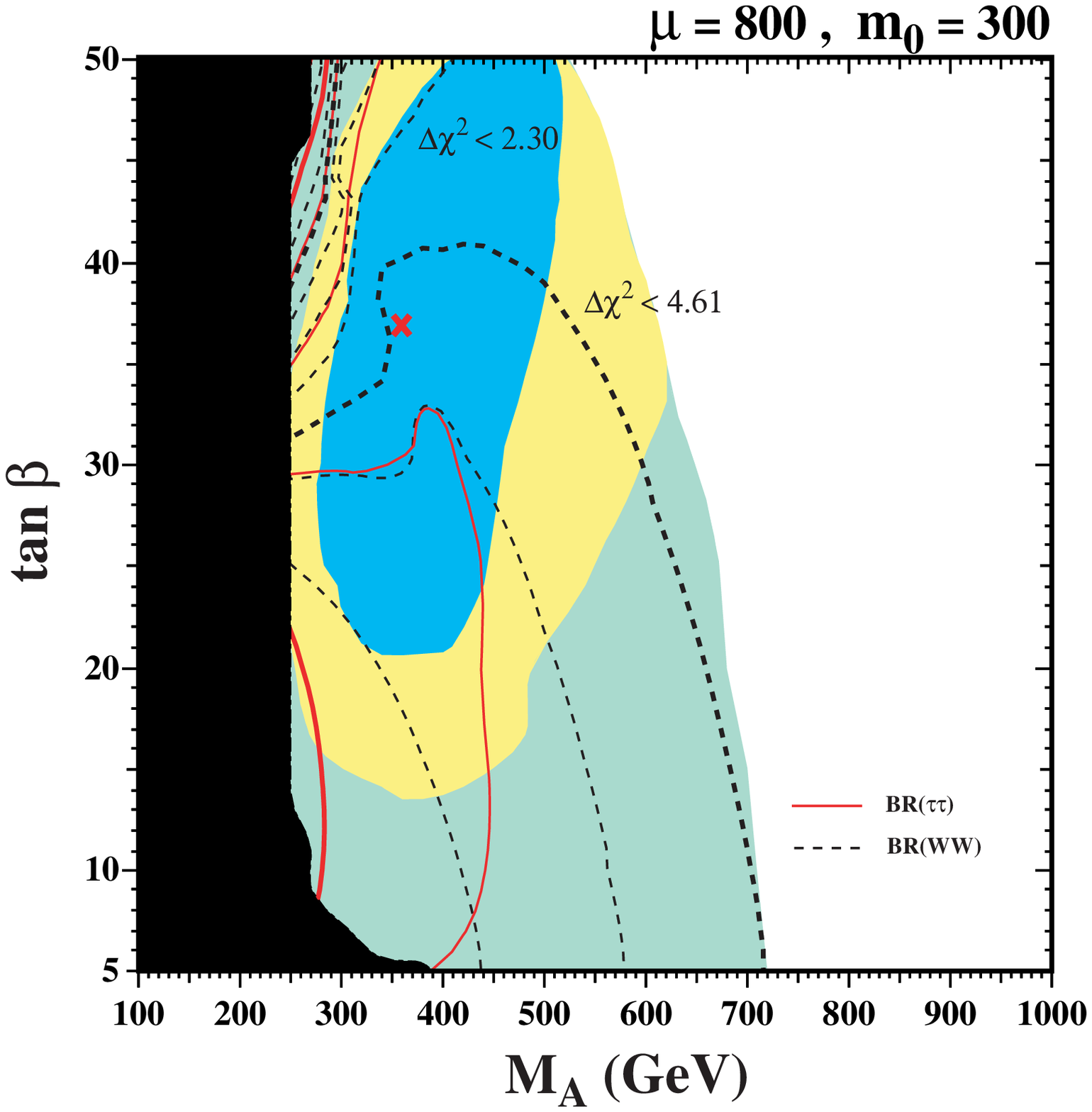}
\includegraphics[width=.46\textwidth]{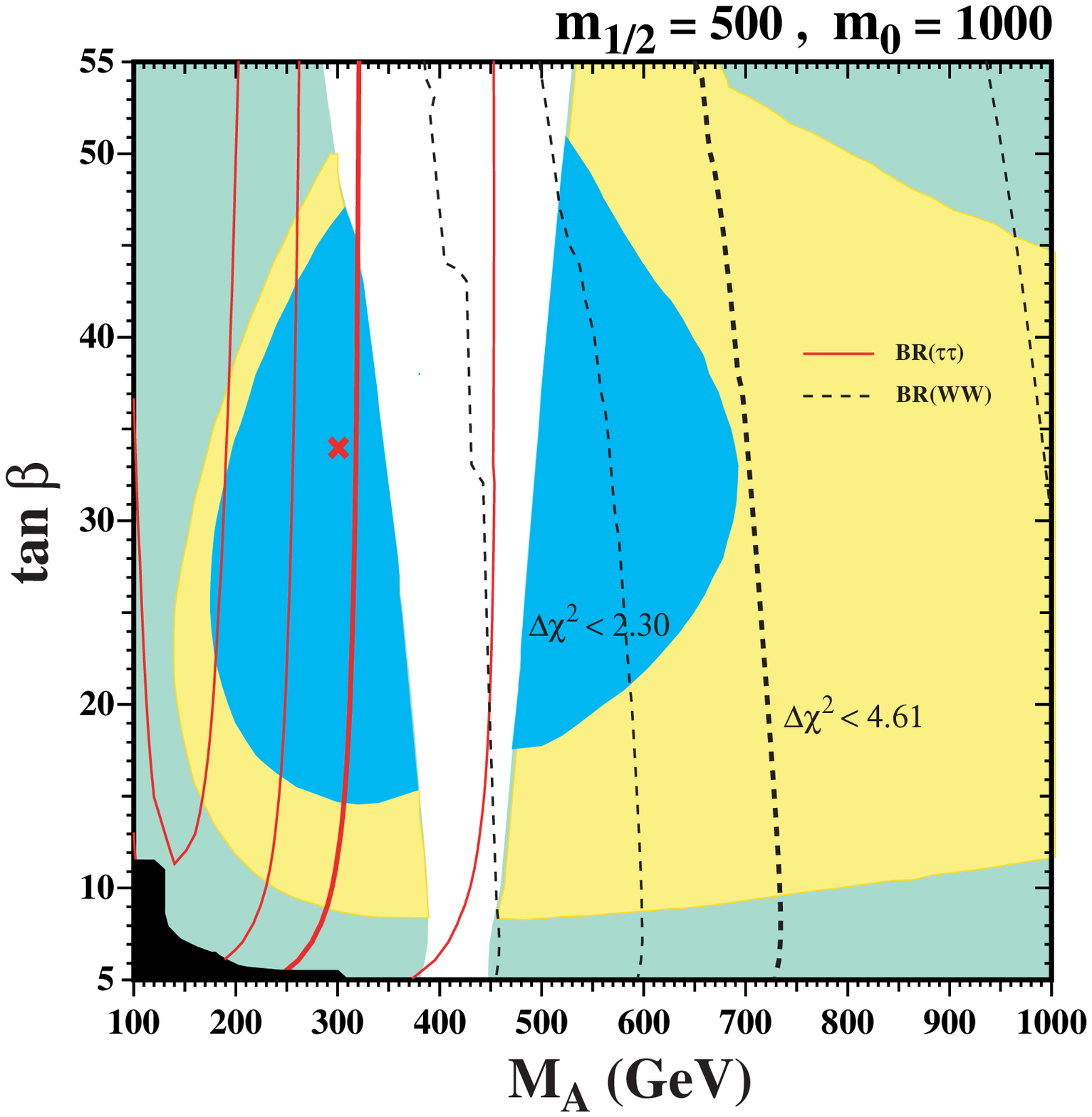}
%\hspace{-25mm}
\includegraphics[width=.46\textwidth]{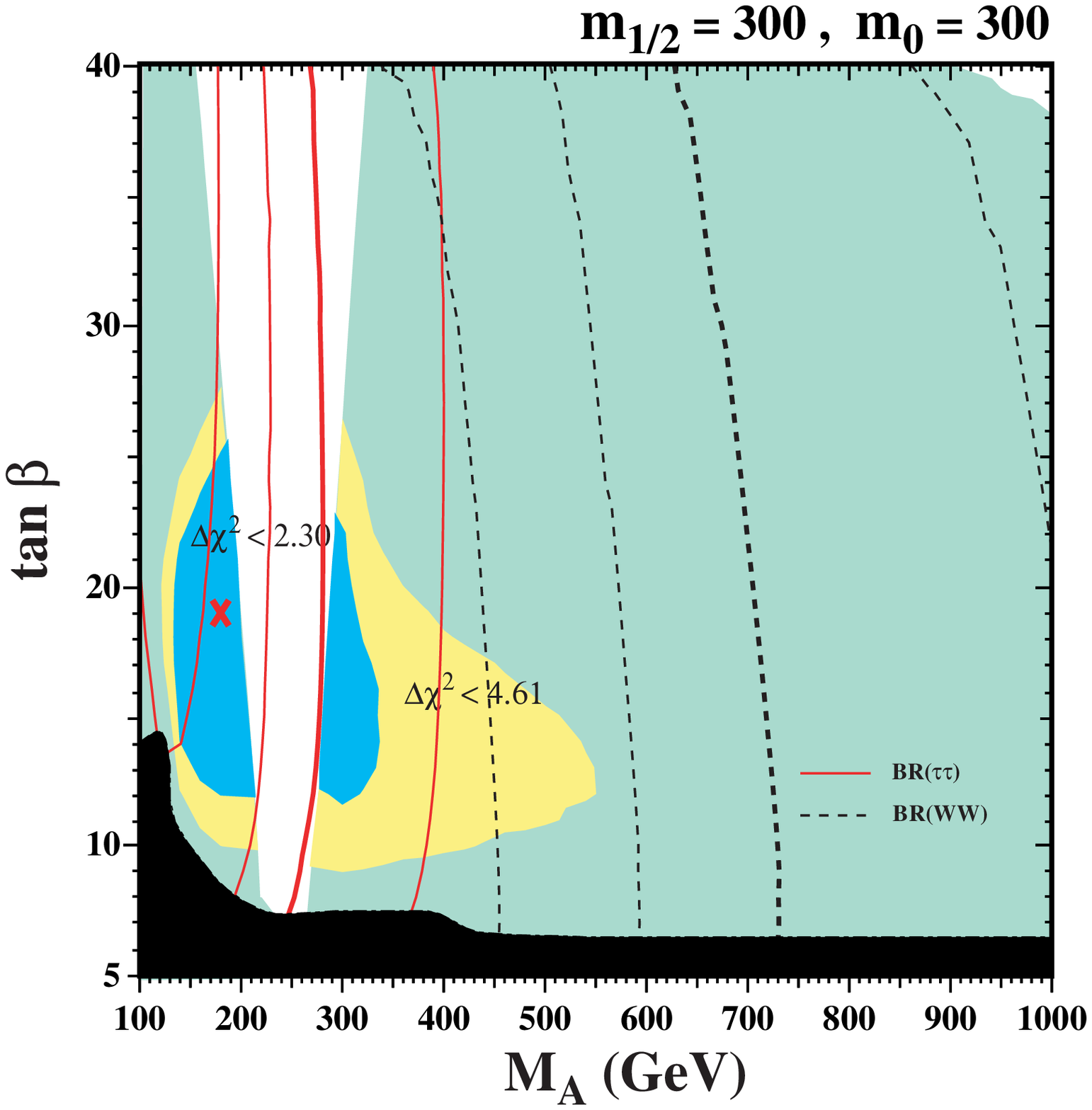}
%\hspace{-5mm}
%\vspace{-25mm}
\caption{%
The same $(\MA, \tb)$ planes for the NUHM benchmark surfaces (a)
\Athree, (b) \Afive, (c) \Atwo\ and (d) \Afour\ as in
Fig.~\protect\ref{fig:Mh}, displaying $5,3,2,1$-$\si$ sensitivity contours 
for SUSY effects on the $\br(h \to \tau^+ \tau^-)$ at the ILC (solid
red lines). Also shown are the $-5,-3,-2,-1$-$\si$ sensitivity contours 
for the SUSY effects on $\br(h \to WW^*)$ at the ILC 
(dashed black lines). Note that for surface \Afive\, $\pm 2, \pm 1$ and 
$0$-$\si$ are shown for $\br(h \to \tau^+ \tau^-)$, and 
$\pm 5, \pm 3, \pm 2, \pm 1$ and $0$-$\si$
are shown for $\br(h \to WW^*)$.}
\label{fig:htt}
\end{center}
\vspace{1em}
\end{figure}
%%%%%%%%%%%%%%%%%%%%%%%%% F I G U R E %%%%%%%%%%%%%%%%%%%%%%%%%%%%%%%%%%%%%%%%%

We have not made a complete study of the combined sensitivity of the ILC
measurements to the benchmark surfaces, but it is clear from this brief
survey that the ILC measurements would in general provide interesting
tests of the MSSM at the loop level. In the absence of detailed
studies, we expect that CLIC measurements would have similar
sensitivities, since $h$ production would be more copious at the
higher CLIC energies, and the CLIC luminosity at lower energies could
be similar to that of the ILC~\cite{clic}. 
In addition to the precision measurements described here, the ILC and
CLIC would be able to produce directly associated $H + A$ pairs above
the kinematic threshold. 

%%%%%%%%%%%%%%%%%%%%%%%%%%%%%%%%%%%%%%%%%%%%%%%%%%%%%%%%%%%%%%%%%%%%%%%%%%%%%%%
%%%%%%%%%%%%%%%%%%%%%%%%%%%%%%%%%%%%%%%%%%%%%%%%%%%%%%%%%%%%%%%%%%%%%%%%%%%%%%%

\section{$B$ Physics}
\label{sec:bpo}

We display in \reffi{fig:BPO01} the results for three BPO
$\br(b \to s \ga)$, $\br(B_s \to \mu^+\mu^-)$, 
$\br(B_u \to \tau \nu_\tau)$, in the four benchmark \plane{\MA}{\tb}s. 

The prediction of $B_s \to \mu^+\mu^-$ is based on 
\citere{ourBmumu1,ourBmumu2}.
The solid (beige) line indicates $\br(B_s \to \mu^+\mu^-) = 10^{-7}$,
corresponding roughly to the current upper bound from CDF~\cite{bsmmexp}
and D0~\cite{bsmmD0}. The latest bound reported by CDF has recently been
lowered to $5.8 \times 10^{-8}$~\cite{bsmmexpLP07}.
The dashed (beige) line indicates a BR of 
$2 \times 10^{-8}$. 
In \reffi{fig:BPO01} we see that the current upper limit on 
$B_s \to \mu^+ \mu^-$ 
already excludes regions of the planes at small $\MA$ and large $\tb$,
starting to cut into the region with $\De\chi^2 < 4.61$. 
The prospective sensitivities would extend as far as the
best-fit points. 

%%%%%%%%%%%%%%%%%%%%%%%%% F I G U R E %%%%%%%%%%%%%%%%%%%%%%%%%%%%%%%%%%%%%%%%%
\begin{figure}[htb!]
\vspace{10mm}
\begin{center}
\includegraphics[width=.49\textwidth]{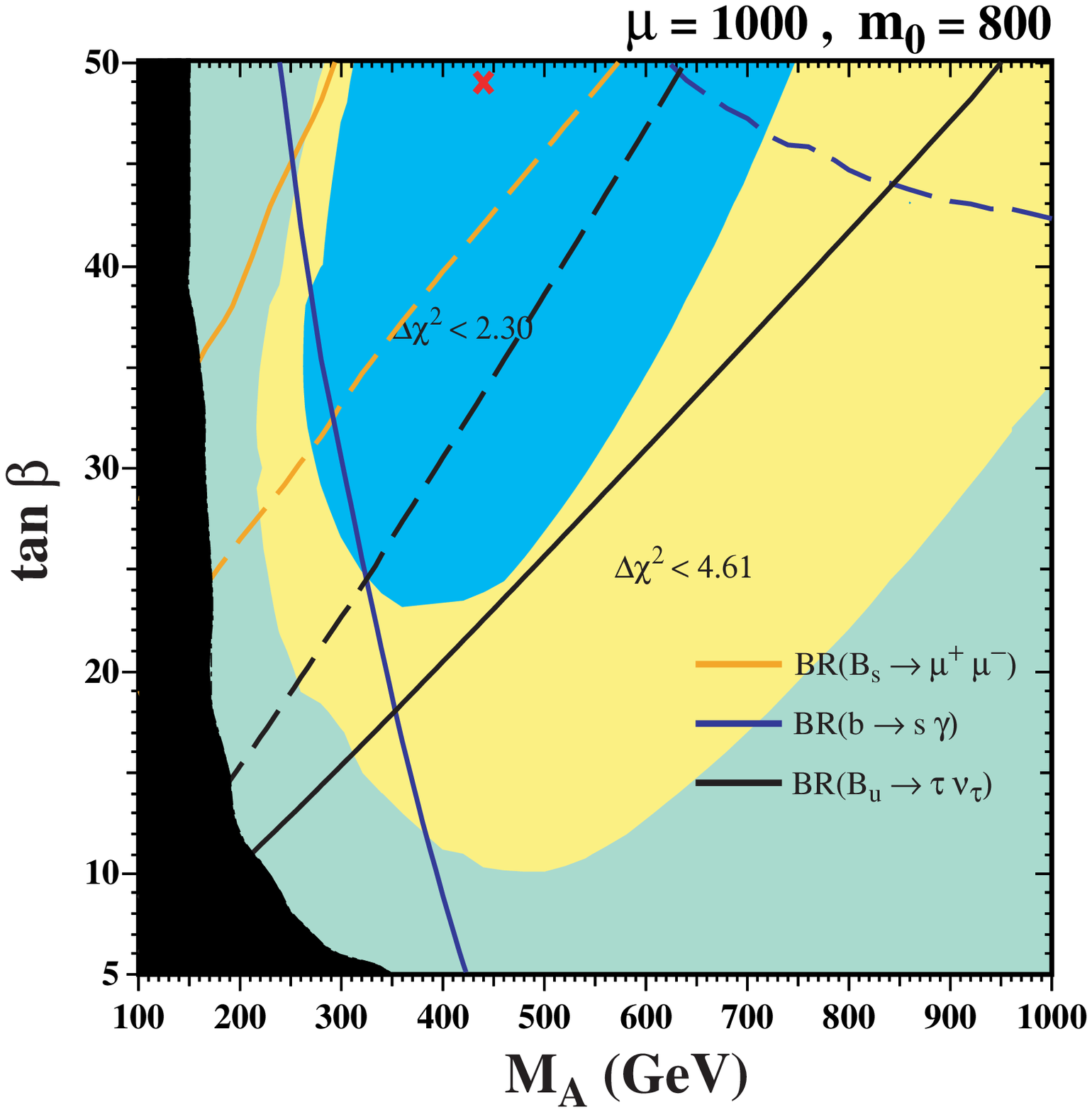}
%\hspace{-25mm}
\includegraphics[width=.49\textwidth]{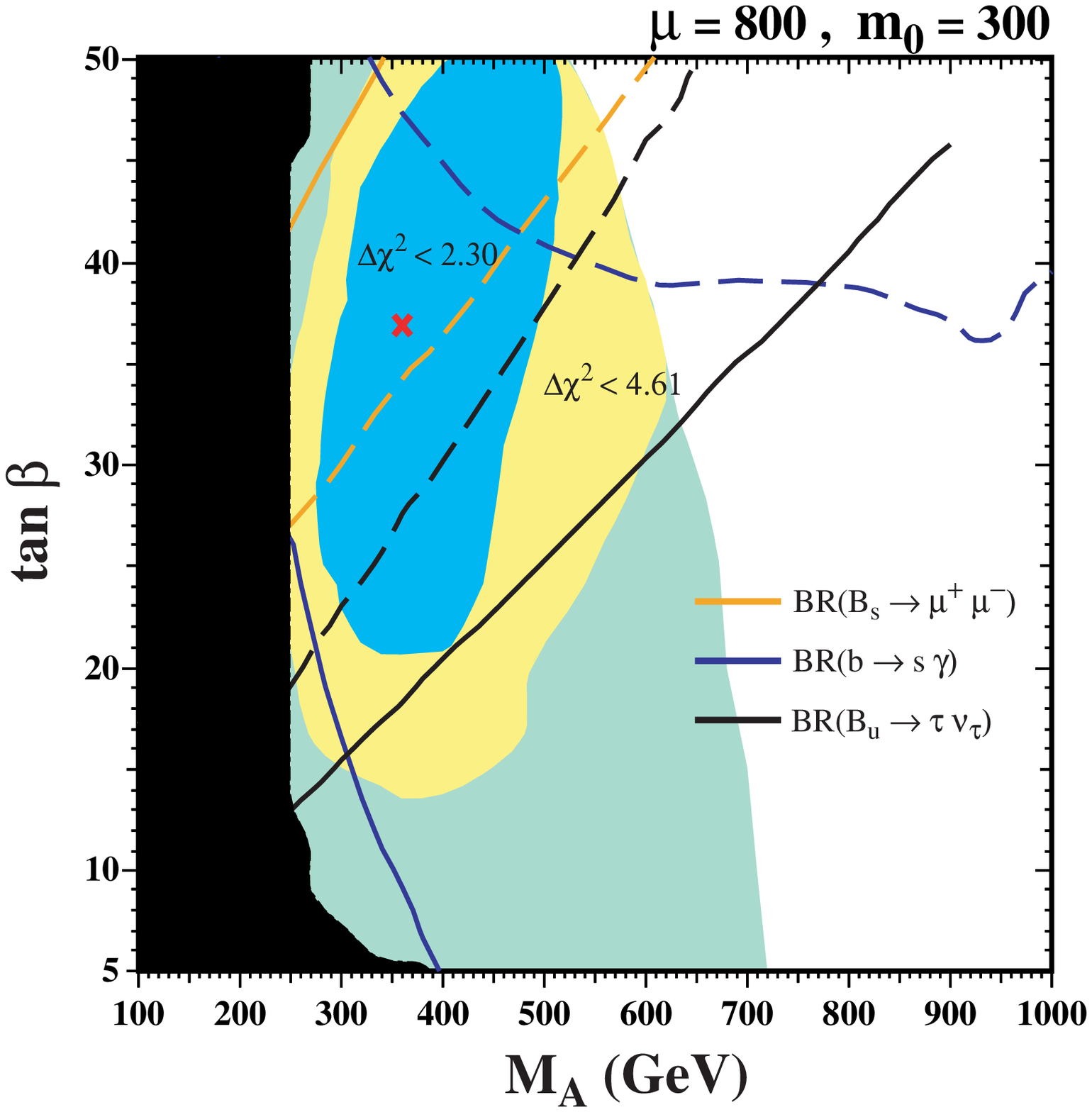}
\includegraphics[width=.49\textwidth]{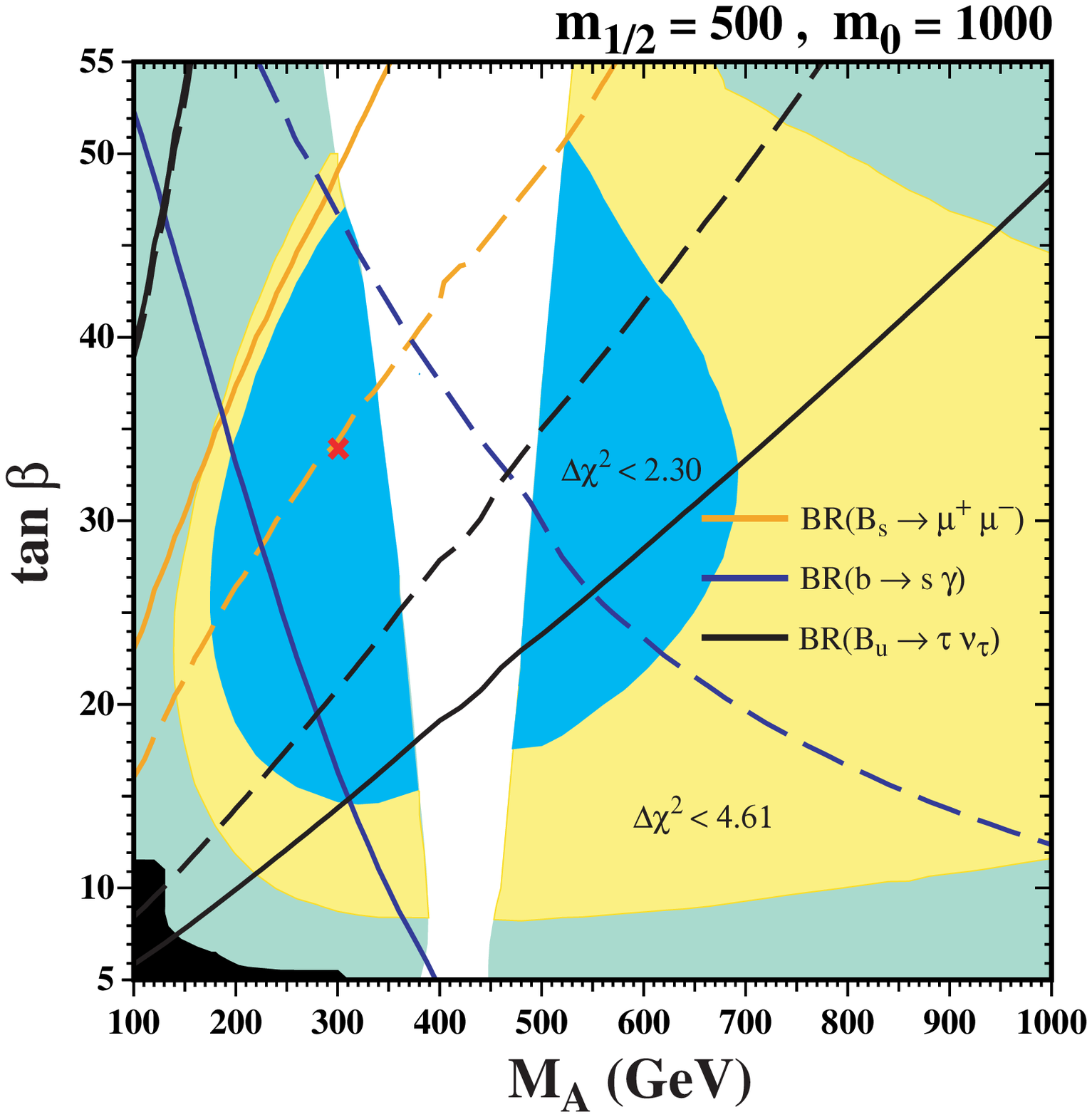}
%\hspace{-25mm}
\includegraphics[width=.49\textwidth]{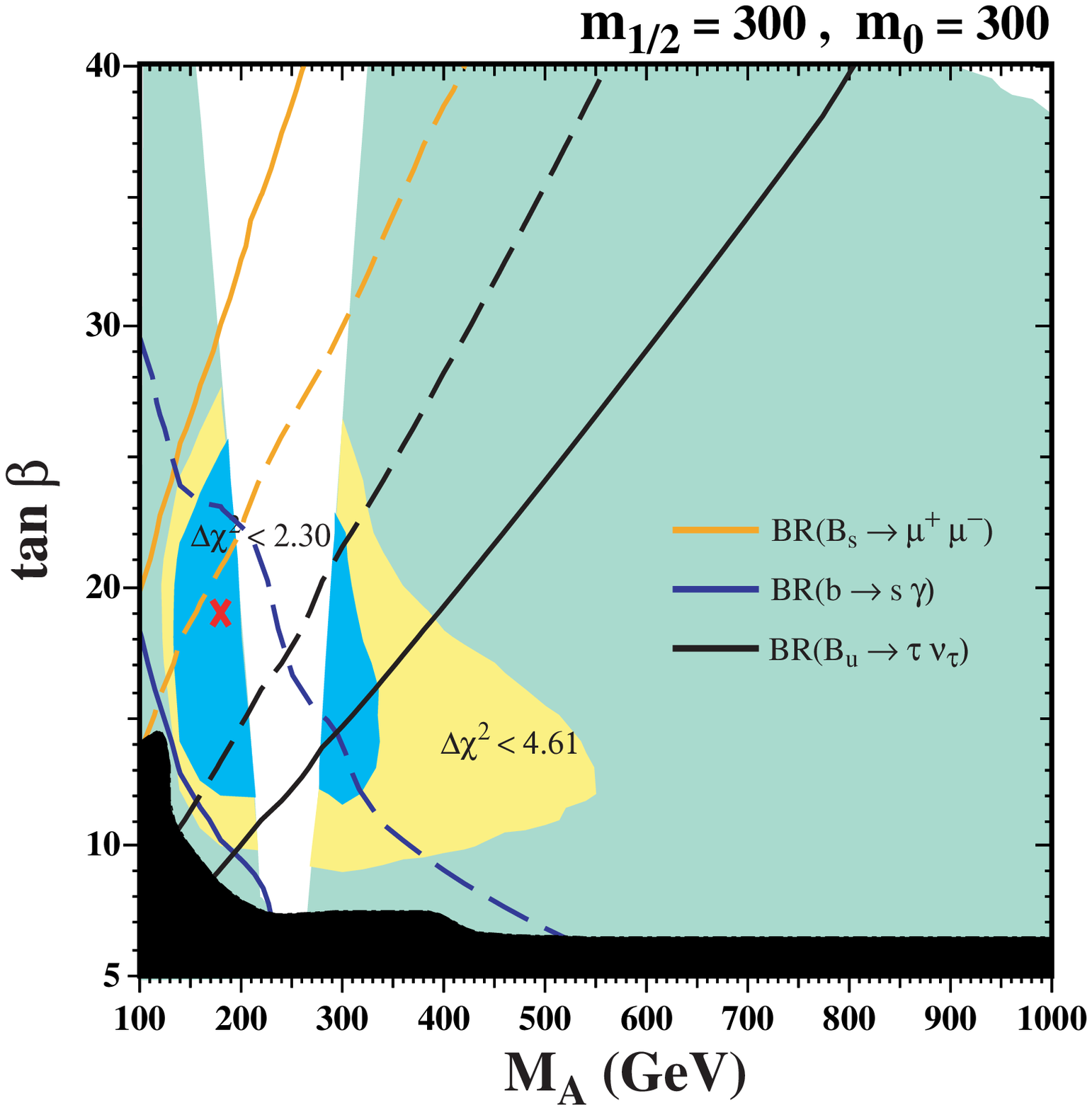}
%\hspace{-5mm}
%\vspace{-25mm}
\caption{%
The same $(\MA, \tb)$ planes for the NUHM benchmark surfaces (a)
\Athree, (b) \Afive, (c) \Atwo\ and (d) \Afour\ as in
Fig.~\protect\ref{fig:Mh}, displaying the expected sensitivities of
the $B$ physics observables $B_s \to \mu^+ \mu^-$, $b \to s \ga$ and
$B_u \to \tau \nu$. The various lines indicate: 
$\br(B_s \to \mu^+\mu^-) = 10^{-7} (2 \times 10^{-8})$ as solid (dashed),
$\br(b \to s \ga) = 4 (3) \times 10^{-4}$ as solid (dashed),
$\br(B_u \to \tau \nu_\tau)_{\rm MSSM/SM} = 0.9 (0.7)$ as solid (dashed).
}
\label{fig:BPO01}
\end{center}
\vspace{-1em}
\end{figure}
%%%%%%%%%%%%%%%%%%%%%%%%% F I G U R E %%%%%%%%%%%%%%%%%%%%%%%%%%%%%%%%%%%%%%%%%

For $b \to s \ga$ our numerical results have been derived with the 
$\br(b \to s \ga)$ evaluation provided in \citeres{bsgKO2},
incorporating also the latest SM corrections provided in~\citere{bsgtheonew}. 
The results in \reffi{fig:BPO01} are shown as the two blue lines
indicating 
$\br(b \to s \ga)$ of $4 \times 10^{-4}$ (solid) and $3 \times 10^{-4}$
(dashed). 
These have to be compared to the experimentally preferred value of 
$\br(b \to s \ga) = 
(3.55 \pm 0.24^{+0.09}_{-0.10} \pm 0.03) \times 10^{-4}$~\cite{hfag}. 
The best-fit point 
together with large parts of the $\chi^2$~preferred regions lie between
the two lines, i.e., large parts of the four benchmark planes are in
good agreement with the current experimental value.

Our results for $\br(B_u \to \tau \nu_\tau)$ are based on
\citere{BPOtheo}. In the four benchmark scenarios of
\reffi{fig:BPO01} the results are shown in form of the NUHM result
divided by the SM prediction as black lines. The solid (dashed) lines
correspond to a ratio of 0.9 (0.7), where the current central value is 
$0.93 \pm 0.41$~\cite{btnexp,btnexp2}. It can be seen that the best fit
value as well as large 
parts of the $\chi^2$~preferred parts of the benchmark planes predict a
value somewhat lower than the current experimental result. However, with
the current precision no firm conclusion can be drawn.

%%%%%%%%%%%%%%%%%%%%%%%%%%%%%%%%%%%%%%%%%%%%%%%%%%%%%%%%%%%%%%%%%%%%%%%%%%%%%%%
%%%%%%%%%%%%%%%%%%%%%%%%%%%%%%%%%%%%%%%%%%%%%%%%%%%%%%%%%%%%%%%%%%%%%%%%%%%%%%%

\section{Direct Detection of Supersymmetric Dark Matter}
\label{sec:DD}

In \reffi{fig:DD} we show how the direct detection of the LSP
via spin-independent scattering on nuclei
probes the four \plane{\MA}{\tb}s. We focus here on the bound
from the XENON10 experiment that was recently
published by the XENON collaboration~\cite{Xenon}, which improves on
the previous CDMS results~\cite{CDMS}. 
We note that the XENON10 experiment has seen some potential signal
events which are, however, interpreted as background. 

The constraint imposed by the limits from direct detection experiments
is sensitive to
two theoretical uncertainties that are independent of the specific
dark matter model. One is the local density of cold dark matter, which
is normally estimated to be $\rho_{\rm CDM} = 0.3$~GeV/cm$^3$, although
smaller values may be consistent with some models of the Galaxy. The
other important uncertainty is that in the nucleonic matrix element of
the local operator responsible for the spin-independent scattering
amplitude. This is related, in particular, to the so-called $\si$
term, $\Si_{\pi N}$, that may be derived from measurements of
low-energy $\pi$-nucleon scattering.

%%%%%%%%%%%%%%%%%%%%%%%%% F I G U R E %%%%%%%%%%%%%%%%%%%%%%%%%%%%%%%%%%%%%%%%%
\begin{figure}[htb!]
\vspace{10mm}
\begin{center}
\includegraphics[width=.49\textwidth]{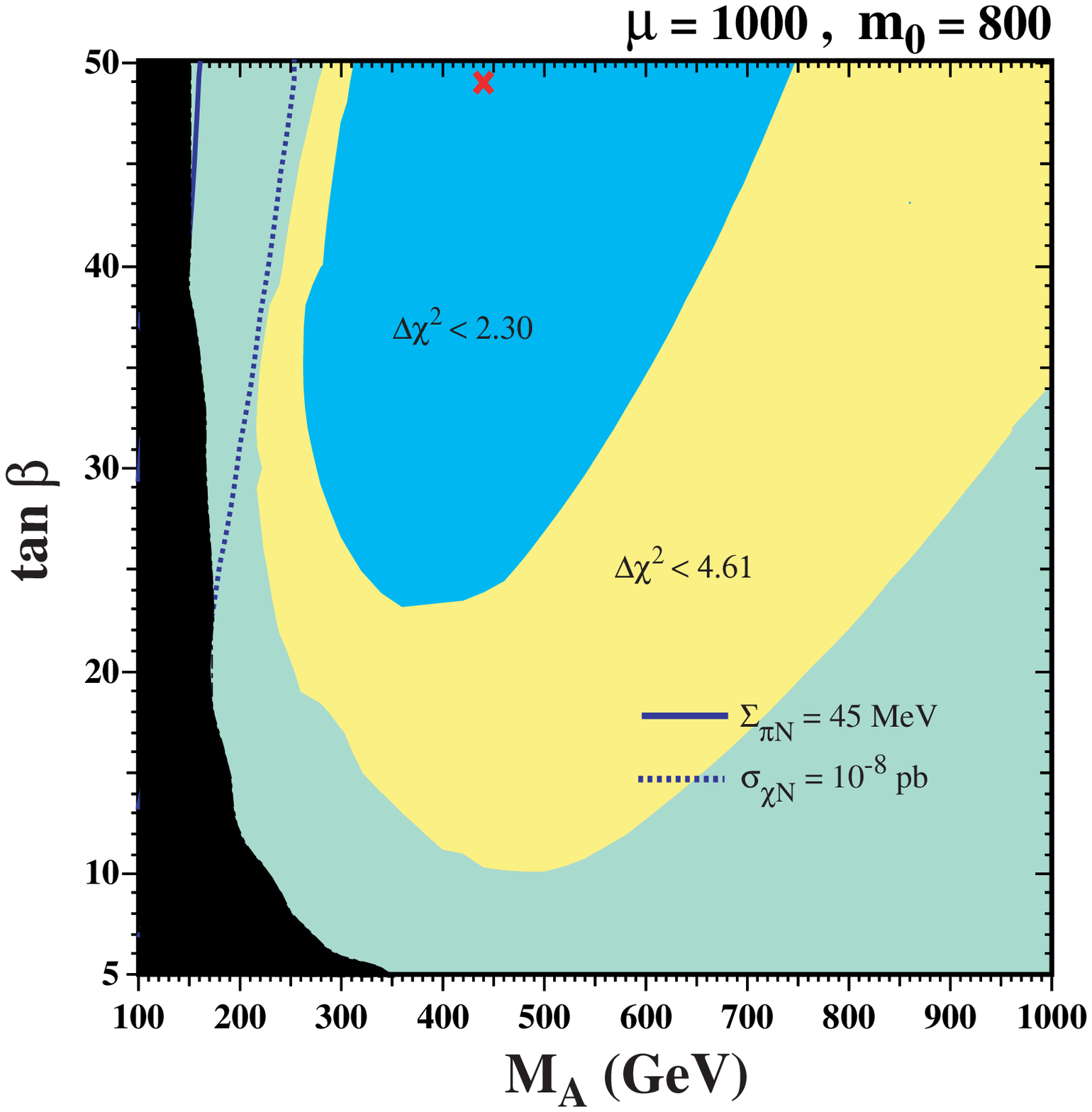}
%\hspace{-25mm}
\includegraphics[width=.49\textwidth]{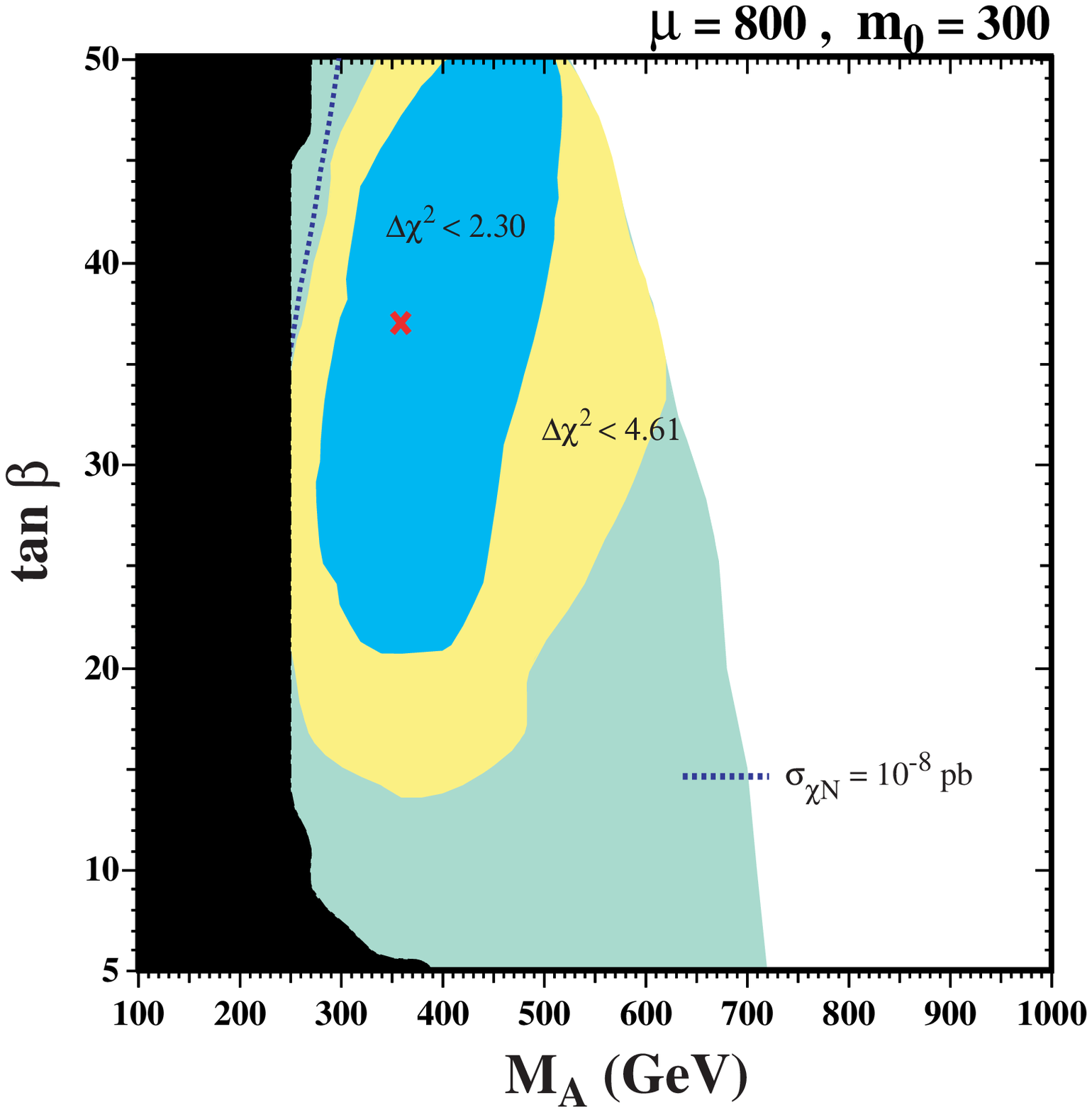}
\includegraphics[width=.49\textwidth]{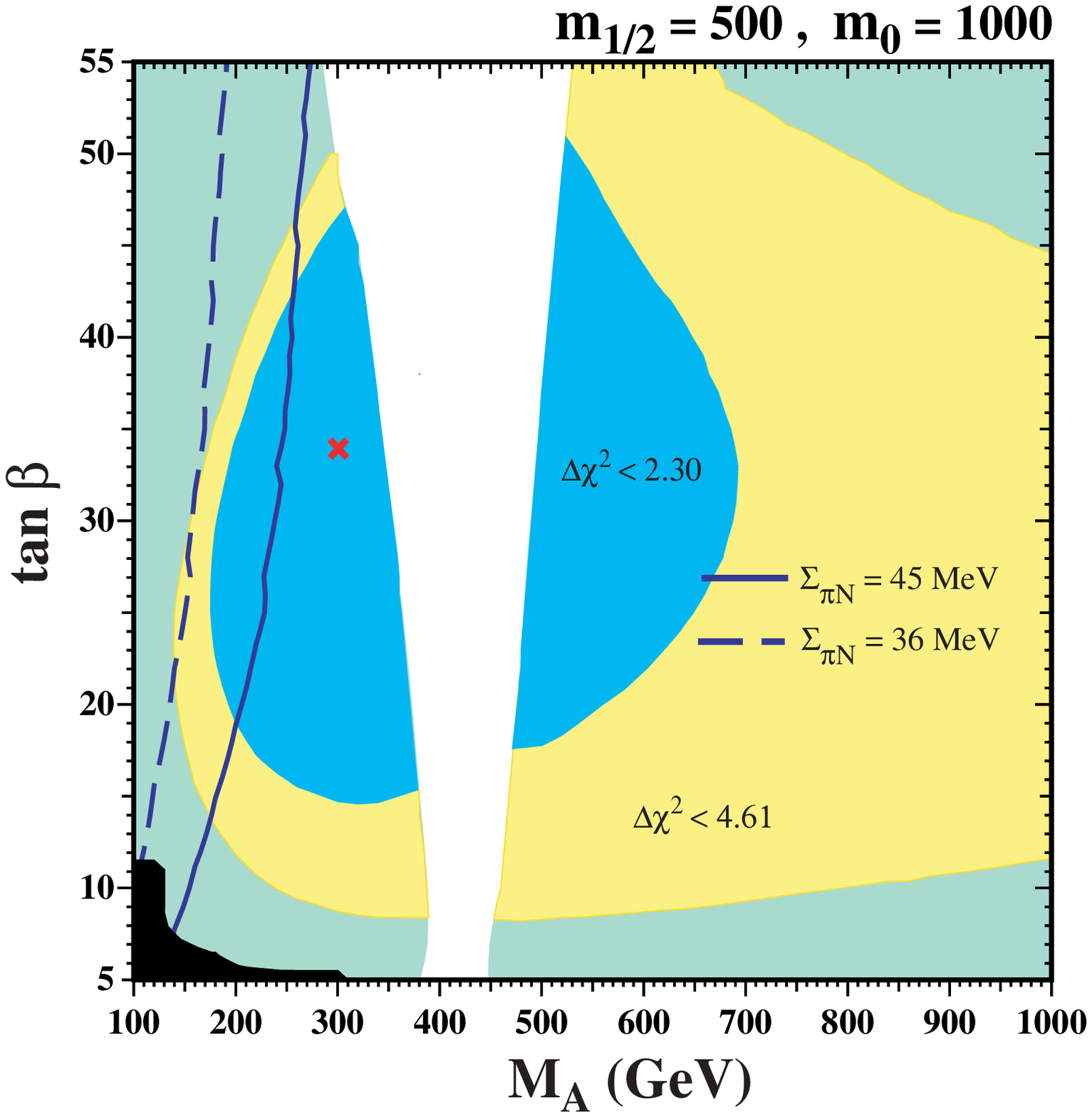}
%\hspace{-25mm}
\includegraphics[width=.49\textwidth]{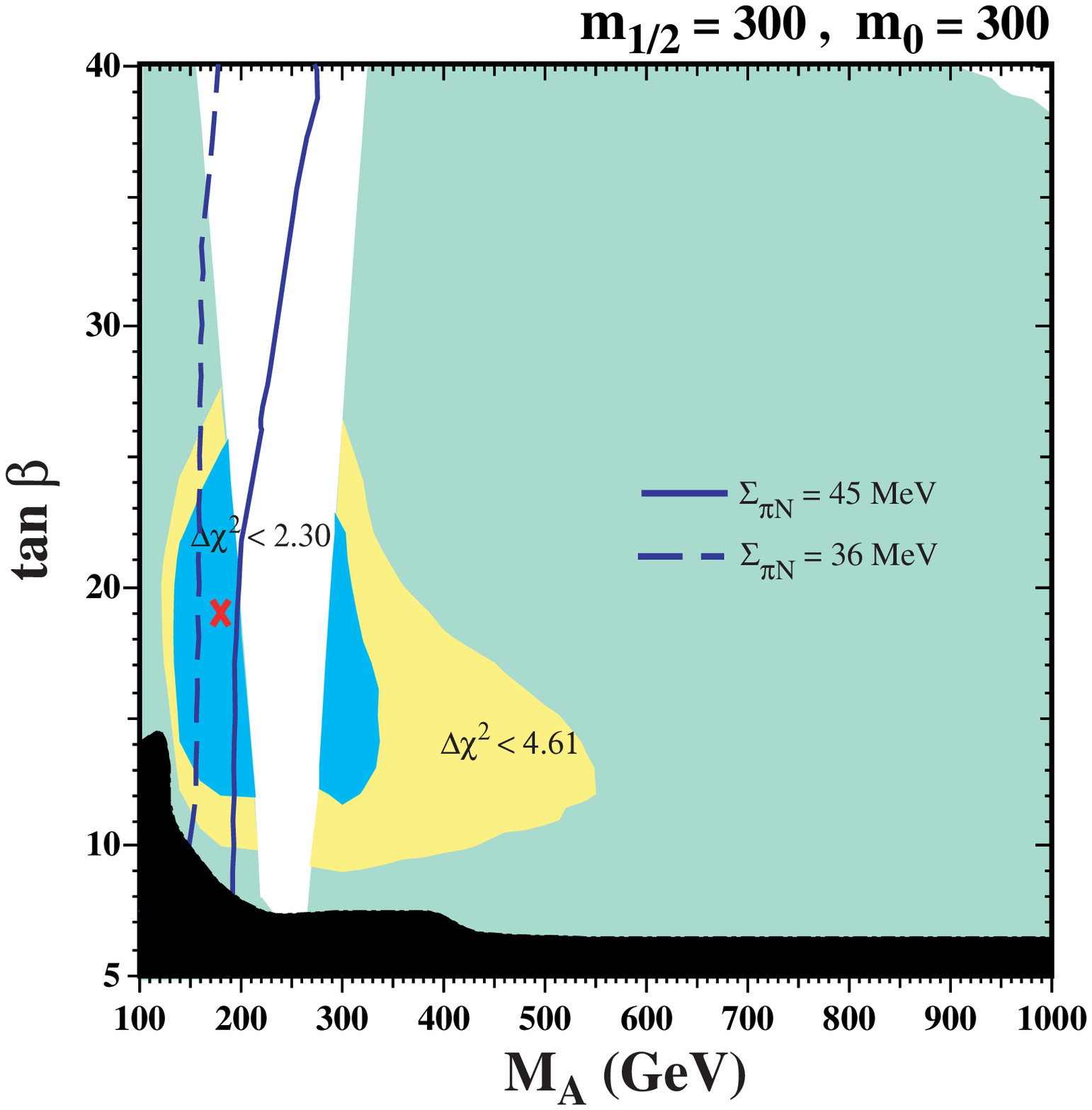}
%\hspace{-5mm}
%\vspace{-25mm}
\caption{%
The same $(\MA, \tb)$ planes for the NUHM benchmark surfaces (a)
\Athree, (b) \Afive, (c) \Atwo\ and (d) \Afour\ as in
Fig.~\protect\ref{fig:Mh}, displaying the expected sensitivities of
present and prospective future direct searches for the scattering
of dark matter particles (see text).
}
\label{fig:DD}
\end{center}
\vspace{1em}
\end{figure}
%%%%%%%%%%%%%%%%%%%%%%%%% F I G U R E %%%%%%%%%%%%%%%%%%%%%%%%%%%%%%%%%%%%%%%%%

The solid lines in \reffi{fig:DD} correspond to the XENON10 bound obtained
assuming $\rho_{\rm CDM} = 0.3$~GeV/cm$^3$ and using $\Si_{\pi N} = 45 \mev$ 
as input, corresponding to a relative strange-quark density 
$y \equiv 2 \langle N | {\bar s} s | N \rangle / 
            \langle N | ({\bar u} u + {\bar d} d) | N \rangle 
= 0.2$~\cite{oldnp}.
These assumptions are realistic, though there is a large uncertainty in
the strangeness contribution which may lead to larger rates if
$\Si_{\pi N}$ is larger or significantly lower rates if
the strangeness contribution to the proton mass is small. 
The dashed lines show the bounds that one would obtain from the
XENON10 experiment assuming the same value of $\rho_{\rm CDM}$, but
with $\Si_{\pi N} = 36 \mev$ corresponding to $y = 0$, and therefore
representing more conservative assumptions. Finally, as an example of the
possible sensitivity of future experiments, the dotted lines
show the contours one would obtain for a spin-independent
cross section of $10^{-8}$~pb, assuming the same
value of $\rho_{\rm CDM}$ and $\Si_{\pi N} = 45 \mev$ as input. 

We see from \reffi{fig:DD} that the surfaces \Athree\ and \Afive\
are not probed by the current limits from the XENON10 experiment. Only the
possible future sensitivity at $10^{-8}$~pb starts to cut into the 
$\De\chi^2 < 4.61$
region. For these planes,  accelerator searches are clearly more powerful.
The situation is different for the planes \Atwo\ and \Afour, due
to the relatively low values of $m_{1/2}$ across these planes.
We recall that, for planes \Athree\ and \Afive, $m_{1/2}$ scales with $\MA$
and the sparticle spectrum is typically heavier at large $\MA$ than at
the corresponding points in planes \Atwo\ and \Afour.  As a result, the
spin-independent $\neu{1}-p$ elastic cross section is
suppressed for planes \Athree\ and \Afive.
On the other hand, we see that the current XENON10 bound  
probes large parts of the $\De\chi^2 < 2.30$ areas of \Atwo\ and
\Afour\ planes, if one uses the moderate values 
of $\Si_{\pi N} =45 \mev$ and the
strange-quark content. Indeed, in the case of the \Afour\ surface, the current
XENON10 bound would even cover the best-fit point for this value of
$\Si_{\pi N}$ and the default value for the local density of cold dark
matter. The more 
conservative analysis, on the other hand, is sensitive only to smaller
$\MA$ values, and probes only a much smaller part of the regions
preferred by the $\chi^2$ analysis. Finally, we note that a future  
sensitivity to a cross section of $10^{-8}$ pb would cover the
entire \Atwo\ and \Afour\ surfaces.

%%%%%%%%%%%%%%%%%%%%%%%%%%%%%%%%%%%%%%%%%%%%%%%%%%%%%%%%%%%%%%%%%%%%%%%%%%%%%%%
%%%%%%%%%%%%%%%%%%%%%%%%%%%%%%%%%%%%%%%%%%%%%%%%%%%%%%%%%%%%%%%%%%%%%%%%%%%%%%%

%%%%%%%%%%%%%%%%%%%%%%%%%%%%%%%%%%%%%%%%%%%%%%%%%%%%%%%%%%%%%%%%%%%%%%%%%%%%%%%
%%%%%%%%%%%%%%%%%%%%%%%%%%%%%%%%%%%%%%%%%%%%%%%%%%%%%%%%%%%%%%%%%%%%%%%%%%%%%%%

\section{Conclusions}

The value of benchmark studies is that they allow one to understand
better the range of possibilities opened up by supersymmetry. It is
therefore desirable that benchmarks be chosen in such a way as to
respect, as far as possible, the definitive experimental constraints,
and also that they be susceptible to systematic study. We have
demonstrated in this paper how  NUHM benchmark surfaces chosen so that
the relic cold dark matter density falls within or below the range
favoured by WMAP and other experiments may be used to probe
supersymmetric phenomenology. 
Our approach based on the NUHM scenario significantly differs from
previous proposals of benchmark scenarios for the MSSM Higgs sector
that were entirely formulated in terms of low-scale parameters and that 
were not suitable for a
phenomenologically acceptable prediction of the cold dark matter
density. 
The analysis of our proposed benchmark surfaces is facilitated by
developments in the {\tt FeynHiggs} code that are described in the
Appendix. These will enable the interested reader to explore the
prospects for her/his favourite experimental probe of supersymmetry in
these benchmark surfaces. 

We have displayed the constraints currently imposed in the new benchmark
surfaces by electroweak precision observables,  and explored the
prospects for Higgs searches at the Tevatron collider, the LHC and the
ILC, and we have also explored indirect effects in $B$ physics
and in dark matter detection. Whereas the Tevatron collider will be able
only to nibble at corners of these NUHM benchmark surfaces, experiments
at the LHC will be able to cover them entirely, and the ILC will have
good prospects for precision measurements. There are good prospects for
$B$ experiments in parts of the benchmark surfaces, and direct dark
matter may be detectable in some cases.

It should of course be noted that benchmark studies may soon
be rendered obsolete -- namely by the discovery of supersymmetry.\\

{\it As we were completing this paper, we heard the sad news of the passing
away of Julius Wess, one of the discoverers 
and founding fathers of supersymmetry. Julius did
so much to develop our understanding of supersymmetry, to awaken our
appreciation of its beauty, and to convince us of its importance for
physics. Humbly and respectfully, we dedicate this paper to his memory. }

%%%%%%%%%%%%%%%%%%%%%%%%%%%%%%%%%%%%%%%%%%%%%%%%%%%%%%%%%%%%%%%%%%%%%%%%%%%%%%%
%%%%%%%%%%%%%%%%%%%%%%%%%%%%%%%%%%%%%%%%%%%%%%%%%%%%%%%%%%%%%%%%%%%%%%%%%%%%%%%

\subsection*{Acknowledgements}
%\vspace{-0.5em}
S.H.\ thanks R.~Kinnunen for data on the charged Higgs-boson search at
CMS, A.~Lath for communication on the CDF projections, A.~Nikitenko
for the CMS data on the Higgs decay to photons, and A.~Korytov and 
E.~Yazgan for information on the $WW$~fusion channels at CMS.\\
The work of K.A.O.\ was partially supported by DOE grant DE-FG02-94ER-40823. 
Work supported in part by the European Community's Marie-Curie Research
Training Network under contract MRTN-CT-2006-035505
`Tools and Precision Calculations for Physics Discoveries at Colliders'

%%%%%%%%%%%%%%%%%%%%%%%%%%%%%%%%%%%%%%%%%%%%%%%%%%%%%%%%%%%%%%%%%%%%%%%%%%%%%%%
%%%%%%%%%%%%%%%%%%%%%%%%%%%%%%%%%%%%%%%%%%%%%%%%%%%%%%%%%%%%%%%%%%%%%%%%%%%%%%%

\clearpage
\pagebreak

\begin{appendix}

\newcommand\Eg{e.g.\ }
\newcommand\Ie{i.e.\ }
\newcommand\Code[1]{\ensuremath{\texttt{#1}}}
\newcommand\Ind[1]{\noindent\hbox{\texttt{~~~~~}}\Code{#1}}
\newcommand\Or[1]{\noindent\rlap{\textit{or}}\Ind{#1}}
\newenvironment{mytt}%
  {\trivlist\item}%
  {\endtrivlist\noindent\ignorespaces}

\section{Evaluation of Benchmark Surfaces with {\tt FeynHiggs}}

The new benchmark surfaces have been implemented into the code
{\tt FeynHiggs}~\cite{feynhiggs,mhiggslong,mhiggsAEC,mhcMSSMlong}.  
In this way, any
user may apply them to perform phenomenological analyses.

From the mathematical point of view, the NUHM/CDM constraints introduce
non-trivial relations between input parameters, which thus cannot be
scanned naively by independent loops.  To solve this in a generic way,
{\tt FeynHiggs}\ 2.6 allows the user to interpolate the inputs from a
Parameter Table into which arbitrary relations can be encoded.  
The tables containting the four benchmark surfaces can be downloaded
from {\tt http://www.feynhiggs.de}.
To implement the new format
of a Parameter Table, significant internal rearrangements were necessary
from which the concept of a {\tt FeynHiggs}\ Record evolved.

A Record is a new data type which captures the entire content of a
parameter file in the native format of {\tt FeynHiggs}.  In this respect
it is akin to the SUSY Les Houches Accord Record \cite{SLHA}, but also encodes
information about parameter loops and has `inheritance rules' for
default values.  Using the routines to manipulate a Record, the
programmer can, among other things, process {\tt FeynHiggs}\ parameter files
independently of the front end.

In addition to containing loops over parameters, a Record can be
associated with a Parameter Table in such a way that values not
explicitly given in the parameter file are interpolated from the table
(as it can be done for the four benchmark scenarios).

The {\tt FeynHiggs}\ Record is conceptually a superstructure `on top' of the
conventional part of {\tt FeynHiggs}.  This means that a Record can be
manipulated 
without any influence on the computation of Higgs observables at first. 
Only when the \Code{FHSetRecord} subroutine is invoked are its current
values set as the inputs for the computation.  So in principle, the 
{\tt FeynHiggs}\ 
Record can be used without doing any computation of Higgs observables at
all.

Technically, a Record is a two-dimensional real array of the form
$$
\begin{array}{l|cccc}
\Code{rec($i_\downarrow$,$j^\rightarrow$)} &
  \Code{iVar  } & \Code{iLower} & \Code{iUpper} & \Code{iStep} \\ \hline
\Code{iTB} & L & U & U & U \\
\Code{iMA0} & L & U & U & U \\
\dots & \dots
\end{array}
$$
\begin{itemize}
\item
The column index $i$ specifies the parameter.  The indices are 
labelled as in the parameter file, but prefixed with an \Code{i} (see 
Table~\ref{tab:fhindices}).

\item
The row index $j$ enumerates the variables that constitute the loop over 
a parameter, \Ie the current, lower, and upper value and the step size.
The loop inferred through these parameters has the form

\texttt{%
do rec($i$,iVar) = rec($i$,iLower), rec($i$,iUpper), rec($i$,iStep) \\
\hbox{~~~}\dots \\
enddo}

\item
$U$ entries indicate fields filled in by the user.  If no loop is 
desired over a particular parameter, the fields \Code{rec($i$,iUpper)} 
and \Code{rec($i$,iStep)} can be omitted.  On top of that there are also 
`inheritance rules' (given in Table~\ref{tab:fhindices}), stating for 
example that \Code{M3SL} defaults to \Code{MSusy} if not given 
explicitly.

\item
$L$ entries indicate fields replaced by the \Code{FHLoopRecord} routine
while working off the loops over parameter space, \Ie these fields are
updated automatically according to the current point in the loop.  
For example, if the Record contains
\begin{verbatim}
   rec(iTB,iLower) = 10
   rec(iTB,iUpper) = 50
   rec(iTB,iStep)  = 10
\end{verbatim}
the first call to \Code{FHLoopRecord} will set \Code{rec(iTB,iVar)} to 
10, the next to 20, etc.
\end{itemize}

\begin{table}
\begin{small}
\caption{\label{tab:fhindices}The parameter index names of a {\tt FeynHiggs}
Record.  Indices of real parameters are listed in the left, of complex
ones in the right column.  Complex quantities, \Eg $A_t$, can be
accessed either through \Code{Re(iAt)} and \Code{Im(iAt)}, or
\Code{Abs(iAt)} and \Code{Arg(iAt)}, with \Code{iAt} alone as a synonym
for \Code{Re(iAt)}.  In cases where both \Code{Re/Im} \emph{and}
\Code{Abs/Arg} are given, the latter take precendence.  Please consult 
the FeynHiggs(1) manual page for more details.}

\begin{center}
\newcommand\Bline{\cline{1-3}\cline{5-7}}
\newcommand\Rline{\cdashline{1-3}[2pt/2pt]}
\newcommand\Cline{\cdashline{5-7}[2pt/2pt]}
\begin{tabular}{|lll|l|lll|}
\Bline
Index name & Parameter & Default value &&
        Index name & Parameter & Default value \\
\Bline
\Code{iAlfasMZ} & $\alpha_s(M_Z^2)$ & $-1$ &&
        \Code{iM1} & $M_1$ & 0 \\
\Rline
\Code{iMC} & $m_c$ & $-1$ &&
        \Code{iM2} & $M_2$ & \\
\Code{iMT} & $m_t$ & &&
        \Code{iM3} & $M_3$ & \\
\Cline
\Code{iMB} & $m_b$(on-shell) & $-1$ &&
        \Code{iAt} & $A_t$ & \\
\Code{iMW} & $M_W$ & $-1$ &&
        \Code{iAc} & $A_c$ & \Code{iAt} \\
\Code{iMZ} & $M_Z$ & $-1$ &&
        \Code{iAu} & $A_u$ & \Code{iAc} \\
\Rline
\Code{TB} & $\tan\beta$ & &&
        \Code{iAb} & $A_b$ & \Code{iAt} \\
\Code{MA0} & $M_{A^0}$ & &&
        \Code{iAs} & $A_s$ & \Code{iAb} \\
\Code{MHp} & $M_{H^+}$ & &&
        \Code{iAd} & $A_d$ & \Code{iAs} \\
\Rline
\Code{iMSusy} & $M_{\text{SUSY}}$ & &&
        \Code{iAtau} & $A_\tau$ & \Code{iAb} \\
\Code{iM3SL} & $M_{\tilde L}^3$ & \Code{iMSusy} &&
        \Code{iAmu} & $A_\mu$ & \Code{iAtau} \\
\Code{iM2SL} & $M_{\tilde L}^2$ & \Code{iM3SL} &&
        \Code{iAe} & $A_e$ & \Code{iAmu} \\
\Cline
\Code{iM1SL} & $M_{\tilde L}^1$ & \Code{iM2SL} &&
        \Code{ideltaLLuc} & $\delta^{LL}_{uc}$ & 0 \\
\Code{iM3SE} & $M_{\tilde E}^3$ & \Code{iMSusy} &&
        \Code{ideltaLRuc} & $\delta^{LR}_{uc}$ & 0 \\
\Code{iM2SE} & $M_{\tilde E}^2$ & \Code{iM3SE} &&
        \Code{ideltaRLuc} & $\delta^{RL}_{uc}$ & 0 \\
\Code{iM1SE} & $M_{\tilde E}^1$ & \Code{iM2SE} &&
        \Code{ideltaRRuc} & $\delta^{RR}_{uc}$ & 0 \\
\Code{iM3SQ} & $M_{\tilde Q}^3$ & \Code{iMSusy} &&
        \Code{ideltaLLct} & $\delta^{LL}_{ct}$ & 0 \\
\Code{iM2SQ} & $M_{\tilde Q}^2$ & \Code{iM3SQ} &&
        \Code{ideltaLRct} & $\delta^{LR}_{ct}$ & 0 \\
\Code{iM1SQ} & $M_{\tilde Q}^1$ & \Code{iM2SQ} &&
        \Code{ideltaRLct} & $\delta^{RL}_{ct}$ & 0 \\
\Code{iM3SU} & $M_{\tilde U}^3$ & \Code{iMSusy} &&
        \Code{ideltaRRct} & $\delta^{RR}_{ct}$ & 0 \\
\Code{iM2SU} & $M_{\tilde U}^2$ & \Code{iM3SU} &&
        \Code{ideltaLLut} & $\delta^{LL}_{ut}$ & 0 \\
\Code{iM1SU} & $M_{\tilde U}^1$ & \Code{iM2SU} &&
        \Code{ideltaLRut} & $\delta^{LR}_{ut}$ & 0 \\
\Code{iM3SD} & $M_{\tilde D}^3$ & \Code{iMSusy} &&
        \Code{ideltaRLut} & $\delta^{RL}_{ut}$ & 0 \\
\Code{iM2SD} & $M_{\tilde D}^2$ & \Code{iM3SD} &&
        \Code{ideltaRRut} & $\delta^{RR}_{ut}$ & 0 \\
\Code{iM1SD} & $M_{\tilde D}^1$ & \Code{iM2SD} &&
        \Code{ideltaLLds} & $\delta^{LL}_{ds}$ & 0 \\
\Rline
\Code{iQtau} & $Q_\tau$ & 0 &&
        \Code{ideltaLRds} & $\delta^{LR}_{ds}$ & 0 \\
\Code{iQt} & $Q_t$ & 0 &&
        \Code{ideltaRLds} & $\delta^{RL}_{ds}$ & 0 \\
\Code{iQb} & $Q_b$ & 0 &&
        \Code{ideltaRRds} & $\delta^{RR}_{ds}$ & 0 \\
\Rline
\Code{iCKMtheta12} & $\theta_{12}$ & $-1$ &&
        \Code{ideltaLLsb} & $\delta^{LL}_{sb}$ & 0 \\
\Code{iCKMtheta23} & $\theta_{23}$ & $-1$ &&
        \Code{ideltaLRsb} & $\delta^{LR}_{sb}$ & 0 \\
\Code{iCKMtheta13} & $\theta_{13}$ & $-1$ &&
        \Code{ideltaRLsb} & $\delta^{RL}_{sb}$ & 0 \\
\Code{iCKMdelta13} & $\delta_{13}$ & $-1$ &&
        \Code{ideltaRRsb} & $\delta^{RR}_{sb}$ & 0 \\
&&&&
        \Code{ideltaLLdb} & $\delta^{LL}_{db}$ & 0 \\
&&&&
        \Code{ideltaLRdb} & $\delta^{LR}_{db}$ & 0 \\
&&&&
        \Code{ideltaRLdb} & $\delta^{RL}_{db}$ & 0 \\
&&&&
        \Code{ideltaRRdb} & $\delta^{RR}_{db}$ & 0 \\
\Bline
\end{tabular}
\end{center}
\end{small}
\end{table}

\subsection{Fortran Use}

\subsubsection{Declaration}

Every subroutine or function which uses a Record must first include 
the definitions:
\begin{verbatim}
#include "FHRecord.h"
\end{verbatim}
Records can then be declared with the preprocessor macro
\Code{RecordDecl}, which hides the declaration details.
For example,
\begin{verbatim}
     RecordDecl(rec)
\end{verbatim}
declares the Record \Code{rec($i$,$j$)}.  When declaring several
records, each needs its own \Code{RecordDecl} statement, \Ie
\Code{RecordDecl(rec1, rec2, ...)} is not permissible.

\subsubsection{Initializing a Record}

A FeynHiggs Record has to be brought into a defined state before its 
first use, either by clearing it with
\begin{verbatim}
     call FHClearRecord(rec)
\end{verbatim}
or by reading it from a file, which similarly overwrites any previous 
content:
\begin{verbatim}
     call FHReadRecord(error, rec, "file")
     if( error .ne. 0 ) stop
\end{verbatim}
where \Code{file} is the name of a parameter file in FeynHiggs' native 
format.

Fields can be set or read out using ordinary Fortran array access, \Eg
\begin{mytt}
\Ind{rec(iTB,iLower) = 10} \\
\Or{print *, "At = ", rec(Re(iAt),iVar), rec(Im(iAt),iVar)}
\end{mytt}%
The `current value' field (\Code{iVar}) should not be set explicitly, as 
it is updated automatically by \Code{FHLoopRecord}.

\subsubsection{Looping over a Record / Setting the {\tt FeynHiggs} input}

The loops over parameters contained in a Record are worked off through 
calls to \Code{FHLoopRecord}, which update the Record's `current value' 
fields (\Code{iVar}).  \Code{FHLoopRecord} is thus usually invoked in 
the context of a looping construct, such as
\begin{verbatim}
   call FHLoopRecord(error, rec)
   do while( error .eq. 0 )
     ...
     call FHLoopRecord(error, rec)
   enddo
\end{verbatim}
The subroutine \Code{FHSetRecord} can be used to set the `current
value' fields (\Code{iVar}) as input parameters for FeynHiggs.  This
works effectively as a combination of \Code{FHSetPara}, \Code{FHSetCKM},
and \Code{FHSetNMFV}, except that the parameters are taken from the
Record.  In a typical application the above loop would be extended to
\begin{verbatim}
   call FHLoopRecord(error, rec)
   do while( error .eq. 0 )
     call FHSetRecord(error, rec, 1D0)
     if( error .ne. 0 ) stop
     call FHHiggsCorr(error, MHiggs, SAeff, UHiggs, ZHiggs)
     if( error .ne. 0 ) stop
     ...
     call FHLoopRecord(error, rec)
   enddo
\end{verbatim}
The third argument in {\tt FHSetRecord} is the same scale factor which
appears in {\tt FHSetPara} and which determines the renormalization
scale as a multiple of the top mass.

\subsubsection{Associating a Record with a Table}

The FeynHiggs Record allows one to interpolate parameters from a data table. 
The table is interpolated in two user-selectable variables which can be 
chosen identical if interpolation in only one variable is desired.

The table first needs to be loaded into internal storage.  At the moment
FeynHiggs has a static allocation for one table of at most 2400 lines. 
This allows the complete implementation in Fortran and seems sufficient
for all present applications.  The table's format is rather 
straightforward: the first line contains the column names (same 
identifiers as in the FeynHiggs input file), followed by the data rows.  
All items are separated by whitespace.

Loading the table can either be done through the input file and is thus 
automatically performed in \Code{FHReadRecord}.  To this end one has to 
add a line
\begin{verbatim}
   table file var1 var2
\end{verbatim}
to the parameter file.  For example, ``\Code{table mytable TB MA0}''
reads the file \Code{mytable} into memory and sets \Code{TB} and
\Code{MA0} as input variables for the interpolation.  The table must 
obviously contain columns for the input variables.

It is also possible to integrate the table file into the parameter 
file.  The \Code{table} statement then takes the form
\begin{verbatim}
   table - var1 var2
\end{verbatim}
and must be the last statement in the parameter file, followed 
immediately by the table data.

Alternately, the table is loaded by
\begin{verbatim}
   call FHLoadTable(error, "file", 5)
   if( error .ne. 0 ) stop
\end{verbatim}
The table is read from \Code{file}, unless that equals ``\Code{-}'', in
which case the table is read from the Fortran unit given in the third
argument (unit 5 is Fortran's equivalent of stdin and hence a good
default argument here).

The table is associated with the record through
\begin{verbatim}
   call FHTableRecord(error, rec, var1, var2)
   if( error .ne. 0 ) stop
\end{verbatim}
where \Code{var1} and \Code{var2} are the indices of the input 
variables, \Eg \Code{iTB} and \Code{iMA0}.  To translate parameter names 
(strings) into indices, one can use the \Code{FHRecordIndex} subroutine, 
as in:
\begin{verbatim}
   call FHRecordIndex(index, name)
\end{verbatim}

\subsection{Mathematica Use}

Using FeynHiggs Records in Mathematica is for the larger part very 
similar to doing so in Fortran.  The main difference is that one does 
not have to declare a Record.  Rather, both initialization routines 
`create' the Record:
\begin{mytt}
\Ind{rec = FHClearRecord[]} \\
\Or{rec = FHReadRecord["file"]}
\end{mytt}%
The Record is represented as an \Code{FHRecord} object in Mathematica.
Access to fields is very similar to the Fortran case, \Eg
\begin{mytt}
\Ind{rec[[iTB,iLower]] = 10} \\
\Or{Print["At = ", rec[[Re[iAt],iVar]], rec[[Im[iAt],iVar]]\,]}
\end{mytt}%
So is the use of \Code{FHLoopRecord}, except that the updated Record is
returned, rather than modified in situ.  In other words,
\Code{FHLoopRecord} returns an \Code{FHRecord} as long as the loop
continues.  The loop would thus look like
\begin{verbatim}
   While[ Head[rec = FHLoopRecord[rec]] === FHRecord,
     ...
   ]
\end{verbatim}
The other routines are used straightforwardly, for example:
\begin{verbatim}
   FHSetRecord[rec, 1]
   FHLoadTable["file"]
   rec = FHTableRecord[rec, var1, var2]
   index = FHRecordIndex[name]
\end{verbatim}

\subsection{Examples}

\subsubsection{Command-line Mode with Parameter Table}

In the simplest case, a Parameter Table can be processed through an 
input file with a \Code{table} statement:
\begin{verbatim}
   MA0   203
   TB    5.7
   table file.dat MA0 TB
\end{verbatim}
The Parameter Table is read from \Code{file.dat} in a format like
\begin{verbatim}
   MT     MSusy  MA0   TB   At     MUE ...
   171.4  500    200   5    1000   761
   171.4  500    210   5    1000   753
   ...
   171.4  500    200   6    1000   742
   171.4  500    210   6    1000   735
   ...
\end{verbatim}
Alternately, the Table can be integrated into the parameter file, as in
\begin{verbatim}
   MA0   203
   TB    5.7
   table - MA0 TB
   MT     MSusy  MA0   TB   At     MUE ...
   171.4  500    200   5    1000   761
   171.4  500    210   5    1000   753
   ...
   171.4  500    200   6    1000   742
   171.4  500    210   6    1000   735
   ...
\end{verbatim}
This minimal setup assumes that all parameters are contained in the 
table.  More generally, the ones not contained in the table have to be 
given in the parameter file.
The interpolation for the parameters given (here {\tt MA0} and {\tt TB})
is performed automatically by {\tt FeynHiggs}.

\subsubsection{Using a Record with Table in Fortran}

In Fortran, the same example might be coded as
\begin{verbatim}
   program record_test
   implicit none

#include "FHRecord.h"

   RecordDecl(rec)
   integer error
   double precision MHiggs(4)
   double complex SAeff, UHiggs(3,3), ZHiggs(3,3)

   call FHClearRecord(rec)
   rec(iMA0,iLower) = 203
   rec(iTB,iLower) = 5.7

   call FHLoadTable(error, "file.dat", 5)
   if( error .ne. 0 ) stop

   call FHTableRecord(error, rec, iTB, iMA0)
   if( error .ne. 0 ) stop

   call FHSetFlags(4, 0, 0, 3, 0, 2, 1, 1, 3)

   call FHLoopRecord(error, rec)
   do while( error .eq. 0 )
     call FHSetRecord(error, rec, 1D0)
     if( error .ne. 0 ) stop

     call FHHiggsCorr(error, MHiggs, SAeff, UHiggs, ZHiggs)
     if( error .ne. 0 ) stop

     print *, "TB, Mh1 = ", rec(iTB,iVar), MHiggs(1)

     call FHLoopRecord(error, rec)
   enddo
   end
\end{verbatim}

\subsubsection{Using a Record with Table in Mathematica}

In Mathematica, the structure and syntax is very similar to Fortran 
(mainly round brackets have to be converted into square ones):
\begin{verbatim}
Install["MFeynHiggs"]

rec = FHClearRecord[]

rec[[iMA0,iLower]] = 203;
rec[[iTB,iLower]] = 5.7

FHLoadTable["file.dat"]

rec = FHTableRecord[rec, iTB, iMA0]

FHSetFlags[4, 0, 0, 3, 0, 2, 1, 1, 3]

While[ Head[rec = FHLoopRecord[rec]] === FHRecord,
  FHSetRecord[rec, 1];
  res = FHHiggsCorr[];
  Print["TB, Mh1 = ", rec[[iTB,iVar]], (MHiggs /. res)[[1]] ];
]
\end{verbatim}

\end{appendix} 

%%%%%%%%%%%%%%%%%%%%%%%%%%%%%%%%%%%%%%%%%%%%%%%%%%%%%%%%%%%%%%%%%%%%%%%%%%%%%%%
%%%%%%%%%%%%%%%%%%%%%%%%%%%%%%%%%%%%%%%%%%%%%%%%%%%%%%%%%%%%%%%%%%%%%%%%%%%%%%%

\clearpage
\pagebreak
%\newpage

%%%%%%%%%%%%%%%%%%%%%%%%%%%%%%%%%%%%%%%%%%%%%%%%%%%%%%%%%%%%%%%%%%%%%%%%%%%%%%%
%%%%%%%%%%%%%%%%%%%%%%%%%%%%%%%%%%%%%%%%%%%%%%%%%%%%%%%%%%%%%%%%%%%%%%%%%%%%%%%

\end{document}